\documentstyle[epsf,preprint,aps]{revtex}

\begin{document}

\draft

\title{Evidence for Neutrino Oscillations from Muon Decay at Rest}

\author{C. Athanassopoulos$^{12}$, L. B. Auerbach$^{12}$,
R. L. Burman$^7$,\\
I. Cohen$^6$, D. O. Caldwell$^3$, B. D. Dieterle$^{10}$, J. B. Donahue$^7$,
A. M. Eisner$^4$,\\ A. Fazely$^{11}$,
F. J. Federspiel$^7$, G. T. Garvey$^7$, M. Gray$^3$, R. M. Gunasingha$^8$,\\
R. Imlay$^8$, K. Johnston$^{9}$,
H. J. Kim$^8$, W. C. Louis$^7$, R. Majkic$^{12}$,
J. Margulies$^{12}$,\\ K. McIlhany$^{1}$, W. Metcalf$^8$, G. B. Mills$^7$,
R. A. Reeder$^{10}$, V. Sandberg$^7$, D. Smith$^5$,\\
I. Stancu$^{1}$, W. Strossman$^{1}$, R. Tayloe$^7$, G. J. VanDalen$^{1}$,
W. Vernon$^{2,4}$, N. Wadia$^8$,\\ 
J. Waltz$^5$, Y-X. Wang$^4$, D. H. White$^7$, 
D. Works$^{12}$, Y. Xiao$^{12}$, S. Yellin$^3$ \\
LSND Collaboration}
\address{$^{1}$University of California, Riverside, CA 92521}
\address{$^{2}$University of California, San Diego, CA 92093}
\address{$^3$University of California, Santa Barbara, CA 93106}
\address{$^4$University of California
Intercampus Institute for Research at Particle Accelerators,
Stanford, CA 94309}
\address{$^{5}$Embry Riddle Aeronautical University, Prescott, AZ 86301}
\address{$^6$Linfield College, McMinnville, OR 97128}
\address{$^7$Los Alamos National Laboratory, Los Alamos, NM 87545}
\address{$^8$Louisiana State University, Baton Rouge, LA 70803}
\address{$^{9}$Louisiana Tech University, Ruston, LA 71272}
\address{$^{10}$University of New Mexico, Albuquerque, NM 87131}
\address{$^{11}$Southern University, Baton Rouge, LA 70813}
\address{$^{12}$Temple University, Philadelphia, PA 19122}
 
\date{\today}
\maketitle
\begin{abstract}
A search for $\bar\nu_{\mu}
\to \bar\nu_{e}$ oscillations 
has been conducted at the Los Alamos Meson Physics 
Facility using $\bar\nu_{\mu}$ from $\mu^+$ decay at rest. 
The $\bar\nu_e$ are detected via the reaction 
$\bar\nu_e\,p \rightarrow e^{+}\,n$, 
correlated with the 2.2 MeV $\gamma$ from $np\rightarrow d\gamma$.  
The use of tight cuts to identify $e^+$ events with 
correlated $\gamma$ rays 
yields 22 events with $e^+$ energy between 36 and
$60\,{\rm MeV}$ and only $4.6 \pm 0.6$ background events.
The probability that this excess is due entirely to a
statistical fluctuation is $4.1 \times 10^{-8}$.  
A $\chi^2$ fit to the entire $e^+$ sample
results in a total excess of
$51.8 ^{+18.7}_{-16.9} \pm 8.0$
events with $e^+$ energy between 20 and
$60\,{\rm MeV}$. If attributed to $\bar \nu_\mu \rightarrow \bar \nu_e$ 
oscillations, this corresponds
to an oscillation probability (averaged over the experimental
energy and spatial acceptance) of 
($0.31^{+0.11}_{-0.10} \pm 0.05$)\%.
\end{abstract}

\section*{1.\ \ Introduction}%

\subsection*{1.1\ \ Motivation}%

This paper describes the evidence for neutrino oscillations
from the Liquid Scintillator Neutrino Detector (LSND) 
apparatus described in reference \cite{bigpaper1}.
The result of a search for $\bar\nu_\mu\to\bar\nu_e$ 
oscillations has been reported \cite{paper1} for data taken in 1993 and
1994 in this experiment, where an excess of events consistent with
neutrino oscillations was observed.
The purpose of the present paper is to provide details of that analysis which 
can not be covered in a Letter publication.
In addition, data taken in 1995 have been included.
Also, further work has shown ways in which the analysis can be made more 
efficient so that the data sample can be increased, with the result that
the beam excess is now sufficiently large that it cannot be due to a 
statistical fluctuation of the beam-off background. The excess must be
due to neutrino oscillations or to an unknown neutrino 
source or interaction with
a very similar signature.

The existence of neutrino oscillations would imply nonconservation of lepton
family number and different neutrino mass eigenstates.
In the standard model the neutrinos are massless. Observation of
neutrino oscillations would require an extension of the standard
model and could help in leading to a more encompassing theory.
In addition, since there are about $10^2\rm cm^{-3}$ 
neutrinos of each family left
over from the initial expansion of the universe, neutrino mass of even a few
eV would have profound effects on the development of structure in the 
universe.

There are hints of neutrino mass from observations of both solar and 
atmospheric neutrinos.
Solar models predict more neutrinos from the sun than are detected in four 
experiments of three quite different types 
\cite{bahcall,davis,Kamioka,gallium}. Solving this problem solely by adjusting
solar models requires disregarding at least two of the three types of
experiment.
Neutrino oscillations provide a quantitative explanation of this deficit of 
electron neutrinos ($\nu_e$), requiring that the difference in the 
square of the masses ($\Delta m^2$) of the 
neutrinos involved be very small, 
$\mathrel{\raise.3ex\hbox{$<$\kern-.75em\lower1ex\hbox{$\sim$}}}
10^{-5}$eV$^2$ 
from the implied energy
dependence of the deficit.
In the atmospheric neutrino case, three experiments 
find the ratio of muon to electron neutrinos ($\nu_\mu/\nu_e$) produced by 
secondary cosmic ray interactions to be about 60\% of that expected 
\cite{atmosphere,atmoskam,soudan,gaisser}, and 
this can be explained by $\nu_\mu\to\nu_e$ or 
$\nu_\mu\to\nu_\tau$ oscillations
with large mixing. One of these experiments \cite{atmoskam} infers a 
$\Delta m^2$ of $\sim 10^{-2}$ eV$^2$.
However, the ($\nu_{\mu}/\nu_e$) ratio observed by the three experiments
can be explained by larger values of $\Delta m^2$.

This experiment deals with a range of $\Delta m^2$ values that
is much larger than
that applied to the solar neutrino case.
It is perhaps possible \cite{fuller} to explain both the
atmospheric neutrino effect and this LSND result by the same $\Delta m^2$.
Although this paper reports strong evidence for neutrino oscillations,
more experimental data will be needed to firmly establish the existence of 
neutrino oscillations and to clarify any relationship among these
several indications of oscillations.

\subsection*{1.2\ \ Experimental Method}%

LSND was designed to detect
neutrinos originating in a proton target and beam stop at the Los Alamos
Meson Physics Facility (LAMPF), and to search specifically for
both $\bar\nu_{\mu}\to\bar\nu_e$ and $\nu_{\mu}\to\nu_e$ transitions
with high sensitivity. 
This paper focuses on the first of these two complementary searches.
The neutrino source and detector are described in detail in \cite{bigpaper1}.
First results on the $\bar\nu_e$ search have been reported in \cite{paper1},
using data collected in 1993 and 1994.

For the experimental strategy to be successful, the beam stop is required 
to be a copious source of $\bar\nu_{\mu}$, while producing relatively few 
$\bar\nu_e$ 
by conventional means 
in the energy range of interest. 
The detector must be able to recognize interactions of $\bar\nu_e$ with 
precision and separate them from other neutrino types, including a
large expected flux of $\nu_e$.
The observation of $\bar\nu_e$ in excess of the number expected from 
conventional sources is interpreted as evidence for neutrino 
oscillations. 
However, although in this paper 
we will concentrate on the oscillation hypothesis,
it must be noted that any exotic process that creates $\bar\nu_e$ 
either at production, in flight, or in detection can produce a 
positive signal in this search.
Lepton number violation in muon decay,  
$\mu^{+}\to e^{+}+\bar\nu_e+\nu_x $, is a good example and would
require an extension of the standard model. 

The high flux of protons on the water target produced pions copiously
\cite{bigpaper1}.
Most of the positive pions came to rest and decayed through the sequence
$$\pi^{+}\to\mu^{+}+\nu_{\mu}\ \ ,$$
$$\mu^{+}\to e^{+}+\nu_e+\bar\nu_{\mu}\ \ ,$$
supplying $\bar\nu_{\mu}$ with a maximum energy of $52.8\,{\rm MeV}$.
The energy dependence of the $\bar\nu_{\mu}$ flux from decay at rest (DAR) is 
very well known, and the absolute value is known 
to 7\% \cite{bigpaper1,burman}.
The open space around the target is short compared to the pion decay
length, so only a small fraction of the $\pi^+$ (3.4\%) decay in flight (DIF) 
through the first reaction.
A much smaller fraction (approximately 0.001\%) of the muons DIF, 
due to the difference in lifetimes and the fact that a $\pi^+$ must first 
DIF.

The symmetrical chain starting with $\pi^-$ might 
lead to an intolerable number of $\bar\nu_e$, but three factors result
in a large suppression of this background.
First, for the LAMPF proton beam and beam stop configuration, 
positive pion production exceeds that 
of negative pions by a factor of about eight.  Second,
negative pions which come to rest in the beam stop are captured 
through strong interactions before they can decay, so only 
the 5\% which DIF can contribute to a $\bar\nu_e$ 
background. (Note that $5\%$ of $\pi^-$ and $3.4\%$ of $\pi^+$ produced
in the beam stop decay in flight.)
Third, virtually all of the negative muons arising from such pion 
DIF come to rest in the beam stop before decaying.
Most are captured from atomic orbit, a process which leads to
a $\nu_{\mu}$ but no $\bar\nu_e$, leaving only 12\% of them to decay 
into $\bar\nu_e$.
Hence one can estimate the relative yield, compared to the positive 
channel, to be $\sim (1/8) * 0.05 * 0.12 \approx 7.5 \times 10^{-4}$.
Thus, it is expected that $\bar{\nu}_{e}$ are present only at
this level in
the isotropic flux of neutrinos from the source. A detailed Monte
Carlo simulation \cite{burman} 
gives a value of $7.8 \times 10^{-4}$
for the ratio of $\bar \nu_e$ from $\mu^-$ DAR to $\bar \nu_{\mu}$
from $\mu^+$ DAR.

It is, however, necessary to deal with the DAR $\nu_e$ produced 
one-for-one with the desired $\bar\nu_{\mu}$.
Although it is not possible to distinguish an $e^-$ from an $e^+$,
the key to rejecting these $\nu_e$ as a background to the $\bar\nu_e$ search
is the presence of free protons (hydrogen) in the detector.
LSND detects $\bar\nu_e$ via
$$\bar\nu_e + p \to e^+ + n\ \ \ ,$$
a process with a well-known cross section \cite{vogel},
followed by the neutron-capture reaction
$$n + p \to d + \gamma (2.2\,{\rm MeV})\ \ \ .$$
Thus the detection signature consists of an ``electron'' signal, followed
by a $2.2\,{\rm MeV}$ photon correlated with the electron signal in both 
position and time.
Detection of $\nu_e$ in LSND is dominated by charged current 
reactions on $^{12}C$.
But an electron from
$\nu_e ^{12}C \to e^- ~^{12}N$ 
with a DAR $\nu_e$ has energy $E_e < 36\,{\rm MeV}$
because of the mass difference of $^{12}C$ and $^{12}N$.
Moreover, the
production of a correlated photon via $\nu_e ^{12}C \to e^- n ^{11}N$ 
can likewise
occur only for $E_e < 20\,{\rm MeV}$ because of the threshold for free
neutron production.
Hence, the $\nu_e$ background is greatly suppressed
by neutron detection for $E_e > 20\,{\rm MeV}$.
In addition, the requirement of
a minimum $e^\pm$ energy of $36\,{\rm MeV}$ eliminates most 
of the $\nu_e$ background due to an {\sl accidental} coincidence with an 
uncorrelated $\gamma$ signal.

The detector is a tank filled with 167 metric tons of dilute liquid 
scintillator, located about $30\,{\rm m}$ from the neutrino source, and 
surrounded
on all sides except the bottom by a liquid scintillator veto shield.
The dilute mixture allows the detection in photomultiplier tubes (PMTs) 
of both 
$\check{\rm C}$erenkov light and isotropic scintillation light, so that 
reconstruction software
provides robust particle identification (PID) for $e^\pm$ along with the 
event vertex and direction.
The electronics and data acquisition systems are designed to detect related
events separated in time.
This is necessary both for neutrino induced reactions and for cosmic 
ray backgrounds.
The response of the detector in the energy range for the 
$\bar\nu_{\mu}\to\bar\nu_e$ search is calibrated using a large sample of 
Michel 
$e^\pm$ from the decays of stopped cosmic ray muons. The response to
$2.2\,{\rm MeV}$ photons is understood by studying the capture of cosmic 
ray neutrons.

Despite $\sim 2\,$kg/{\rm cm}$^2$ of ``overburden'' shielding above the 
detector,
there remains a very large background to the
oscillation search due to cosmic rays, which needs to be suppressed by 
about nine orders of magnitude to reach a sensitivity limited by the 
neutrino source itself.
The cosmic ray muon rate through the tank is $\sim 4$\,{\rm kHz}, of which 
$\sim 10\%$ stop and decay in the scintillator, whereas even if every
$\bar \nu_{\mu}$ oscillated to $\bar \nu_e$, the total rate of
$\bar \nu_e p$ interactions in the entire tank would be $<0.01$ Hz.
There are five lines of attack in removing this background.
First, an in-time veto rejects muons, but decay $e^\pm$ remain,
along with neutrons and a small fraction of unvetoed muons due to veto shield 
inefficiency.
Second, these $e^\pm$ are greatly reduced by imposing a veto on any event
that occurs soon after a specific number of PMT hits in the detector or veto 
shield.
A trigger threshold at 7 muon lifetimes is increased in analysis to as
much as 18 muon lifetimes.
Third, cosmic ray-induced neutrons are strongly suppressed by use of $e^\pm$ 
PID criteria, based upon timing, vertex, and direction information from the 
detector. 
Fourth, the requirement of a correlated capture $\gamma$ 
discriminates  against cosmic ray particles other than neutrons.
Fifth, the level of remaining cosmic ray background is very well measured 
because 
about 14 times as much data are collected when the beam is off as on.
The result of these procedures is to reduce cosmic ray particles 
to a small background for the DAR oscillation search.

\subsection*{1.3\ \ Outline of This Paper}%

We present a brief description of the 
detector system and data collection in chapter two.
Chapter three describes the methodology of identifying 2.2 MeV $\gamma$s 
associated with neutron capture on free protons.
Chapter four describes event selection and acceptance.
Chapter five contains an assessment of neutrino backgrounds.
Distributions of data are shown in chapter six, and fits to the data are 
discussed in chapter seven.
An interpretation of the data in terms of neutrino oscillations
is given in chapter eight, together with
a comparison with other neutrino oscillation experiments.

\section* {2. \ \ Detector and Data Collection}%

\subsection*{2.1\ \ Overview}%

Reference \cite{bigpaper1} contains a detailed description of the neutrino 
source and detector and a discussion of detector performance.
Here the detector is described briefly in section 2.2 and the veto 
shield in section 2.3.

\subsection*{2.2\ \ Detector and Data Collection}%

This experiment is carried out at 
LAMPF using 800$\,{\rm MeV}$ protons from the linear accelerator.
Pions were produced from 14772 Coulombs 
of proton beam at the primary beam stop 
over three years of operation between 1993 and 1995. There were 1787 Coulombs
in 1993, 5904 Coulombs in 1994, and 7081 Coulombs in 1995.
The fraction of the total
DAR neutrino flux produced
in each of the three years was 12\% in 1993, 42\% in 1994,
and 46\% in 1995 and varied slightly from the Coulomb fractions due to
small variations in the beam stop geometry. 
The duty ratio is defined to be the ratio of data
collected with beam on to that with beam off. It
averaged 0.070 for the entire data
sample, and was 0.076, 0.080, and 0.060 for the years 1993, 1994,
and 1995, respectively.
The primary beam stop consists of a 30 cm water target surrounded by 
steel shielding and followed by a copper beam dump. 
The DAR neutrino flux varies approximately as $r^{-2}$ from the average
neutrino production point, where $r$ is the distance
traveled by the neutrino.
The detector is located 30 m from this main production target, while two 
thinner subsidiary targets are located approximately 75 m and
100 m farther upstream.

The detector is a roughly cylindrical tank containing 167 tons of liquid
scintillator and viewed by 1220 uniformly spaced $8 ''$ Hamamatsu
PMTs.
The digitized time and pulse height of each of 
these PMTs (and each of the 292 
veto shield PMTs) are 
recorded when the deposited energy in the tank exceeds a threshold of 
about $4\,{\rm MeV}$ (electron equivalent energy) with less than 4 veto PMT 
hits.
Activity in the detector or veto shield during the 51.2 $\mu s$ preceding
a primary trigger is also recorded,
provided there are $>17$ detector PMT hits or $>5$ veto PMT hits.
Data after the primary are recorded for 1 ms with a threshold of 
about $0.7 \,{\rm MeV}$.
The detector operates without reference to the beam spill, but 
the state of the beam is 
recorded with the event. Approximately $93\%$
of the data is taken between beam spills. This allows an accurate
measurement and subtraction of cosmic ray background surviving the event
selection criteria.
The detector scintillator consists of mineral oil ($CH_2$) in which is 
dissolved a small concentration (0.031 g/l) of b-PBD 
\cite {reeder}.
This mixture allows the separation of $\check{\rm C}$erenkov light and 
scintillation light
and produces about 33 photoelectrons per MeV of electron energy deposited
in the oil.
The combination of the two sources of light provides direction information
and makes PID 
possible for relativistic particles.
Identification of neutrons is accomplished through the detection of 
the $2.2\,{\rm MeV}$ 
$\gamma$ from neutron capture on free protons. Note that the oil consists
almost entirely of carbon and hydrogen. The fractional mass of oxygen
and nitrogen in the oil from the b-PBD (0.031 g/l of $C_{24}H_{22}N_2O$)
and the vitamin E added as a preservative 
(0.010 g/l of $C_{19}H_{28}O$) is about 
$2\times 10^{-6}$ and $3\times 10^{-6}$, respectively.   
Also, nitrogen is bubbled through the oil continually to remove
oxygen that can decrease the oil's attenuation length. However, the
fractional mass due to this nitrogen is $<<10^{-6}$.

\subsection*{2.3\ \ Veto Shield}%

The veto shield encloses the detector on all sides except the bottom.
Additional counters were placed below the veto shield 
after the 1993 run to reduce cosmic 
ray background entering through the bottom support structure.
The main veto shield \cite{veto} consists of a 15-cm layer of liquid 
scintillator in an external tank and 15 cm of lead shot in an internal 
tank.
This combination of active and passive shielding
tags cosmic ray muons that stop in the lead shot.
A veto inefficiency $<10^{-5}$ is achieved with this detector
for incident charged particles. The veto inefficiency is larger for
incident cosmic-ray neutrons.

\section*{3.\ \ Correlated Photons from Neutron Capture}%

\subsection*{3.1\ \ Overview}%

The performance of the detector in the detection of 
2.2 $\,{\rm MeV}$ $\gamma$s 
from neutron capture on free protons is discussed in this chapter.
Neutrons produced in the reaction $\bar \nu_e p \rightarrow e^+ n$ 
are identified by detection of the subsequent 2.2 MeV 
$\gamma$ from the reaction $n+p\to d+\gamma$.
These recoil neutrons are produced with kinetic energy in the 0 
to 5.2 $\,{\rm MeV}$ energy
range and typically travel about $10$ cm before capture.
The expected mean capture time of 186 $ \mu s$ is essentially independent of
the initial neutron energy because the time taken for the neutron to degrade 
to less than $1\,{\rm MeV}$ is small compared to 186 $\mu s$.
The detector performance is measured empirically from a large 
sample of cosmic 
ray neutron events which appear in the main detector
and are discussed in section 3.2.
The energy and position reconstruction of 2.2 $\,{\rm MeV}$ $\gamma$s is
discussed in section 3.3.
Properties of 2.2 $\,{\rm MeV}$ $\gamma$ candidates and accidental $\gamma$
background are discussed in section 3.4. 
A Monte Carlo simulation for correlated low-energy neutrons is discussed
in section 3.5.
The likelihood parameter that is used to separate correlated and accidental 
$\gamma$s is described in section 3.6.

\subsection*{3.2\ \ Cosmic Ray Neutron Sample}%

A cosmic ray neutron sample is obtained with the following selection 
criteria: deposited electron equivalent energy between 36 and 60 MeV, 
PID consistent with a neutron 
(satisfying $\chi_{tot} > 0.8$ and $\chi_r <0.75$, to reduce events
with multiple neutrons, where $\chi_{tot}$ and $\chi_r$ are described
in section 4.2), less than 4 veto hits within the 0.5 $\mu s$
event window, beam off,
and at least one triggered $\gamma$ event within 1 ms of the primary event.
Charged particles below $\check{\rm C}$erenkov threshold produce less light 
per energy
deposited than do $\beta \sim 1$ electrons. 
Also, neutrons deposit
much of their energy by scattering from protons and nuclei.
The energy scale used in this
paper is
based on the light-to-energy ratio for electrons.
Fig. 1 shows the time difference between neutrons that satisfy
the above criteria and a subsequent $\gamma$ with 21 to 50 hit PMTs.
The distribution is fit to a sum of an exponential for correlated 
$\gamma$s and a flat background for accidental $\gamma$s (solid
curve).
The fitted time constant of $188 \pm 3$ $\mu$s agrees well with the
186 $\mu s$ capture time for neutrons in mineral oil.
The $\gamma$s in the last 250 $\mu s$ of the 1 ms window are almost entirely 
accidental $\gamma$s and are used to define the 
characteristics of the ``accidental $\gamma$'' sample.
Similarly, a ``correlated $\gamma$'' sample is defined to contain a
$\gamma$ in the first 
250 $\mu s$ of the 1 ms window after subtraction of the accidental $\gamma$
contribution (see section 3.4).

\subsection*{3.3\ \  Gamma Reconstruction Algorithm}%

Activities with 21 to 50 hit PMTs, with average charge per PMT hit greater
than 0.8 photoelectrons, and which occur within 1 ms of the
primary event are defined to be $\gamma$ candidates and are fit for
position with a special reconstruction algorithm.
The algorithm defines the $\gamma$ position to be
the average of the 
position of all hit PMTs weighted by the pulse height of each
PMT.
This algorithm, although simple, results in a position error which
is comparable to (or better than) some more elaborate methods
(see section 3.4).

\subsection*{3.4\ \  Properties of Photon Candidates}%

\subsubsection*{3.4.1\ \ Correlated and Accidental Photons}%

In Fig. 2 is shown the observed distributions of photons from the
``correlated $\gamma$'' (solid curve) and ``accidental $\gamma$'' 
(dashed curve)
samples.
The distributions are: (a) the time of the $\gamma$ after the primary;
(b) the number of photon PMT hits; (c) the 
distance of the reconstructed $\gamma$ from the primary.

These three ``correlated $\gamma$''
distributions are found to be approximately independent of the primary event
location in the fiducial volume. 
As expected from the uniformity of the oil, 
there is no correlation between the neutron capture time 
and the other two variables.
Furthermore, the number of photon PMT hits is observed to be independent of 
distance from the primary, except for a small correlation for distances
beyond 2 m. 
Events with fewer PMT hits have a slightly broader distance distribution,  
which is expected because the position correlation of the $\gamma$ and primary
vertex is dominated by reconstruction errors. 
However, the observed distance distribution from the cosmic ray neutron
sample is broader than expected for neutrons from the reaction 
$\bar\nu_e + p \to e^+ + n$ 
because the primary arises from an initial neutron interaction of higher 
energy and thus travels slightly farther before reaching thermal energies.
Monte Carlo studies (see section 3.5)
indicate that the mean measured distance distribution is up to 
$20$ cm larger on average than for the low-energy neutrons of interest.

The dependence of the three ``accidental $\gamma$'' distributions on
primary event locations was also investigated. For this study the
reconstructed $\gamma$ position was required to be within 2.5 m of the
primary vertex, a criterion imposed in the $\gamma$ identification
procedure described later. The three distributions are uniform over
the fiducial volume except near the bottom, upstream corner of the
detector (see section 3.4.2), 
where there is a higher rate of accidental $\gamma$s. 
For primary events in this region, both the number of photon
PMT hits
and the distance distribution have lower average values 
than elsewhere in
the detector.

\subsubsection*{3.4.2\ \ Spatial Distributions of Accidental Photons}%

The reconstructed position for 
accidental $\gamma$s in the X - Z and Y - Z projections is shown
in Fig. 3. The coordinate system 
is defined such that Y is pointing up in the vertical direction and
Z is pointing downstream along the cylindrical axis of the detector.
These distributions are non-uniform and show a concentration near the 
upstream, bottom portion of the detector.
This concentration may be due to steel shielding underneath the
detector with a high level of radioactivity or to a cable penetration 
though the veto system in that region.
This non-uniformity is taken into account in the fit analyses of chapter 7.
The average accidental $\gamma$ rate over the entire detector is 
$1.07 \pm 0.01$ kHz in 1993, $1.19 \pm 0.01$ kHz
in 1994, and $1.14 \pm 0.01$ kHz in 1995. 

\subsection*{3.5\ \  Monte Carlo Simulation of Photons From Neutron Capture}%

Cosmic ray neutrons selected are of higher energy than those from
the neutrino oscillation reaction $\bar \nu_e p \rightarrow e^+ n$.
Thus, the distance of the reconstructed photon from the 
primary is on average shorter 
for the neutrino oscillation reaction than it is for cosmic ray neutrons.
In order to compute the expected distance distribution between
the reconstructed
$e^+$ and the 2.2 $\,{\rm MeV}$ $\gamma$, three Monte Carlo distance 
distributions were  
used.
(1) Positrons of the expected energy distribution were generated and passed 
through the Monte Carlo detector simulation \cite{bigpaper1,lsndmc}
and reconstruction to find the distribution of
distances between the $e^+$ point of origin and reconstructed position.
(2) A separate Monte Carlo program designed to track low energy neutrons was  
 used to find the distribution in distance between neutron production and 
capture.
This program simulated elastic scattering from the carbon and hydrogen atoms 
according to tabulated neutron cross sections.
Neutrons were tracked even after they have thermalized, at which point it 
becomes important that neutron absorption with resulting $\gamma$ production 
on 
hydrogen and carbon was also included.
(3) The detector simulation is used to 
simulate scintillation light produced 
by the 2.2 MeV $\gamma$.
Two extra single photoelectron hits were randomly added to the hit 
PMTs to simulate PMT noise, which is based on the average PMT noise rate
of about 3 kHz.
The photon reconstruction algorithm described in section 3.3 was used to 
compute the $\gamma$ position, from which the distance between the 
generated and reconstructed photon is obtained.

The expected distribution in distance between the reconstructed $e^+$ and 
the 2.2 $\,{\rm MeV}$ $\gamma$ is the convolution of these three 
distributions and is 
shown as the solid histogram in Fig. 4.
This distribution is dominated by reconstruction errors in the $\gamma$ 
position.
The travel distance of low energy neutrons, 
$e^+$ reconstruction position error, 
and PMT noise contribute little to the overall distance 
distribution. Hence, the distribution is narrower, as expected,
but not vastly different from that
obtained in section 3.4 from cosmic ray neutrons, shown as the
dashed histogram
in Fig. 4. We use both distributions for the fits described in 
chapter 7 and obtain similar results.

\subsection*{3.6\ \  Photon Identification Parameter (R)}%

The three ``correlated $\gamma$'' 
distributions in Fig. 2
are used to determine the likelihood, ${\cal L}_c$, that the $\gamma$ is 
correlated 
with the primary event. Similarly, the three ``accidental $\gamma$''
distributions in Fig. 2
are used to determine the likelihood, ${\cal L}_a$, that the $\gamma$ is 
accidental
and uncorrelated with the primary event.
Each likelihood, therefore, is the product of the three probability
densities, ${\cal L} = P(hits)\times P(\Delta r) \times P(\Delta t)$.
A likelihood ratio, $R$, for the event is then defined as the ratio
of these likelihoods, $R \equiv {\cal L}_c/{\cal L}_a$.
Because of the small correlations described 
in section 3.4.1 and the adjustment to the $\Delta r$ distribution
discussed in section 3.5, these ${\cal L}$s are only approximate likelihoods.
Moreover, $R$ does not allow for the variation of accidental rates with
the position of the primary particle. Nonetheless, $R$ is a very
powerful tool for separating correlated from uncorrelated $\gamma$s,
and the $\Delta r$ and rate effects are fully allowed for in the fitting
procedures to be described later in this paper.

Fig. 5 shows the measured $R$ distribution for events
with the $\gamma$ correlated (solid) and uncorrelated (dashed)
with the primary event. 
As expected, the uncorrelated events are concentrated at low values of $R$.
For events with multiple $\gamma$s, the $\gamma$ 
with the maximum R is used. R
is set to 0 for events without a $\gamma$ that reconstructs within 2.5 m of 
the 
primary, has 21 to 50 PMT hits, and occurs within 1 ms of the primary event.
The definition of $R$ is always based on the spectra
of Fig. 2, using the $\Delta r$ distribution measured from cosmic
ray neutrons.  However, if $\Delta r$ for correlated photons is
actually distributed as given by the Monte Carlo calculation of
Section 3.5, then the distribution of $R$ for correlated
photons will be given by the dotted curve in Fig. 5 instead of the
solid histogram.  Both versions are tried for the fits to be described
in Section 7.1.  It should also be noted that while the accidental
photon spectrum shown in Fig. 5 is averaged over primary event
locations in the fiducial volume, those fits actually use 
a spectrum which takes the local accidental rate into account.

The efficiency for producing and detecting a 2.2 $\,{\rm MeV}$ correlated 
$\gamma$ within 2.5 m, with 21 to 50 PMT hits, and within 1 ms 
was determined to be $63\pm 4\%$ (using the solid curve of Fig. 5. 
This efficiency is the product of the 
probability that the $\gamma$ trigger is  
not vetoed by a veto shield signal within the previous
15.2 $\mu s$ ($82 \pm 1\%$), the data acquisition livetime ($94 \pm 3\%$, 
lower for $\gamma$s than for primary events), the requirement that the 
$\gamma$
occurs between 8 $\mu s$ and 1000 $\mu s$ after the primary event 
($95\pm 1\%$),
the requirement that the $\gamma$ has between 21 and 50 hit PMTs 
($90\pm 4\%$),
and the requirement that the $\gamma$ reconstructs within 2.5 m of the
primary event ($96 \pm 2\%$).
>From the cosmic-ray Michel electron
data, the average probability of finding an accidental 
uncorrelated $\gamma$ within the same cuts is $28 \pm 2\%$.
The $R$ distributions shown in Fig. 5 are then used to determine 
the efficiencies for finding a correlated or uncorrelated $\gamma$ satisfying 
a particular $R$ criterion.
For example, the efficiency that an accidental $\gamma$ satisfies 
$R>30$ ($1.5$) 
is $0.6\%$
($9.0\%$), while the efficiency for a correlated $\gamma$ is 
$23\%$ ($58\%$).
The accidental rate depends on the position of the primary event within the 
fiducial volume, as seen in Fig. 3. However, Fig. 6
shows that the R distributions are almost identical for $R>0$ 
in each of the four quadrants
of the Y - Z plane for correlated $\gamma$s (solid) and accidental $\gamma$s
(dashed). 

\section*{4.\ \ Event Selection and Efficiency}%

\subsection*{4.1\ \ Overview}%

The signature for the principal oscillation search is two-fold -- a positron
and a correlated 2.2 $\,{\rm MeV}$ $\gamma$.
The analysis is performed for two ranges of positron energy. In order to
establish the presence of an excess, the positron is required to be
in the energy range $36<E_e<60$ MeV, where the known neutrino backgrounds
are small. A looser energy requirement, $20<E_e<60$ MeV, provides a larger 
range of $L/E_{\nu}$ and is used to determine the oscillation probability
and the $\Delta m^2$ vs $\sin^2 2\theta$ allowed range.
Isolation of an oscillation signal in this experiment thus consists of 
PID of the positron from the reaction $\bar \nu_e p \rightarrow e^+ n$ 
(without distinguishing between positrons and electrons)
and positive identification of the associated neutron by the presence
 of a correlated 2.2 $\,{\rm MeV}$ $\gamma$ from the reaction 
$n p \rightarrow d \gamma$.
Backgrounds then fall into three categories. Two of them are  
beam-related, the first involving events which include a primary
particle identified as an $e^\pm$ plus a correlated neutron-capture
signal, and the second involving events with an accidental $\gamma$
signal instead of a correlated neutron. The largest category of
backgrounds is from beam-unrelated
cosmic rays. While the latter are eventually
subtracted statistically using beam-off data, the strategy for
positron selection is to reduce these backgrounds to a low level before
making the subtraction. These positron selection criteria are
described in this chapter. The tools for selecting associated neutrons
have been presented in chapter 3 and are applied to event selection in
chapters 6 and 7.

\subsection*{4.2\ \ Positron Selection}%

The positron selection criteria and efficiencies are summarized in Table 
I for two different selections. Selection I is identical to what
has been used previously \cite{paper1}, while selection VI makes use of
additional criteria which reduce the beam-off background
and increase the acceptance. Selections II - V are variations of selections I
and VI and
are discussed at the end of the chapter.

To establish an event excess, 
positrons are required to have an energy in the $36<E_e<60$ MeV range.
The narrow energy range is chosen, as shown in Fig. 7, 
because it is  
above the $\nu_e ~^{12}\mbox{C} \rightarrow e^- ~^{12}\mbox{N}$ endpoint 
and in the range expected for oscillation events.

The primary particle is required to have
a PID consistent with a positron. 
Particles with velocities well above $\check{\rm C}$erenkov threshold are 
separated from
particles below $\check{\rm C}$erenkov threshold by making use of the four 
$\chi$ parameters
defined in reference \cite{bigpaper1}. Briefly, $\chi_r$ and $\chi_a$ are
the quantities minimized for the determination of the event position and
direction, $\chi_t$
is the fraction of PMT hits that occur more than 12 ns after the fitted
event time, and $\chi_{tot}$ is proportional to the product of $\chi_r$,
$\chi_a$, and $\chi_t$.
Fig. 8 shows the four $\chi$ parameters for samples of Michel 
electrons (solid) and cosmic-ray neutrons (dashed) 
with electron energies in the $36<E_e<60$ MeV range. For a neutron, $E_e$ is
the equivalent electron energy corresponding to the observed total charge.
The Michel electrons are identified by their correlation with a parent muon,
while the neutrons are identified by their correlation with a 
2.2 $\,{\rm MeV}$
$\gamma$ from $np$ capture.
By requiring that the $\chi$ parameters satisfy $0.3<\chi_{tot}<0.66$,
$\chi_r<0.61$, $\chi_a<0.20$, and $\chi_t<0.26$
($0.3<\chi_{tot}<0.65$,
$\chi_r<0.60$, $\chi_a<0.19$, and $\chi_t<0.25$ for selection I),
optimal separation is obtained between electrons and particles below 
$\check{\rm C}$erenkov threshold.
(For example, neutrons are reduced by a factor of $\sim 10^3$.)
The lower limit on $\chi_{tot}$ is imposed to eliminate any 
laser calibration events that are not correctly identified.
The overall PID efficiencies for positrons in the $36<E_e<60$ MeV
energy range are $0.77 \pm 0.02$
and $0.84 \pm 0.02$ for selections I and VI, respectively. The PID
efficiencies increase with energy, as shown in Fig. 9.
The PID efficiency in the
$20<E_e<36$ MeV energy range is $0.62 \pm 0.02$ for selection VI. 
There is some variation of PID efficiency with position in the detector,
and the efficiencies above are averaged over the detector fiducial volume.

In order to eliminate Michel electrons from muon decay,
the time to the previous triggered event, $\Delta t_p$, is required to be
greater than 40 $\mu s$ for selection I and 
greater than 20 $\mu s$ for selection VI. For selection VI,
all activities between 20 $\mu s$ and
34 $\mu s$ before the event trigger time are required to be
uncorrelated with the positron by having fewer than 50 PMT hits and a
reconstructed position more than 2m from the positron position.
Fig. 10 shows the $\Delta t_p$ 
distribution of beam-off events that 
satisfy the other 
positron selection criteria for (a) events with no $\Delta t_p$ 
requirement and (b) events after imposing the above criteria for
no correlated activities within 34 $\mu$s.
Note the reduction in the beam-off events shown in the figure
between 20 and 34 $\mu$s.
The locations of the 20 $\mu s$ and 34 $\mu s$ requirements are shown on
the figure. Note that the 20 $\mu s$ requirement, corresponding to 10
$\mu^-$ lifetimes and 9 $\mu^+$ lifetimes in oil, 
allows a negligible amount of background from
$\nu_{\mu} C \rightarrow \mu^- X$ scattering. The remaining 
small cosmic ray
background after these cuts 
is eliminated by beam on-off subtraction. The selection I and
VI efficiencies are $0.50 \pm 0.02$ and $0.68 \pm 0.02$, respectively.

It is required that the number of veto shield hits associated
with the events is less than $2$ for selection I ($0.84 \pm 0.02$ efficiency)
and less than $4$ (the hardware trigger requirement) for selection VI 
($0.98 \pm
0.01$ efficiency) to reduce cosmic ray backgrounds.

The reconstructed positron location is required to be a distance $D$ of
at least 35 cm from the surface tangent to the faces of the PMTs.  
This
cut provides assurance that the positron is in a region of the tank
in which the energy and PID responses vary smoothly and are well
understood; charge response, energy resolution and PID efficiencies all
degrade near and behind the PMTs.
(For the 1993 data a 40 cm requirement is used due to the absence of
additional veto counters placed below the veto shield.)
Fig. 11 shows that, for Michel electrons generated
behind the PMT surface by the Monte Carlo simulation, 
no more than $\approx 1\%$ are reconstructed with
$D > 35$ cm and with more than 150 PMT hits.  This results in a negligible
background of $\nu_{\mu} C$ scattering events in which the muon is missed 
because
it is behind the PMT surface.  The 35 cm cut also avoids the region of
the tank with the highest cosmic ray background, thus reducing the statistical
error from having to subtract that background.

The time to any subsequent triggered event is required to be $>8$ $\mu s$
to remove events that are muons that decay. (A high energy muon above 
$\check{\rm C}$erenkov 
threshold has a small probability for satisfying the PID criteria.) By
requiring no subsequent event within four $\mu^-$ lifetimes, this
background is almost completely eliminated.

To further suppress cosmic ray neutrons, the number of associated $\gamma$s
with $R>1.5$ (see chapter 3) is required to be less than 
$3$ for selection I ($0.99 \pm
0.01$ efficiency) and less than $2$ for selection VI ($0.94 \pm 0.01$ 
efficiency).
Cosmic ray neutrons that enter the detector often produce one or 
more additional neutrons, while recoil neutrons from the $\bar \nu_e 
p \rightarrow e^+ n$ reaction are too low in energy to knock out additional 
neutrons.
Fig. 12 shows the number of associated $\gamma$s with $R>1.5$ 
for beam-off background events of $R>30$ in the 
$36<E_e<60$ MeV energy range with at least one $\gamma$ (dashed), 
compared to the expectation (based on the measured rate of accidental
$\gamma$s in the tank) for oscillation
events (solid).
About $94\%$ of the expected oscillation events and only 
$60\%$ of the beam-off 
background
events with $R>30$ have less than $2$ associated $\gamma$s.

For events that pass the electron selection criteria above, 
beam-off data are different from the expected neutrino interaction 
signal in two respects. 
The first of these is the distribution of $\vec{r} \cdot \hat{dr}$,
where  $\vec{r}$ is the location of the reconstructed event
with respect to the center of the tank, and  $\hat{dr}$ is the 
unit direction of the
event in the same coordinate system.  
This scalar product gives large negative values for events near the edge 
of the fiducial volume that head toward the center of the tank.  In the
dashed line of
Fig. 13, the $\vec{r} \cdot \hat{dr}$ distribution for the
beam-off sample is shown.  As expected for events originating outside
the fiducial volume, the distribution is peaked at large negative values.
For neutrino
events on the other hand, the distribution is much more 
symmetric about the origin. This is illustrated by the solid line of
Fig. 13,
which shows the  $\vec{r} \cdot \hat{dr}$ distribution for a 
sample of $\nu_{e} C$ scattering events. (Note that 
$\vec{r} \cdot \hat{dr}$ does not depend on energy.)
 
The number of hits in the veto system
is also observed to be different from that expected 
from signal. 
The number of veto hits in the beam-off sample is displayed in the dashed line
of Fig. 
14, while the number expected (from accidental coincidences)
in the signal is shown in the solid line. (This last distribution is
measured by looking at the number of veto hits in coincidence with 
random firing of the laser flasks \cite{bigpaper1}.)

Using the distributions of these two variables, the
likelihoods ${\cal L}_{off}$ and ${\cal L}_{on}$ are calculated 
that a given event is due to beam-off background
or to beam-on signal, respectively. 
The ratio of these likelihoods ($S={\cal L}_{on}/{\cal L}_{off}$)
is plotted for the $\nu_e C$
and beam-off samples in Fig. 15. (Note that the bias caused by
using the beam-off data sample
for both the S determination and to correct for cosmic-ray background
in the beam-on sample has been shown by Monte Carlo simulations to be
negligible.)  A cut at $S>0.5$ is 87\% efficient for 
neutrino-induced events, while eliminating 33\% of the beam-off
background. This cut is used only for selection VI and 
completes the positron selection criteria.

\subsection*{4.3  Efficiencies of Positron Selection Criteria}%

The efficiencies for selection VI are summarized below.
The efficiency of the PID selection criteria for positrons is measured using 
the Michel electron sample.
The resulting PID selection efficiency is about $84 \pm 2\%$.
The requirement that the time to the previous triggered event is greater than 
20 $\mu s$ and the time to any correlated activity is greater than 34 $\mu s$
has an efficiency of $68 \pm 2 \%$.
The veto shield hit requirement has an efficiency of $98 \pm 1\%$, as
determined from laser calibration events. 
Because all event yield calculations are based upon the number of
target atoms inside the 35 cm fiducial volume cut, an efficiency
correction of $85 \pm 5\%$ is applied to allow for the tendency of
the position reconstruction algorithm to push events toward the PMT
surfaces \cite{bigpaper1}. 
Additional efficiencies result from 
the requirement of no triggered event
within 8 $\mu$s in the future
after the primary event to eliminate muon decays ($99\pm 1\%$),
the requirement of $<2$ associated $\gamma$s with $R>1.5$ ($94\pm 1\%$),
the 
$S>0.5$ requirement ($87 \pm 2\%$), 
and the data acquisition
system livetime ($97\pm 1\%$).
The overall positron selection efficiency 
is $37 \pm 3\%$, and is higher
than the $26 \pm 2\%$ efficiency (see Table I) 
obtained with selection I. 
Selection V is defined to
be the same as selection VI but without the $S>0.5$ requirement, while
 selection IV is defined to be the same as selection V but without 
the $<2$ associated $\gamma$ requirement. Selection II is the same as 
selection I
but with the looser PID criteria, and selection III is the same as 
selection II
but with the looser veto hits less than 4 requirement. Selections II - V have
positron selection efficiencies of $28 \pm 2\%$, $33 \pm 3\%$, $45 \pm 3\%$, 
and $43 \pm 3\%$,
respectively.
 
\section*{5.\ \ Beam-Related Backgrounds}%

\subsection*{5.1\ \ Beam-Related Backgrounds with a Correlated $\gamma$}%

Beam-related backgrounds with neutrons are 
estimated individually in the $36<E_e<60$ MeV energy
range before the correlated $\gamma$ requirement is imposed.
Table II lists the backgrounds in the above energy range for 
$R \ge 0$ (the full positron sample)
and $R>30$, while Table III lists the backgrounds for 
the $20<E_e<60$ MeV energy range. Selection criterion VI, 
defined in chapter 4,
is used, and backgrounds for other selection criteria can be obtained by
multiplying by the relative efficiencies. 
The DAR and DIF neutrino fluxes have been estimated by a 
detailed beam Monte Carlo simulation \cite{burman}. Uncertainties in the
efficiency, cross section, and DIF $\nu$ flux lead to systematic errors
of between 20 and 50\% for the backgrounds discussed below.

\subsubsection*{5.1.1  Neutrons Entering the Detector}%

Despite the amount of shielding between the beam dump and the detector,
one must consider the 
possibility, nonetheless, that neutrons from the target could 
find their way into the tank.
A limit on the beam neutron background relative to the cosmic neutron 
background
is set by looking for a beam-on minus beam-off excess of neutron events in 
the $40 - 180\,{\rm MeV}$ electron equivalent energy range.
This comparison is made by examining neutron candidates which pass
neutron, rather than $e^+$, PID criteria.
For events with $\chi_{tot}>0.75$ and an associated $2.2\,{\rm MeV}\ \gamma$ 
within $1.5\,{\rm m}$ and 
$0.5\,{\rm ms}$, 89700 beam-off events and 6915 beam-on events
are observed in a partial data set with a duty ratio of 0.075,
implying an excess of $187.5 \pm 86.1$ events.
This excess of events is consistent with the $\sim 200$ events expected from 
$\nu C \rightarrow \nu n X$ scattering.  
However, even if the entire excess is interpreted as beam neutrons entering
the tank, fewer than  $187.5/6915 = 3\%$ of the beam-on events
are actually beam-related.  Applying this same ratio for
neutrons {\sl passing\/} the $e^+$ PID criteria, 
the beam-related
neutron background in the $e^+$ sample is less than 0.03 times the
number of beam-unrelated neutrons.  Based upon the R distribution of the
beam-off data sample,
less than $15\%$ of beam-unrelated events in the
selected $e^+$ sample are due to neutrons.
Hence the beam-related neutron background is less than 0.005 times
the total beam-unrelated background, and is negligible.

\subsubsection*{5.1.2  $\ \bar\nu_e$ from Standard Processes}%

The largest beam-related background with a correlated neutron is
due to $\bar\nu_e$ produced in the beam stop by conventional processes.
Such events are detected in the same way as oscillation candidates, via
$\bar\nu_e p \rightarrow e^+ n$.
Their most important source is the DAR of $\mu^-$ in the
beam stop.  As outlined in section~1.2,
the $\bar \nu_e$ flux from $\mu^-$ decay is suppressed by more than
three orders of magnitude compared to the $\bar \nu_{\mu}$ flux from $\mu^+$
decay.  Another possible source of $\bar\nu_e$, the direct decay of
$\pi^- \to e^- \bar\nu_e$, is negligible, as a consequence of its low
branching ratio ($1.2\times 10^{-4}$), the 1/8 ratio of
$\pi^-$ to $\pi^+$ in the target, and the capture of $\pi^-$ in the
material of the beam dump.

The product of neutrino flux ($6.1 \times 10^{-13}\bar \nu_e /cm^2/p$), 
number of protons on target ($9.2 \times 10^{22}$,
corresponding to 14772 C), average cross section over the entire energy
range ($0.72 \times 10^{-40}\,{\rm cm}^2$) \cite{vogel}, 
the number of free protons in the fiducial volume ($7.4 \times 10^{30}$),
the fraction of events with $E>36$ MeV (0.45), and
the average positron reconstruction efficiency after cuts (0.36), 
gives a total background in the full positron sample of 
$4.8 \pm 1.0$ events. Note that the positron efficiency is energy dependent.
The systematic uncertainty is largely due to that in the $\bar\nu_e$ 
flux \cite{bigpaper1}, but also includes contributions for the 
efficiency (section 4.3). 

The energy-dependence of this background is determined by folding the
$\bar\nu_e$ spectrum from $\mu^-$ DAR (softer than the $\bar\nu_{\mu}$
DAR spectrum and hence of potential
oscillation events) with the detection cross section.  It is shown
in Fig. 16.

A related background is due to $\bar \nu_e~^{12} C 
\rightarrow e^+ ~^{11}B n$ scattering. The
cross section to the $~^{12}B$ ground state is calculated to be
$6.3 \times 10^{-42}$ cm$^2$ \cite{fuku2} and the cross section to the
$~^{11}B n$ final state must be at least a factor of two smaller,
especially because the first four excited states of $~^{12}B$
are stable against neutron emission.
Therefore, we estimate that this background is $<2\%$ of the 
$\bar \nu_e p \rightarrow e^+ n$ background and is negligible.
Furthermore, the maximum positron energy from this background
is 36.1 MeV, so that almost all of the positron energy
spectrum is $<36$ MeV.
 
\subsubsection*{5.1.3  Misidentification of $\bar\nu_{\mu}$ Events}%

The second most important source of beam-related background events with 
correlated
neutrons is the misidentification of $\bar\nu_{\mu}$ charged-current
interactions in the tank as $\bar\nu_e$ events.  Because of the energy
needed to produce a $\mu^+$, such a $\bar\nu_{\mu}$ must arise from a
$\pi^-$ that decays in flight. In the tank it interacts by
either $\bar\nu_{\mu} p \rightarrow \mu^+ n$ or (less often)
$\bar\nu_{\mu} \mbox{C} \rightarrow \mu^+ n X$, followed by
$\mu^+ \rightarrow e^+ \nu_e \bar\nu_{\mu}$.  There are four 
possible reasons for
the misidentification.

First, the muon can be missed because the deposited energy
is below the 18 phototube threshold for activity triggers.
This is either because
the muon is too low in energy or is produced behind the phototube surfaces.
The detector Monte Carlo simulation is used to show that this threshold
corresponds to a $\mu^-$ kinetic energy $T_{\mu}$ of 3 to $4\,{\rm MeV}$.  
Since the associated neutron also produces a little light in the tank,
the background will be quoted for the case of muons below $3\,{\rm MeV}$.  
Their
yield is computed by folding the DIF $\bar\nu_{\mu}$ flux 
with the charged-current cross sections.
The background rate from $\bar\nu_{\mu} p \rightarrow \mu^+ n$ is written
as the product of the number of protons on target ($9.2 \times 10^{22}$),
the total $\bar\nu_{\mu}$ flux ($8.7 \times 10^{-12}\,\bar \nu_{\mu}
/{\rm cm}^2/p$),
the average flux-weighted cross section ($0.70 \times 10^{-40}\,{\rm cm}^2$,
including the $\bar \nu_{\mu}$ energy range below threshold)
\cite{vogel},
the fraction of $\mu^+$ having $T_{\mu} < 3\,{\rm MeV}$ (0.0215), the
number of free protons in the fiducial volume ($7.4 \times 10^{30}$), the
positron  efficiency (0.37), 
and the fraction of events with $36<E<60\,{\rm MeV}$ 
(0.58), for a background of 1.9 events. (Note that the positron efficiency
varies with energy.) 
Similar estimates for the backgrounds from 
$\bar\nu_{\mu} \mbox{C} \rightarrow \mu^+ n X$ and 
$\nu_{\mu} \mbox{C} \rightarrow \mu^- n X$ \cite{kolbe}
add 0.1 and 0.4 events, respectively,
for a total of $2.4 \pm 1.2$ events. It is estimated \cite{kolbe} that
about 80\% of the $\bar \nu_{\mu} C \rightarrow \mu^+ X$ and 6\% of the 
$\nu_{\mu} C \rightarrow \mu^- X$ scattering events will 
have a recoil neutron.
The 50\% systematic error includes the uncertainty in the threshold, 
as well as
smaller contributions from the $\bar\nu_{\mu}$ flux and efficiency.

Second, a $\mu^+$ above the hit threshold 
can be missed if a prompt decay to $e^+$ caused
the muon and electron to be collected in a single event which is then 
identified as
an $e^\pm$.  This effect is considerably suppressed by the $\chi$ cuts and
the requirement that the reconstructed time be consistent with the
triggered event time.
The detector Monte Carlo simulation shows that this misidentification
only occurs for $\mu^+$ decays within $100\,{\rm ns}$, decreases with 
$T_{\mu}$,
and is almost zero above $10\,{\rm MeV}$.  Using the Monte Carlo 
misidentification
probabilities, a calculation similar to that above implies a background of
$0.20 \pm 0.10$ events. 

Third, the $\mu^+$ can be lost because it is produced behind the 
PMT surface and the electron radiates a hard $\gamma$ that reconstructs 
within the
fiducial volume. A background of $0.1 \pm 0.1$ 
events is estimated from the Monte Carlo simulation. 

Fourth, a muon can be missed by trigger inefficiency. In 1995, we
acquired for many online positron triggers complete digitization 
information for all veto and detector phototubes over the 6 $\mu$s interval
prior to the positron. Analysis of these data, discussed in section
7.3.2, shows the trigger inefficiency for low-energy muons to be
negligible.

The total background due to misidentified muons is thus $2.7 \pm 1.3$
events. 
It has a detected energy spectrum
which is very close to that for positrons from $\mu^+$ decay.

\subsubsection*{5.1.4  Other Backgrounds Considered}

Additional backgrounds are from $\bar\nu_e$ produced by 
$\mu^- \rightarrow e^- 
\nu_{\mu} \bar\nu_e$ and $\pi^- \rightarrow e^- \bar\nu_e$ DIF.
These $\bar\nu_e$ can interact on either $C$ or a free proton to yield the
oscillation signature of a positron and a recoil neutron.
For $36 < E_e < 60\,{\rm MeV}$, $0.1 \pm 0.1$ events are estimated.
The reactions $\nu_e ~^{12}C \rightarrow e^- n X$ and $\nu_e~^{13}C
\rightarrow e^- n X$ are negligible ($<0.1$ events) 
for $E_e<36$ MeV and cannot occur
for $E_e>36$ MeV.
Other backgrounds, for example $\nu_{\mu} C \rightarrow \nu_{\mu} n \gamma X$
with $E_{\gamma} >20$ MeV and $\nu_e C \rightarrow e^- p X$ followed
by $~^{13}C(p,n)~^{13}N$, are also negligible.

\subsection*{5.2\ \ Beam-Related Backgrounds Without a Correlated $\gamma$}%

There are eight beam-related backgrounds without neutrons that are  
considered (see Tables II and III). 
Although their total is determined empirically by a fit
involving the photon parameter R (see section 7.1), 
they are also estimated individually in the 
$36<E_e<60$ $\,{\rm MeV}$ energy range before the associated $\gamma$ 
requirement 
is imposed. These estimates are outlined below, using positron selection 
criterion VI as defined in chapter 4.

\subsubsection*{5.2.1 \ \ DIF Backgrounds Without a Correlated $\gamma$}%

The first background is due to $\pi$ DIF in the beam
 stop, followed by $\nu_{\mu} C \rightarrow \mu^- X$ and $\mu^- \rightarrow 
e^- \nu \bar \nu$ in the detector.
This background occurs if the muon is missed because it is below the 
18 PMT threshold, either because the muon is produced at too low an energy 
(.005 probability) or behind the PMT surface (.001 probability) or the muon 
decays promptly so that the muon and electron are considered one particle 
that pass the PID (.001 probability). 
The estimated number of events is the sum of the above contributions
(.007 probability) multiplied by the $\nu_{\mu}$ flux ($6.5 \times 10^{-11} 
\nu_{\mu}$/cm$^2$/p), 
the number of protons on target ($9.2 \times 10^{22}$), the
flux-average cross section 
($2.3 \times 10^{-40}$ cm$^2$) \cite{albert}, the electron 
efficiency(0.39), the fraction of events with $36<E_e<60$ MeV (0.58), and the
 number of $~^{12}C$ atoms in the fiducial volume ($3.7 \times 10^{30}$), 
which results in a total of $8.1 \pm 4.0$ events. 

Another background from $\pi$ DIF is $\nu_{\mu} e 
\rightarrow \nu_{\mu} e$ elastic scattering.
The product of neutrino flux, number of protons on target given above, 
the flux-averaged cross section ($1.4 \times 10^{-43}$ cm$^2$), 
the electron reconstruction
efficiency (0.38), the fraction of events with $36<E_e<60$ MeV (0.16), and the
number of electrons in the fiducial volume ($3.0 \times 10^{31}$) gives 
$1.5 \pm 0.3$ events. 

Other backgrounds are due to $\mu^+ \rightarrow e^+ \bar \nu_{\mu} \nu_e$ 
and $\pi^+ \rightarrow e^+ \nu_e$ DIF followed by $\nu_e C \rightarrow e^- X$ 
scattering ($0.6 \pm 0.1$ events), and $\pi^+ \rightarrow  \mu^+ \nu_{\mu}$
DIF followed by $\nu C \rightarrow \nu C \pi^o$ coherent scattering 
\cite{pi0} 
($0.2 \pm 0.1$ events).

\subsubsection*{5.2.2 \ \ DAR Backgrounds Without a Correlated $\gamma$}%
 
The next background we consider is $\nu_e e \rightarrow \nu_e e$ and 
$\bar\nu_{\mu} e
\rightarrow \bar\nu_{\mu} e$ elastic scattering from $\mu^+$ DAR 
in the beam stop.
Note that $\nu_{\mu}$ from $\pi^+$ DAR are too low in energy to produce 
electrons above 36 MeV.
The number of events from this source is estimated as the product of
the neutrino flux 
($7.8 \times 10^{-10} \nu$/cm$^2$/p), the number of protons
 on target ($9.2 \times 10^{22}$), the average cross section sum for 
$\nu_e e$ and 
$\bar\nu_{\mu} e$ scattering ($3.5 \times 10^{-43}$ cm$^2$), the electron 
reconstruction efficiency (0.38), the fraction of events with 
$E>36 \,{\rm MeV}$
 (0.042), and the number of electrons in the fiducial volume 
($3.0 \times 10^{31})$, which results in $12.0 \pm 1.2$ events. 

Another background from $\mu^+$ DAR in the beam stop is $\nu_e C$
scattering.
For $\nu_e ~^{12}C \rightarrow e^- X$ scattering (including the
transition to the $~^{12}N$ ground state) an average
cross section of $1.5 \times 10^{-41}$ cm$^2$ \cite{kolbe} 
is used. For an electron 
reconstruction efficiency of 0.36, the fraction of events with $E>36$ MeV
of 0.014 (as determined by the Monte Carlo simulation and which dominates
the systematic error), 
and the number of $~^{12}C$ atoms in the fiducial volume of 
$3.7 \times 10^{30}$, a total of $20.1 \pm 4.0$ events is obtained.
As shown in Fig. 16, this is the dominant background for $E_e<36$ MeV.
For $\nu_e ~^{13}C \rightarrow e^- X$ scattering, an average
cross section \cite{fuku}
of $5.3 \times 10^{-41}$ cm$^2$ is used, an electron reconstruction
efficiency of 0.37, the fraction of events with $E>36$ MeV of 0.39, and the 
number of $~^{13}C$ nuclei in the fiducial volume of $4.1 \times 10^{28}$ 
(1.1\% of the carbon nuclei are $~^{13}C$) to
obtain a total of $22.5 \pm 4.5$ events. 
Note that the highest energy electron that
can be produced with a recoil neutron from $~^{13}C$ is 30 MeV.

Finally, there is a background from $\pi^+ \rightarrow e^+ \nu_e$ DAR in 
the beam stop followed by $\nu_e C \rightarrow e^-X$ scattering.
An average cross section of $2.9 \times 10^{-40}$ cm$^2$ \cite{kolbe}
is used with an 
electron reconstruction efficiency of 0.39, a branching ratio of $1.2 
\times 10^{-4}$, and a number of $~^{12}C$ atoms in the fiducial volume 
of $3.7 \times 10^{30}$ to obtain a total of $3.6\pm 0.7$ events.

\subsection*{5.3\ \ Total Beam-Related Background and Maximal 
Oscillation Signal}%

Summing all of the above backgrounds, a total beam-related
background of $76.2 \pm 9.7$ events is obtained in the $36<E_e<60$ MeV 
energy range with no $\gamma$ requirement ($R \ge 0$).
Using efficiencies for correlated and accidental $\gamma$s with $R>30$ 
(0.23 and 0.006, respectively), 
the total beam-related background for $R>30$ is $2.1 \pm 0.4$ events in
the $36<E_e<60$ MeV energy range.
The total beam-related background is
shown as a function of energy in Fig. 16 for (a) $R \ge 0$ and 
(b) $R>30$.

Table II also gives the number of events expected for $100 \%$ 
$\bar\nu_{\mu} \rightarrow \bar \nu_e$ transmutation, where the total 
due to $\bar \nu_e p \rightarrow e^+ n$ 
is $12500 \pm 1250$ events for $R \ge 0$, including a systematic error
of 10\%. 
This number is  
the product of neutrino flux ($7.8 \times 10^{-10} \nu$/cm$^2$/p),
number of protons on target ($9.2 \times 10^{22}$), 
the average cross section \cite{vogel} over the entire energy
range ($0.95 \times 10^{-40}$ cm$^2$), the average positron reconstruction
efficiency (0.37), the fraction of events with $E>36$ MeV (0.67), 
and the number of free protons in the fiducial volume 
($7.4 \times 10^{30}$).
The number implied for $R>30$ is then $2875\pm 345$ events, where a 12\% 
systematic error is used (see section 7.1).
Table III gives the expected number of events for the $20<E_e<60 \,{\rm MeV}$
 energy range.

\section*{6.\ \ Data Signal}%

\subsection*{6.1\ \ Event Excess}%

Table IV lists the number of signal, beam-off background and 
neutrino-background events for the various selections described in chapter 4.
Excess/Efficiency is the excess number of events divided by the
total efficiency. Also shown in the table are the probabilities that the 
event excesses are entirely due to  
statistical fluctuations. 
With selection criterion VI and no correlated $\gamma$ requirement,
$139.5 \pm 17.7$ beam-excess events are observed in the $36<E_e<60$ MeV 
energy 
range, which is more than the $76.2 \pm 9.7$ events expected from 
conventional 
processes and which results in a total excess of $63.3 \pm 20.1$ events.
To determine whether a $\gamma$ is a 2.2-MeV $\gamma$ correlated with an 
electron or 
from an accidental coincidence, the approximate likelihood ratio, $R$, 
is employed, as described in chapter 3.
As listed in Table IV, 22 events beam-on and $36 \times 0.07 = 2.5$
events beam-off, 
corresponding to a beam on-off excess of $19.5 \pm 4.7$ events, are observed 
for $R>30$, a region in which backgrounds with an accidental $\gamma$
are greatly suppressed.
When each of the electron selection criteria is relaxed, the
background increases slightly, but the beam-on minus beam-off event excess
does not change significantly.
The total estimated neutrino background for $R>30$ is $2.1 \pm 0.4$ events,
which results in a net excess, beam-on minus total background, of 
$17.4 \pm 4.7$ events in the $36 <E_e < 60 \,{\rm MeV}$ energy range.
The probability
that this excess is due entirely 
to a statistical fluctuation of a $4.6\pm 0.6$ event 
expected total background
is $4.1 \times 10^{-8}$.
The corresponding excess for the cuts used in reference \cite{paper1}
(selection I) is $8.7 \pm 3.6$ events.
Table IV lists the results for this and all other selections described
in chapter 4. 
The Excess/Efficiency numbers are all statistically consistent.

\subsection*{6.2\ \ Alternative Geometric Criteria}%

Two alternative geometric criteria discussed in reference \cite{paper1}
were also studied to minimize cosmic-ray
background, although it is reliably measured from beam-off data. 
The first criterion, defined as selection VIa, removes 6\% of the acceptance
by requiring $Y>-120$ cm for events with $Z<0$ cm. The second criterion,
defined as selection VIb, removes 55\% of the acceptance by requiring
$Y>-50$ cm, $Z>-250$ cm, and $D>50$ cm. The relative acceptances were  
determined with the sample of $\nu_e C \rightarrow e^- X$ scattering
events.
As shown in Table IV, the resulting Excess/Efficiency numbers are
consistent with the other selections.

\subsection*{6.3\ \ Distributions of Data}%

Table V lists the 26 beam-on events from selection IV with
$R>30$ and energy in the range
$36<E_e<60$ MeV.
For each event the energy, position, and distance from the PMT surfaces
are given.
Also given are the selections that each event satisfies.
Fig. 17 shows the beam-on minus beam-off energy distributions over
an extended energy range, for both $R \ge 0$ (the full positron sample)
and $R>30$ samples that 
satisfy selection VI. 
The dashed histograms show the total estimated 
beam-related backgrounds.  
In order to
illustrate compatibility of the energy distribution with one example of
an oscillation hypothesis, a contribution from high-$\Delta m^2$
($\Delta m^2 \rightarrow \infty$)
oscillations has been added to the backgrounds, resulting in the solid 
histograms in the two plots.  The shape of this contribution is of
course sensitive to $\Delta m^2$.
Fig. 18 shows
the X, Y, Z spatial distributions for the $R\ge 0$ and $R>30$ samples,
while Figs. 19 and 20 are two-dimensional
plots showing the distribution of events in the Y - X and Y - Z planes
for (a,b) the beam-on events
and (c,d) the beam-off events. 

Figures 21 to 27 
show a variety of other beam-on minus beam-off 
distributions for the $R>30$ selection VI sample, all restricted to
$36<E_e<60\,{\rm MeV}$.  For the $cos\theta_b$ distribution shown in 
Fig. 21,
where $\theta_b$ is the angle between the neutrino direction 
and the reconstructed positron direction,
the solid histogram also illustrates expectations from a
high-$\Delta m^2$ oscillation hypothesis.  The observed average
value of $cos\theta_b$ is $0.20\pm 0.13$, in agreement with the
expected value of 0.16 for $\bar \nu_e p$ interactions. Electrons
from muon decay and $\nu_e C$ scattering have expected values
of 0 and $<0$, respectively. 
For the remaining plots, the expected
distributions are those for any neutrino-induced reactions.
These distributions are obtained from samples of $\nu_e C
\rightarrow e^- X$
scattering events in the $20<E_e<36$ MeV energy range 
and electrons from muon decay with the same selection as for
the oscillation sample.
Fig. 22 displays the
distance from the PMT surfaces, $D$.
The $\chi$, $\Delta t_p$, 
and veto shield multiplicity 
distributions are shown in Fig. 23,
Fig. 24 and 
Fig. 25, where the $\chi$ variables 
are the particle ID parameters
discussed in chapter 4, $\Delta t_p$
is the time to the previous event, and the veto shield multiplicity 
is the number of hit veto PMTs in time with the event.
Finally, the
$\vec r \cdot \hat {dr}$ and the S distributions,
discussed in chapter 4, are shown in Fig. 26
and  Fig. 27.

\subsection*{6.4\ \ Tests of Spatial, Energy, and Time Distributions}%

\subsubsection*{6.4.1 Spatial Distribution of Beam-Related Data}

Cosmic-ray background is larger in the outer regions of the detector and
where the veto has gaps -- beneath the detector (large negative Y), 
and around the
periphery of the upstream end at large negative Z.
Because the beam-on data includes cosmic-ray background, it is  
expected to show concentrations in the same regions of the detector.
In fact, any effect from strong or electromagnetic interactions coming from 
outside the detector should be concentrated near the detector boundary.

The source of neutrinos is concentrated in the region of the beam stop
described in reference \cite{bigpaper1}.
The distance from the beam stop to the center of the detector is 29.8
m, and the angular distribution of the neutrinos is isotropic.
The neutrino flux from targets A1 and A2, which are 105 m and 130 m away,
respectively,
imposes a small variation on the flux distribution calculated using the A6
location.
Neutrino event distributions in the detector are expected to reflect
the varying solid angle of the detector with small effects from the
finite extent of the source.
This is simulated in detail, although the deviation from uniformity is small,
and these fluxes are used in estimating rates.

It is important to test whether the 
spatial distributions of beam excess events are
compatible with neutrino oscillations.
To this end, a Kolmogorov statistic is computed for each distribution being 
tested for consistency.
For a given variable V, an observed cumulative probability distribution, 
$F_{on}$, is computed for beam-on data.
If $N_{on}$ is the number of beam-on events, then $N_{on}F_{on}(w)$ is the 
number of beam-on events with V less than $w$.
$F_{on}$ is a step function.
If the distribution in V is consistent with beam-off background plus a 
contribution from neutrino interactions, then $F_{on}$ should be approximately
 equal to an expected cumulative probability distribution, $F$, that is a
combination of these two contributions.
The Kolmogorov statistic, K, is the maximum distance between $F_{on}$ and $F$.
The probability distribution of K is computed for the case of the beam-on 
excess coming from neutrino interactions.

One contribution to $N_{on}F(w)$ is the expected number of events from the
cosmic background.  If there are $N_{off}$ beam-off events, and $r$ is the
ratio between beam-on and beam-off time, then the expected total number from
cosmic background is $rN_{off}$.  If the step function, $F_{off}$, is defined
the same as $F_{on}$, except for beam-off events instead of beam-on ones, then
the expected contribution to $N_{on}F(w)$ from cosmic background is equal to
$rN_{off}F_{off}(w)$.  The remaining contribution to $N_{on}F(w)$ is from
the $N_{on}-rN_{off}$ excess of presumably neutrino events, which should be
distributed according to a smooth cumulative probability distribution, 
$F_\nu$.
For each variable, V, we take $F_\nu(w)$ to be the expected fraction of
neutrino interactions in our acceptance with V below $w$.  It
is computed with a Monte Carlo program that includes the position dependent
neutrino flux and position dependent positron detection efficiency, and of
course includes the requirement that positrons be reconstructed at least
35 cm from the photomultiplier tube faces.  Then $(N_{on}-rN_{off})F_\nu(w)$
is the expected contribution to $N_{on}F(w)$ from the beam-on excess if that
excess is from neutrino interactions.  Thus
$$F(w) = {rN_{off}\over N_{on}}F_{off}(w) + \left(1 - {rN_{off}\over N_{on}}
\right) F_\nu(w).$$

The Kolmogorov statistic, K, is easily determined, given the functions,
$F_{on}$ and $F$.  Each computation of K involves comparing a cumulative
distribution of data ($F_{on}$) with a function that is a
linear combination of a distribution of other data ($F_{off}$) and a smooth
theoretical function ($F_\nu$).  The probability distribution of K is not
given in standard tables for such a case.  We therefore perform a Monte Carlo
computation of the probability of K accidentally being at least as large as
is measured.

One complication is that the Z distribution for neutrino oscillation
events depends on the value of $\Delta m^2$. In the limit of large
$\Delta m^2$, however, the distribution has the same $L^{-2}$ 
dependence as for other neutrino interactions, where $L$ is the
distance from the neutrino production point to the neutrino
interaction location. The consistency checks
are calculated for this case.
If K is measured to be especially high, i.e., if there is an especially 
low probability of K accidentally being higher, then the 
observed distribution 
is inconsistent with the assumption that the beam-on excess 
comes from neutrino
 interactions in the detector tank.
The consistency checks on the spatial distribution of the data amount to 
finding such probabilities for each of several distributions, including
those shown in Figs. 18 and 22. A high
probability near one means that the distribution is very similar to the
expected distribution, while a probability near zero means that the
distribution is not very similar. 
Results are presented in Table VI with various cuts for identification 
of the $\gamma$ from $np\rightarrow d\gamma$.

If the gap in the veto beneath the detector is responsible for the beam-on 
excess, the Y distribution would be expected to show an especially low
probability.
If the holes in the veto at its upstream end are responsible for 
the excess, the Z distribution 
or the distribution in distance from the bottom upstream corner with
 low Y,Z would show a low probability.
If the events are anomalously concentrated towards the outer part of the 
detector, then there would be a low probability for the variable 
that measured
 the distance from the PMT faces.
These probabilities are computed for various cuts on R, the photon 
discrimination parameter. The probabilities for $R\ge 0$ are observed to
be smaller than the probabilities for $R>1.5$ or $R>30$. 
This is due to the high
statistics of the $R\ge 0$ sample, which makes this sample very sensitive to
uncertainties in the expected position distribution.
For example, the Z distribution
and low YZ distribution
probabilities for $R\ge 0$ increase from 0.047 to 0.331 and from 0.016 to
0.074, respectively, when one assumes that
the events are uniformly distributed in the detector instead of having a
$L^{-2}$ position dependence.  Although
this assumption is unrealistic for the large expected beam-associated
neutrino background with $R\ge 0$ (see Table IV), a contribution
from neutrino oscillations at low $\Delta m^2$ would have a uniform position
dependence.

\subsubsection*{6.4.2  Kolmogorov Test on the Energy Distribution}%

The energy distribution of events with $R>30$ has been subjected to the 
same Kolmogorov test as in the previous section on the geometric
distribution of events.
Events near and above 60 MeV provide incentive for this test. 
The cumulative distribution for neutrinos, $F_\nu(E)$, is taken to be
the expected energy distribution for neutrino oscillations in the limit of
high $\Delta m^{2}~$.
The contribution shown in Fig. 17 from known
neutrino interactions is ignored, as well as possible contributions from
DIF oscillation events.
The probability that the energy distribution for $R>30$ is consistent with 
this hypothetical distribution is 35\% for $36<E_e<60\,{\,{\rm MeV}}$ 
and 37\% 
for $36<E_e<80\,{\,{\rm MeV}}$. There is no evidence of an 
excess of events above
60 MeV. For the $60<E_e<80$ MeV interval there are 4 events beam on and
62 events beam off, corresponding to an excess of $-0.3 \pm 2.1$ events.
The solid curve in Fig. 17b shows that there is no 
incompatibility between the oscillation hypothesis and the data excess, 
given present statistical errors.

\subsubsection*{6.4.3 Time Distribution of Beam-Related Data}

Another consistency check on our data and analysis methods is whether the
evidence for neutrino oscillations is reasonably uniform from one year of
data collection to the next.  Small
problems with the apparatus, corrected as the
experiment progressed, can make spurious signals appear only in data
collected before hardware repairs.  Unconscious prejudices can lead
experimenters to tune cuts until a selection is found that accidentally gives
a spurious signal.  Such a selection would not show a signal for data
collected after the cuts have been tuned.

In order to test for time variation of our data, we bin beam-on and beam-off
data for $R>30$ by the year in
which it is collected.  Most changes in apparatus and procedures are made
in the periods between the running periods of different years.  We consider
two selections for the data: Selection I, which is the 
same as was used before
beginning the 1995 runs and on which a previous publication
\cite{paper1} is based; and Selection VI, which includes the the most recent
analysis improvements.  The excesses 
(beam-on minus duty ratio times beam-off)
should be roughly proportional to the $\bar\nu_\mu$ fractions of integrated
beam intensity during each time period.  The consistency checks
test how probable are the observed deviations from rough proportionality.

Table VII shows the results of two types of consistency checks.
``Prob 1'' is the probability of beam-on data accidentally being distributed
in an equally likely or less likely way than is observed, given the beam-off
numbers of events in each year and the duty ratios.  ``Prob 2'' is
the probability of the 1995 beam-on number accidentally being as low as is  
observed given the beam-off numbers in each year, the total beam-on
number of events, and the duty ratios.  No probability is so low as to
demonstrate a serious inconsistency.

\section*{7.\ \ Fits to the Data}%

\subsection*{7.1\ \ Fits to Determine the Oscillation Probability}

For the observed excess, the overall oscillation
probability is found by fitting the R distribution to determine the
fraction of events with a correlated $\gamma$. 
The overall oscillation probability is the 
number of excess correlated 
events divided by the total number of events expected
for 100\% $\bar \nu_{\mu} \rightarrow \bar \nu_e$ transmutation.
Note that for any experiment the oscillation probability is
dependent on the experiment's
geometry and energy range in addition to $\sin^22\theta$ and 
$\Delta m^2$.
The one-parameter $\chi^2$ 
fit to the R distribution takes into account the position dependence
of the $\gamma$ rates by using the actual beam-on and beam-off events that 
satisfy the oscillation criteria. The accidental 
$\gamma$ spatial distributions are  
determined from laser calibration events.
Fig. 28 shows the R distribution,
beam-on minus beam-off excess, for events that satisfy selection VI (see
chapter 4) and that have energies in the range $20<E_e<60$ MeV. 
There are 1763 beam-on events and 11981 beam-off events in this energy
range, corresponding
to a beam on-off excess of 924.3 events.

The R distribution is fit to the two different R shapes discussed in
chapter 3 and illustrated in Fig. 5. 
The fit using the R shape from cosmic ray neutrons
has a $\chi^2 = 6.9/9$ D.O.F. and determines 
that $68.5 ^{+19.4}_{-17.6}$ events have a $\gamma$ that is correlated with
the primary, while the fit using the Monte Carlo R shape 
has a $\chi^2 = 5.4/9$ D.O.F. and determines
that $60.1 ^{+17.6}_{-15.7}$ events have a $\gamma$ that is correlated with
the primary. 
Averaging these numbers and subtracting the neutrino 
background with
a correlated $\gamma$ ($12.5 \pm 2.9$ events) results in a net excess of 
$51.8 ^{+18.7}_{-16.9}$ events. 
(If the number of events with a correlated $\gamma$ is set to the
background estimate of 12.5 events, the $\chi^2$ increases by 
15.0 and
14.1, respectively, compared to the above two fits.) 
This corresponds to an oscillation probability
of $(0.31^{+0.11}_{-0.10} \pm 0.05)\%$, where the first error is statistical
and the second error is the systematic error arising from uncertainties
in the neutrino flux (7\%), $e^+$ efficiency (7\%),
and $\gamma$ efficiency (7\%). The latter two uncertainties are lower than
in our previous publication \cite{paper1} due to improved understanding
of the detector performance. Note that the statistical error is
non-Gaussian and corresponds to an increase of the $\chi^2$ by one over
the minimum $\chi^2$ fit. 
The systematic error is for both the background estimate
and the expected number of oscillation events.
Also, $860.8 ^{+18.5}_{-16.7}$ 
events do not have a correlated $\gamma$, which
agrees with the estimated neutrino background of $795.0 \pm 133.9$ events
from Table VIII. 
The solid curve in
Fig. 28 is the best fit to the data, while the dashed curve is the component
of the fit with an uncorrelated $\gamma$. Table VIII summarizes the
results of the $\chi^2$ fit. Also shown in Table VIII is
the result of a likelihood fit that uses for each positron event the
local accidental R distribution rather than a weighted average, and
the number of signal and background events in the $20<E_e<60$ MeV energy
range with $R>30$. 

\subsection*{7.2\ \ Favored Regions of $\Delta m^2$ vs. $\sin^22\theta$}%

Assuming that the observed event excess is due to neutrino oscillations,
a likelihood fit is performed to determine favored regions in
the $\Delta m^2$ vs. $\sin^2 2\theta$ plane, where $\Delta m^2$
is the difference of the squares of the approximate mass eigenstates
and $\theta$ is the mixing angle. 
A general formalism for neutrino oscillations would involve all three 
generations and the possibility of CP violation.
In fact, any pair of neutrinos 
($\bar \nu_e$, $\bar \nu_{\mu}$, $\bar \nu_{\tau}$,
or more properly $\nu_1$, $\nu_2$, or $\nu_3$) with a $\Delta m^2$ in the
region of experimental sensitivity could lead to a signal in a
$\bar \nu_{\mu} \rightarrow \bar \nu_e$ oscillation search.
However, here the formalism is simplified by assuming that only 
two generation mixing is important.
Then the oscillation probability can be written
$$P=\sin^2(2\theta) \sin^2 \left(1.27\Delta m^{2} L/E_{\nu}\right)\ \ , $$
where 
$L$ is the distance 
from neutrino production to detection in meters 
and $E_{\nu}$ is the neutrino energy
in MeV.
The discussion is limited to this restricted formalism solely as a basis for 
experimental parameterization, and no judgement is made as to the simplicity
of the actual situation.

Four measured quantities are used to separate oscillation candidates from
background and determine the parameters of the oscillation.  These are
$E_{e}$ (the measured energy of the positron), $R$ 
(the gamma likelihood ratio),
$\cos \theta_b$ (the cosine of the angle between the
$e$ and $\nu$ directions),
and $L$ (the measured distance from the
$\bar \nu_{\mu}$ source).  The 1763 beam-on events passing selection VI are
binned in four dimensions according to these measured quantities.
Using the background
estimates from chapter 5, 
the distributions of beam-related background
events in these variables are calculated.  To calculate the 
beam-unrelated background, the measured beam-off
data set is smoothed and scaled by the duty ratio.
  
A likelihood function, ${\cal L}$, is constructed:
\[
{\cal L}(n_1,n_2,\ldots | \Delta m^2, \sin^2 2\theta) =
\prod_{i=1}^{N} \frac{1}{n_{i}!} \nu_i^{n_{i}}
e^{-\nu_i},
\]
where $N$ is the total number of bins, $n_i$ is the number of
beam-on events in bin $i$, and $\nu_i$ is the expected number
in bin $i$.  The expected number in bin $i$ may be written
\[
\nu_i =
\nu_{i,\rm BUB} + \nu_{i,\rm BRB} + \nu_{i,\rm osc}
(\Delta m^2,\sin^2 2\theta),\]
where $\nu_{i,\rm BUB}$ is the calculated number of events in bin $i$ due
to beam-unrelated background,  $\nu_{i,\rm BRB}$ is that due to
beam-related background, and $\nu_{i,\rm osc}(\Delta m^2,\sin^2 2\theta)$
is the expected number of events for a particular pair of 
$\Delta m^2,\sin^2 2\theta$
values.  This likelihood function reaches 
its
maxima at $15$ and $19 eV^{2}, \sin^2 2\theta = 0.006$.
The individual distributions of $E_{e}$, $R$, $\cos \theta_b$, and $L$ for 
the data are compared with projections of the expected four-dimensional 
distribution (including oscillations at $19 eV^{2}, \sin^2 2\theta = 0.006$)
in Fig. 29. Note that most of the data in Fig. 29
is from beam-unrelated or neutrino-induced background.

The $\log$ of this likelihood function is calculated for a range
of  $\Delta m^2, \sin^2 2\theta$ values.  
Regions within 2.3 and 4.5 log-likelihood units of 
vertical distance from the peak are identified.
These regions are called  ``90\%'' and ``99\%'' likelihood regions. 
(They do not define confidence limits, 
but do show the regions favored by the experiment.)
These favored regions are calculated several times while varying inputs to 
reflect systematic uncertainties. 
The systematic effects varied included: 
the method used for smoothing the beam-off data,
the method used for calculation of the correlated $R$ distribution, 
and the normalization of the backgrounds 
(both beam-related and beam-unrelated are shifted by $\pm 1 \sigma$). 
Also, the product of neutrino flux and detection efficiency was allowed to
change by $\pm 10\%$. 
Regions which are favored in any of these systematic investigations are shown
in Fig. 30, where the darkly-shaded and lightly-shaded regions correspond
to 90\% and 99\% likelihood regions, respectively.  
This figure shows discrimination against some values of $\Delta m^2$
which would be allowed in an analysis that simply took the 
size of the oscillation signal into account.  This discrimination may be 
understood from the energy plot of Fig. 17b.  The presence of 
relatively high-energy oscillation candidates tends to exclude
$\Delta m^2$ near integral multiples of $4.3 eV^2$.  (These values
of $\Delta m^2$ give 
$\sin^2 \left(1.27\Delta m^{2} L/E_{\nu}\right)$ near 0 for the highest
energy $\bar\nu_{\mu}$.)  

Some of the
favored region is excluded by the ongoing KARMEN experiment \cite{karmen} 
at ISIS, E776 at BNL \cite{wonyong}, 
and the Bugey reactor experiment \cite{bugey} (see section 8.2).
However, there remains a region at small values of
$\Delta m^2$ and $\sin^22\theta$ where our oscillation parameters 
are not in conflict with any other experiment. 

It is difficult to place additional constraints
on $\Delta m^2$ with the few events collected to date.  Fig. 31 
shows the $L/E_{\nu}$ distribution of the high-$R$ data 
(from the top 3 $R$ bins of Fig. 29) compared
with expectations for several pairs of 
$\Delta m^2, \sin^2 2\theta$. ($E_{\nu}$ is calculated
from the measured values $E_e$ and $\cos \theta_b$.)  This plot gives an
indication of the statistical precision needed to distinguish between
high and low values of $\Delta m^2$.  It also shows the expected
$L/E_{\nu}$ distribution for the disfavored $4.3 eV^2$.

\subsection*{7.3\ \ Neutrino Backgrounds with a Correlated  $\gamma$}%

In this section we discuss in more detail the two major 
neutrino backgrounds
with a correlated $\gamma$: (1) $\mu^-$ DAR in the beam stop followed by
the reaction $\bar \nu_e p \rightarrow e^+ n$ in the detector; and
(2) $\pi^-$ DIF in the beam stop followed by the reaction
$\bar \nu_{\mu} p \rightarrow \mu^+ n$ in the detector. As described
in section 5.1, these backgrounds are each estimated to be about
an order of magnitude smaller than the observed excess. Additional
arguments, however, can be made to demonstrate that these
backgrounds are not likely to explain the signal.

\subsubsection*{7.3.1 \ \ $\mu^-$ DAR Background}%

Because the $\bar \nu_e$ spectrum from $\mu^-$ decay is softer than the 
$\bar \nu_{\mu}$ spectrum from $\mu^+$ decay, one can, in principle,
distinguish between $\bar \nu_{\mu} \rightarrow \bar \nu_e$
oscillations and $\mu^-$ DAR background by fitting the energy distribution.
This is accomplished by allowing the $\mu^-$ background to float and
determining how good a fit (see section 7.2)
can be obtained without neutrino oscillations. The best such fit
has a $\mu^-$ DAR background contribution that is 8 times larger than
the estimated background of $8.6 \pm 1.7$ events (see Table VIII).
However, even with such an increase, this best fit has the log of
the likelihood function 2.2 units 
less than the best oscillation fit. Therefore, our observed
excess is less compatible with the shape of the $\mu^-$ DAR background. 

\subsubsection*{7.3.2 \ \ $\pi^-$ DIF Background}%

As mentioned in chapter 2, the nominal trigger threshold for past
activity in LSND is 18 hit
PMTs. This allows a background to arise from $\pi^-$ DIF in the beam stop
followed by $\bar \nu_{\mu} p \rightarrow \mu^+ n$ scattering, where the
$\mu^+$ is below the 18 PMT threshold. (Background contributions also arise
from $\bar \nu_{\mu} C \rightarrow \mu^+ n X$ and
$\nu_{\mu} C \rightarrow \mu^- n X$ scattering.) We are confident of our 
calculation of this background in chapter 5. However, to ensure that such
events do not explain our observed signal, the trigger was modified for
the 1995 running so that all hit PMTs within $0-3$ and $3-6$ 
$\mu$s of selected events are recorded as two extra events. 
Fig. 32 shows the total number of hit
PMTs in the detector tank for those extra events 
that occur $0 - 3 \mu$s and $3 - 6 \mu$s
prior to oscillation candidate events. 
The candidates are in the $25<E_e<60$ MeV energy range
with (a) $R\ge0$ and (b) $R>30$. The data points are the beam on 
events, while the solid curve is what is expected from random PMT hits
as determined from the sample of laser calibration events. There is
good agreement between the data and the laser events and little evidence
of candidates from $\pi^-$ DIF background, which the Monte Carlo simulation
estimates would hit an additional 10 PMTs on average. 
This also confirms that the trigger operated correctly.

The sample of $\nu_{\mu} C \rightarrow \mu^- X$ scattering events also has
been studied to check that the observed hit PMT distribution from the
recoil $\mu$ and $X$
agrees with our Monte Carlo simulation. This sample is cleanly obtained
by requiring a coincidence between the $\mu$ and the decay electron and
by performing a beam-on minus-off subtraction.
Fig. 33 shows the
observed hit PMT distribution for all $\nu_{\mu} C$ scattering events
(including $\nu_{\mu} C \rightarrow \mu^- X$, $\bar \nu_{\mu} C 
\rightarrow \mu^+ X$,
and $\bar \nu_{\mu} p \rightarrow \mu^+ n$) for events with (a) $R\ge 0$ and
(b) $R>30$. The solid histogram in each case is the prediction from the
Monte Carlo simulation, normalized to the data. The agreement is excellent
and serves as a check of our background estimate from chapter 5.

\section*{8.\ \ Interpretation of Results}%

\subsection*{8.1  Possible Explanations}%
This paper reports an excess of events that is consistent with the
reaction $\bar \nu_e p \rightarrow e^+ n$ and is an order of 
magnitude larger than what is expected from conventional physics processes.
This excess is, therefore, evidence for $\bar\nu_{\mu} \rightarrow \bar\nu_e$
oscillations within the allowed range of Fig. 30.
Note that for three neutrino flavors there must be three-generation mixing,
so that the oscillation probability is in general a sum of three terms,
where each term has an oscillation wavelength determined by one of
the three different $\Delta m^2$ values.
However, there are other exotic physics explanations of the observed 
excess.
One example is the lepton-number-violating decay $\mu^+ \rightarrow
e^+ \bar\nu_e \nu_{\mu}$, which can explain these observations with a
branching ratio of $(0.31 ^{+0.11}_{-0.10} \pm 0.05)\%$. The published
upper limit on this ``wrong-sign'' muon decay mode is $1.2\%$ \cite{e645};
however, a preliminary report from the KARMEN experiment \cite{karmen2}
gives a much stricter limit, $\mu^+\rightarrow e^+ \bar \nu_e \nu_{\mu}/
\mu^+ \rightarrow e^+ \nu_e \bar \nu_{\mu} < 0.25\%$ at 90\% C.L.
If an excess similar to that reported in the present paper
is observed also in the $\pi^+$ DIF $\nu_{\mu} \rightarrow 
\nu_e$ search from LSND or from some other experiment, then the oscillation
hypothesis will be favored and the allowed region in Fig. 30 will
be constrained.

\subsection*{8.2  Review of Other Experiments}%

In this section the evidence restricting neutrino oscillation parameters is 
briefly reviewed.
Three experiments using the BNL wide-band beam have searched for $\nu_
{\mu} \rightarrow \nu_e$ oscillations.
They are an experiment primarily designed 
to measure neutrino electron scattering, E734 \cite{e734}, a follow up on a 
previous indication of neutrino oscillations at the CERN PS, E816 
\cite{e816}, and a specifically designed long baseline 
oscillation experiment, 
E776 \cite{wonyong}.

The BNL neutrino beam is a horn focused beam composed mainly of 
$\nu_{\mu}$ and 
$\bar\nu_{\mu}$ from pion and kaon DIF.
The principal $\nu_e$ background for all of the experiments comes from the 
pion-muon decay sequence and from charged and neutral kaon decay.  
Integrated over the entire spectrum, this $\nu_e$ flux is about 1\% of the 
$\nu_{\mu}$ flux with a minimum $\nu_e$ flux of about 0.6\% near a neutrino 
energy of 1 GeV.
Each experiment also has a photon background from $\pi^0~$ production,
where one $\gamma$ is confused as an electron and where the 
second $\gamma$ is  
not seen.
The first two experiments separate photons by observing the primary 
vertex and using the spatial separation of the photon from 
this primary vertex
 to distinguish electrons and photons.
The third experiment relies on a Monte Carlo method to calculate 
the background
 from $\pi^0~$ production.
In each case, the systematic errors dominate the limits reported by E734 
and E776, as shown in Figs. 30 and 34a.

The difference in limits in Fig. 34a is almost completely 
accounted for by the different distances from the target (E734 is at 
120 m and E776 at 1000 m from the neutrino source) because the beam is  
common to both measurements.
The E816 experiment observed an excess of electron events $1.6 \pm  0.9$ 
times that expected.
The average E816 neutrino energy was about 1.2 GeV, although individual 
electron event energies were not reported.
The CCFR experiment \cite{ccfr} provides the most stringent limit on 
$\nu_{\mu} \to \nu_e$ oscillations
near $\Delta m^2 \sim 350 eV^2$, but their 
limits are not as restrictive as E776 
for values of $\Delta m^2 <300$ eV$^2$.

The KARMEN experiment \cite{karmen} has searched for $\nu_{\mu} \rightarrow 
\nu_e$ oscillations using neutrinos from pion DAR. 
These neutrinos are monoenergetic, and the signature for oscillations is  
an electron energy peak at about $12$ $\,{\rm MeV}$.
This method has very different backgrounds and systematics compared
to the previous three experiments but, unfortunately, does not yet have 
statistical precision sufficient to affect the exclusion region of 
Fig. 34b.
The KARMEN experiment also has searched for $\bar\nu_{\mu} \to \bar\nu_e$
oscillations and has produced the exclusion plot shown in 
Figs. 30 and 34b.
KARMEN is located 18m from the neutrino source, compared with 30 m for LSND.
The experiments have sensitivities, therefore, that peak at different values 
of $\Delta m^{2}~$. Experiments E225 and
E645 at LAMPF also searched for 
$\bar\nu_{\mu} \rightarrow \bar\nu_e$ oscillations 
and set less restrictive limits \cite{e225}, \cite{e645}.

The most recent experiments searching for $\bar\nu_e$ disappearance are 
Gosgen \cite{gosgen}, Bugey \cite{bugey}, and Krasnoyarsk \cite{krasnoyarsk}.
Power reactors are prolific sources of $\bar\nu_e$, and the detection method
is similar in the three cases.
The Bugey measurement has the highest reported sensitivity.
The detectors observe both the positron from the primary neutrino 
interaction and the capture energy (4.8 $\,{\rm MeV}$) from 
neutron absorption on $^{6}$Li.
This capture time is about 50 $\mu s$ and, after saturation effects in the 
scintillator are included, the capture energy 
yields $0.5 \,{\rm MeV}$ ~electron equivalent energy.
The positron energy is $1.8 \,{\rm MeV}$ ~below the 
neutrino energy and allows an 
event-by-event measure of neutrino energy.
Detectors are placed at 15m, 40m, and 95 m from the nearest reactor.
Two methods are used to search for oscillations.
The first uses the ratio of events seen in the three detectors and the second
uses an absolute prediction of flux from the reactor as a further constraint.
The resulting limit is shown in Figs. 30 and
34c.

Searches for $\nu_{\mu}$ disappearance have been conducted at both 
CERN and Fermilab by the CDHS \cite{cdhs} and CCFR \cite{ccfr}
experiments.
In each case two detectors are placed at different distances from the 
neutrino source, which is a DIF $\nu_{\mu}$ beam without focusing.
The limits obtained by these experiments are shown in Fig. 34d.
Also shown in this figure are limits derived from 
the E531 Fermilab experiment 
\cite{reay} which searches for the appearance of tau decay from charged 
current interactions in a high energy neutrino beam.
Experiments 
which probe $\nu_e$ disappearance and $\nu_{\mu}$ disappearance have 
given limits which are not sensitive enough to constrain the results here, 
except at the lowest $\Delta m^{2}~$. 

\section*{9.\ \ Conclusions}%

The LSND experiment observed 22 electron events
in the $36 <E_e < 60$ $\,{\rm MeV}$ energy range that 
were correlated in time and space
with a low-energy $\gamma$, and
the total estimated background from conventional processes is $4.6 \pm 0.6$
events. The probability that this excess is due to a statistical fluctuation
is $4.1 \times 10^{-8}$. 
The observed excess is consistent with $\bar\nu_{\mu} \rightarrow \bar\nu_e$
oscillations, and a fit to the entire electron sample with electron energy
in the range $20 <E_e < 60$ $\,{\rm MeV}$
results in an oscillation probability of $(0.31^{+0.11}_{-0.10} \pm 0.05)\%$. 
The
allowed regions of $\sin^22\theta$ vs. $\Delta m^2$ are shown in Fig. 
30.

\paragraph*{Acknowledgements}

The authors gratefully acknowledge the support of Peter Barnes,
Cyrus Hoffman, and John McClelland. 
It is particularly pleasing that a number of undergraduate students
from participating institutions were able to contribute to the experiment.
We acknowledge many interesting and helpful discussions with Dharam
Ahluwalia, Terry Goldman, Peter Herczeg, Petr Vogel, and Geoffrey West.
This work was conducted under the auspices of the US Department of Energy,
supported in part by funds provided by the University of California for
the conduct of discretionary research by Los Alamos National Laboratory.
This work is also supported by the National Science Foundation.
We are particularly grateful for the extra effort that was made by these
organizations to provide funds for running the accelerator at the end of
the data taking period in 1995.

\clearpage

\begin{table}
\caption{The positron selection criteria and corresponding efficiencies
for selections I and VI.
The variables are defined in the text.}
\label{I}
\begin{tabular}{lccr}
Selection I&Efficiency&Selection VI&Efficiency\\
\tableline
PID&$0.77 \pm 0.02$&PID&$0.84 \pm 0.02$\\
$<2$ Veto Hits&$0.84 \pm 0.02$&$<4$ Veto Hits&$0.98 \pm 0.01$\\
$\Delta t_p>40 \mu$s&$0.50 \pm 0.02$&$\Delta t_p>20 \mu$s, $34 \mu$s
&$0.68 \pm 0.02$\\
DAQ Deadtime&$0.97 \pm 0.01$&DAQ Deadtime&$0.97 \pm 0.01$\\
35 cm Fiducial Volume&$0.85 \pm 0.05$&35 cm Fiducial Volume&$0.85 \pm 0.05$\\
No event within $8 \mu$s&$0.99 \pm 0.01$&No event within $8 \mu$s&$0.99 \pm 
0.01$\\ 
$<3$ Associated $\gamma$s&$0.99\pm 0.01$&$<2$ Associated 
$\gamma$s&$0.94\pm 0.01$\\
--&$1.00$&$S>0.5$&$0.87 \pm 0.02$\\
\tableline
Total&$0.26\pm 0.02$&Total&$0.37 \pm 0.03$\\
\end{tabular}
\end{table}

\begin{table}
\caption{A list of all backgrounds with the expected
number of background events in the $36<E_e<60$ MeV energy
range that satisfy selection VI for $R\ge 0$ (the full
positron sample) and $R>30$. 
The neutrinos are from either $\pi$ and $\mu$ 
decay at rest (DAR) or decay in flight (DIF). Also shown are
the number of events expected for $100 \%$ 
$\bar \nu_{\mu} \rightarrow \bar \nu_e$ transmutation.}
\label{II}
\begin{tabular}{lccr}
Background&Neutrino Source&Events with $R \ge 0$&Events with $R>30$\\
\tableline
\tableline
Beam Off&&$160.5 \pm 3.4$&$2.52\pm 0.42$\\
\tableline
Beam-Related Neutrons&&$<0.7$&$<0.1$\\
$\bar \nu_e p \rightarrow e^+ n$&$\mu^- \rightarrow e^- \nu_{\mu} \bar \nu_e$
DAR&$4.8\pm 1.0$&$1.10 \pm 0.22$\\
$\bar \nu_{\mu} p \rightarrow \mu^+ n$&$\pi^- \rightarrow
\mu^- \bar \nu_{\mu}$ DIF&$2.7\pm 1.3$&$0.62\pm 0.31$\\
$\bar \nu_e p \to e^+ n$&$\pi \rightarrow
e \nu$ and $\mu \rightarrow e \nu \bar \nu$
DIF&$0.1\pm 0.1$&$0$\\
\tableline
Total with Neutrons&&$7.6 \pm 1.8$&$1.72\pm 0.41$\\
\tableline
$\nu_{\mu} \mbox{C}\to \mu^- X$&$\pi^+ \rightarrow \mu^+
\nu_{\mu}$ DIF&$8.1\pm 4.0$&$0.05\pm 0.02$\\
$\nu_e ~^{12}C \to e^-~^{12}N$&$\mu^+ \rightarrow
e^+ \bar \nu_{\mu} \nu_e$ DAR&$20.1 \pm 4.0$&$0.12 \pm 0.02$\\
$\nu_e ~^{13}C \to e^-~^{13}N$&$\mu^+ \rightarrow
e^+ \bar \nu_{\mu} \nu_e$ DAR&$22.5 \pm 4.5$&$0.14\pm 0.03$\\
$\nu e \rightarrow \nu e$&$\mu^+ \rightarrow
e^+ \bar \nu_{\mu} \nu_e$ DAR&$12.0 \pm 1.2$&$0.07\pm 0.01$\\
$\nu e \rightarrow \nu e$&$\pi \rightarrow
\mu \nu_{\mu} $ DIF&$1.5\pm 0.3$&$0.01\pm 0.01$\\
$\nu_e \mbox{C}\to e^- X$&$\pi \rightarrow e \nu_e$
DAR&$3.6\pm 0.7$&$0.02\pm 0.01$\\
$\nu_{\mu} \mbox{C}\to \pi X$&$\pi \rightarrow
\mu \nu_{\mu} $ DIF&$0.2 \pm 0.1$&$0$\\
$\nu_e \mbox{C}\to e^- X$&$\pi \rightarrow
e \nu$ and $\mu \rightarrow e \nu \bar \nu$ 
DIF&$0.6\pm 0.1$&$0$\\
\tableline
Total without Neutrons&&$68.6\pm 9.5$&$0.41\pm 0.06$\\
\tableline
Grand Total &&$236.7 \pm 10.2$&$4.65\pm 0.59$\\
\tableline
\tableline
$100\%$ Transmutation&$\mu^+ \rightarrow
e^+ \bar \nu_{\mu} \nu_e$ DAR&$12500 \pm 1250$&$2875 \pm 345$\\
\end{tabular}
\end{table}

\begin{table}
\caption{A list of all backgrounds with the expected
number of background events in the $20<E_e<60$ MeV energy
range that satisfy selection VI 
for $R\ge 0$ (the full positron sample) and $R>30$. 
The neutrinos are from either $\pi$ and $\mu$ decay at
rest (DAR) or decay in flight (DIF). Also shown are
the number of events expected for $100 \%$ 
$\bar \nu_{\mu} \rightarrow \bar \nu_e$ transmutation.}
\label{III}
\begin{tabular}{lccr}
Background&Neutrino Source&Events with $R \ge 0$&Events with $R>30$\\
\tableline
\tableline
Beam Off&&$782.0 \pm 7.4$&$9.2\pm 0.8$\\
\tableline
Beam-Related Neutrons&&$<3.8$&$<0.5$\\
$\bar \nu_e p \rightarrow e^+ n$&$\mu^- \rightarrow e^- \nu_{\mu} \bar \nu_e$
DAR&$8.6\pm 1.7$&$2.0 \pm 0.4$\\
$\bar \nu_{\mu} p \rightarrow \mu^+ n$&$\pi^- \rightarrow
\mu^- \bar \nu_{\mu}$ DIF&$3.8\pm 1.9$&$0.9\pm 0.4$\\
$\bar \nu_e p \to e^+ n$&$\pi \rightarrow
e \nu$ and $\mu \rightarrow e \nu \bar \nu$
DIF&$0.1\pm 0.1$&$0$\\
\tableline
Total with Neutrons&&$12.5 \pm 2.9$&$2.9 \pm 0.6$\\
\tableline
$\nu_{\mu} \mbox{C}\to \mu^- X$&$\pi^+ \rightarrow \mu^+
\nu_{\mu}$ DIF&$11.3\pm 5.6$&$0.1\pm 0.1$\\
$\nu_e ~^{12}C \to e^-~^{12}N$&$\mu^+ \rightarrow
e^+ \bar \nu_{\mu} \nu_e$ DAR&$666.7 \pm 133.3$&$4.0\pm 0.8$\\
$\nu_e ~^{13}C \to e^-~^{13}N$&$\mu^+ \rightarrow
e^+ \bar \nu_{\mu} \nu_e$ DAR&$45.6 \pm 9.1$&$0.3\pm 0.1$\\
$\nu e \rightarrow \nu e$&$\pi^+ \rightarrow \mu^+ \nu_{\mu}$,
$\mu^+ \rightarrow
e^+ \bar \nu_{\mu} \nu_e$ DAR&$56.7 \pm 5.7$&$0.3\pm 0.1$\\
$\nu e \rightarrow \nu e$&$\pi \rightarrow
\mu \nu_{\mu} $ DIF&$8.4\pm 1.7$&$0.1\pm 0.1$\\
$\nu_e \mbox{C}\to e^- X$&$\pi \rightarrow e \nu_e$
DAR&$5.1\pm 1.0$&$0$\\
$\nu_{\mu} \mbox{C}\to \pi X$&$\pi \rightarrow
\mu \nu_{\mu} $ DIF&$0.3 \pm 0.1$&$0$\\
$\nu_e \mbox{C}\to e^- X$&$\pi \rightarrow
e \nu$ and $\mu \rightarrow e \nu \bar \nu$ 
DIF&$0.9\pm 0.2$&$0$\\
\tableline
Total without Neutrons&&$795.0 \pm 133.9$&$4.8 \pm 0.8$\\
\tableline
Grand Total &&$1589.5\pm 134.1$&$16.9\pm 1.3$\\
\tableline
\tableline
$100\%$ Transmutation&$\mu^+ \rightarrow
e^+ \bar \nu_{\mu} \nu_e$ DAR&$16670 \pm 1667$&$3830\pm 460$\\
\end{tabular}
\end{table}

\begin{table}
\caption{The number of signal and background events
in the $36<E_e<60$ MeV energy range. Excess/Efficiency is the excess
number of events divided by the total efficiency.
The beam-off background has been scaled to the beam-on time. Also
shown in the table is the probability that the observed excess is due 
entirely to a statistical fluctuation. Results
are given for $R \ge 0$ (the full positron sample) and for $R>30$.
The different selection criteria are described in section 4.3.
(Note that selections VIa and VIb are restricted-geometry tests
described in section 6.2.)}
\label{IV}
\begin{tabular}{lcccccr}
Selection&Signal&Beam Off&$\nu$ Bkgd.&Excess&Excess/Efficiency&
Fluct. Prob.\\
\tableline
I, $R \ge 0$&221&$133.6 \pm 3.1$&$53.5 \pm 6.8$&$33.9 \pm 16.6$
&$130\pm 64$&\\
I, $R>30$&13&$2.8 \pm 0.4$&$1.5 \pm 0.3$
&$8.7 \pm 3.6$&$146\pm 61$&
$1.0 \times 10^{-3}$\\
II, $R \ge 0$&245&$156.3 \pm 3.3$&$57.6 \pm 7.3$&$31.1\pm 17.6$
&$111\pm 63$&\\
II, $R>30$&14&$4.1 \pm 0.5$&$1.6 \pm 0.3$
&$8.3\pm 3.8$&$129\pm58$&
$3.8 \times 10^{-3}$\\
III, $R \ge 0$&285&$187.3 \pm 3.6$&$67.9 \pm 8.6$&$29.8\pm 19.3$
&$90\pm58$&\\
III, $R>30$&17&$5.3 \pm 0.6$&$1.9 \pm 0.3$
&$9.8\pm 4.2$&$129\pm 54$&
$2.1 \times 10^{-3}$\\
IV, $R \ge 0$&407&$260.3 \pm 4.3$&$93.2 \pm 11.9$&$53.5\pm 23.8$
&$119\pm 53$&\\
IV, $R>30$&26&$6.5 \pm 0.7$&$2.6 \pm 0.5$
&$16.9\pm 5.1$&$163\pm 51$&
$1.2 \times 10^{-5}$\\
V, $R \ge 0$&401&$255.3 \pm 4.2$&$87.6 \pm 11.2$&$58.1\pm 23.3$
&$135\pm 54$&\\
V, $R>30$&25&$4.5 \pm 0.6$&$2.4 \pm 0.4$
&$18.1\pm 5.0$&$183\pm 50$&
$3.8 \times 10^{-7}$\\
VI, $R \ge 0$&300&$160.5 \pm 3.4$&$76.2 \pm 9.7$&$63.3\pm 20.1$
&$171\pm 54$&\\
VI, $R>30$&22&$2.5 \pm 0.4$&$2.1 \pm 0.4$
&$17.4\pm 4.7$&$205\pm 54$&
$4.1 \times 10^{-8}$\\
\tableline
VIa, $R \ge 0$&269&$122.0 \pm 2.9$&$71.6 \pm 9.1$&$75.4\pm 19.0$
&$217\pm 55$&\\
VIa, $R>30$&21&$2.0 \pm 0.4$&$2.0 \pm 0.4$
&$17.0\pm 4.6$&$211\pm 57$&
$2.5 \times 10^{-8}$\\
VIb, $R \ge 0$&99&$33.5 \pm 1.5$&$34.3 \pm 4.4$&$31.2\pm 11.0$
&$187\pm 66$&\\
VIb, $R>30$&6&$0.8 \pm 0.2$&$0.9 \pm 0.2$
&$4.3\pm 2.5$&$110\pm 63$&
$1.1 \times 10^{-2}$\\
\tableline
\end{tabular}
\end{table}

\begin{table}
\caption{The 26 beam-on events with $R>30$ and energy in the 
$36<E_e<60$ MeV range that satisfy selection IV. 
For each event is given the year recorded, 
energy, spatial position, and distance from the PMT surfaces.
Also given are the selections that each event satisfies.}
\label{V}
\begin{tabular}{lccccccr}
Event&Year&E(MeV)&X(cm)&Y(cm)&Z(cm)&D(cm)&Selections\\
\tableline
1&1993&47.6&-66&-84&-77&115&I-VI\\
2&1993&51.1&56&-96&53&103&I-VI\\
3&1994&40.1&-36&196&-203&53&I-VI\\
4&1994&44.2&69&-146&153&53&I-VI\\
5&1994&39.4&-169&96&-347&39&II-VI\\
6&1994&36.3&-156&-79&-207&84&I-VI\\
7&1994&56.8&-221&-24&-309&36&I-V\\
8&1994&52.9&21&106&71&143&IV-VI\\
9&1994&37.0&31&156&-105&93&IV-VI\\
10&1994&42.4&-14&-121&-239&78&IV-VI\\
11&1994&37.7&-91&119&209&109&I-VI\\
12&1994&54.3&-91&191&269&47&III-VI\\
13&1994&55.8&71&-99&-259&100&I-VI\\
14&1994&43.8&6&211&173&38&I-VI\\
15&1995&50.5&153&-159&-193&38&IV-V\\
16&1995&59.9&-132&-164&339&35&III-V\\
17&1995&49.2&-184&10&58&75&I-VI\\
18&1995&56.5&128&-150&199&49&I-VI\\
19&1995&37.4&45&-92&-239&107&IV-VI\\
20&1995&45.1&-186&105&-126&45&IV-VI\\
21&1995&46.7&179&-93&-108&57&III-VI\\
22&1995&40.2&-37&-71&160&128&I-VI\\
23&1995&47.7&-126&-135&-263&64&IV\\
24&1995&45.9&-161&87&-337&49&I-VI\\
25&1995&36.3&46&150&107&100&IV-VI\\
26&1995&37.6&-73&107&-257&129&IV-VI\\
\tableline
\end{tabular}
\end{table}

\begin{table}\centering
\caption{Kolmogorov consistency probability for the
distribution of various spatial quantities for events with $36<E_e<60$
MeV that satisfy selection VI. 
The expected Z distribution is sensitive to $\Delta m^2$ for
oscillation events; we used $\Delta m^2 = 100$ eV$^2$. $D$ is the distance
from the phototube surfaces and $D_{YZ}$ is the distance from the
bottom, upstream end of the detector.}
\label{VI}
\begin{tabular}{|c||c|c|c|}
Distribution &Probability For All R&Probability for $R > 1.5$
&Probability for $R > 30.$ \\
\hline\hline
       X          &0.074    &0.763  &0.147 \\
       Y          &0.129    &0.196  &0.131 \\
       Z          &0.047    &0.713  &0.889 \\
       $D$        &0.314    &0.739  &0.620 \\
       $D_{YZ}$   &0.016    &0.535  &0.891 \\
\hline
\end{tabular}
\end{table}

\begin{table}
\caption{Consistency check on the time dependence of numbers of events with
$R>30$ and $36<E_e<60$ MeV.  
``Prob 1'' is the probability of a worse inconsistency being observed.
``Prob 2'' is the probability of the 1995 excess accidentally being as low
as observed given the overall excess.} 
\label{VII}
\begin{tabular}{|c||c|c||c|c||c|c|c|}
 &\multicolumn{2}{c||}{Selection I }&\multicolumn{2}{c||}{Selection VI}
& Coulombs & $\bar\nu_\mu$ & duty \\
\cline{1-5}
              & On & Off & On & Off   &     &fraction  &ratio \\
\hline\hline
1993          & 2  & 8   & 2  & 7     & 1787    & 0.12 & 0.076 \\
\hline
1994          & 7  & 9   & 11 & 15    & 5904    & 0.42 & 0.080 \\
\hline
1995          & 4  & 23  & 9  & 14    & 7081    & 0.46 & 0.060 \\
\hline
\hline
Prob 1 & \multicolumn{2}{c||}{0.47 }  &\multicolumn{2}{c||}{ 0.80}  &
\multicolumn{3}{c|}{ }\\
\cline{1-5}
Prob 2 & \multicolumn{2}{c||}{0.19 }  & \multicolumn{2}{c||}{0.46 } &
\multicolumn{3}{c|}{ }\\
\end{tabular}
\end{table}

\begin{table}
\caption{The number of signal and neutrino background events
in the $20<E_e<60$ MeV energy range with selection VI, 
together with the oscillation
probability if the observed excess is due to neutrino oscillations. Results
are given for $\chi^2$ and ${\cal L}$ fits to the R distribution for 
all positrons
and for the $R>30$
sample.}
\label{VIII}
\begin{tabular}{lccccr}
Selection&Signal&Beam Off&$\nu$ Bkgd.&Excess&Oscillation Prob.\\
\tableline
$\chi^2$ 
R Fit&$64.3 ^{+18.5}_{- 16.7}$&--&$12.5 \pm 2.9$&$51.8 ^{+18.7}_{- 16.9}$
&$(0.31^{+0.11}_{-0.10}
\pm 0.05 )\%$\\
${\cal L}$ R Fit&$57.2 ^{+19.3}_{- 18.0}$&--&$12.5 
\pm 2.9$&$44.7 ^{+19.5}_{- 18.2}$
&$(0.27^{+0.12}_{-0.11}
\pm 0.04 )\%$\\
$R>30$&38&$9.2 \pm 0.8$&$7.7 \pm 1.0$
&$21.1 \pm 6.3$&$(0.55\pm 0.16 \pm 0.07 )\%$\\
\tableline
\end{tabular}
\end{table}

\begin{figure}
\epsfbox{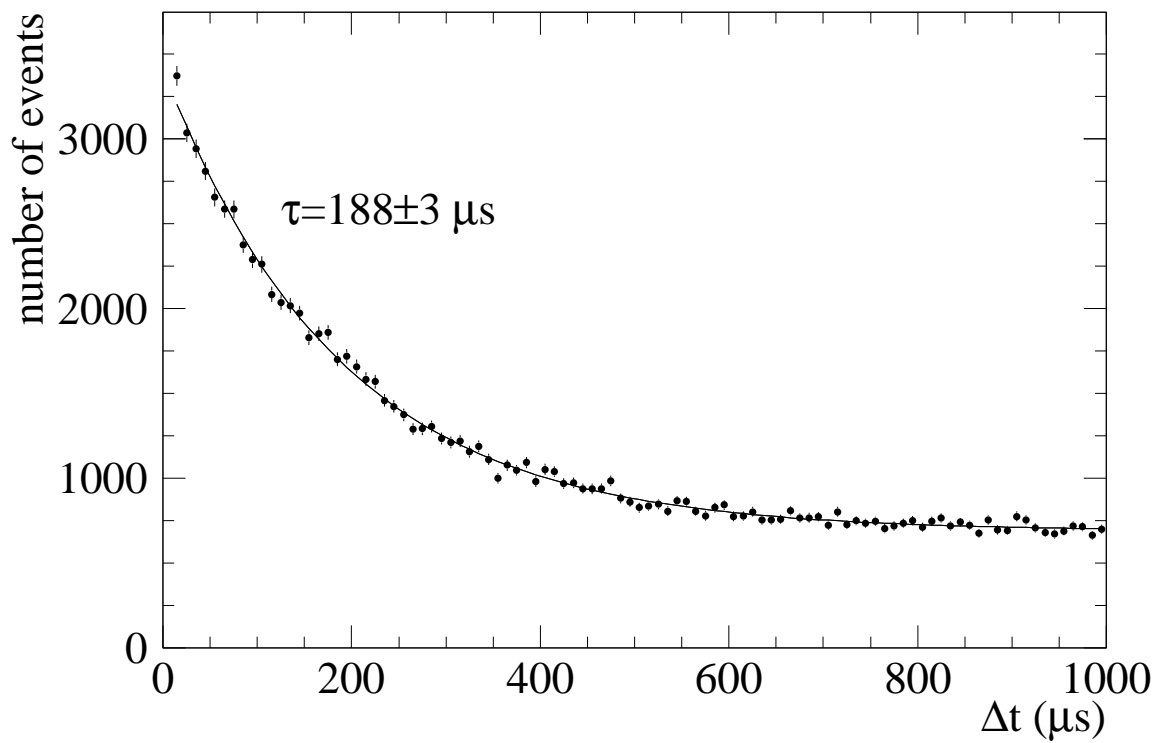}
\caption{Time difference between neutrons and subsequent photons
for correlated plus accidental $\gamma$s. The solid curve is a fit
to a sum of an exponential for correlated $\gamma$s and a flat
background for accidental $\gamma$s.}
\label{Fig. 1}
\end{figure}\newpage

\begin{figure}
\epsfbox{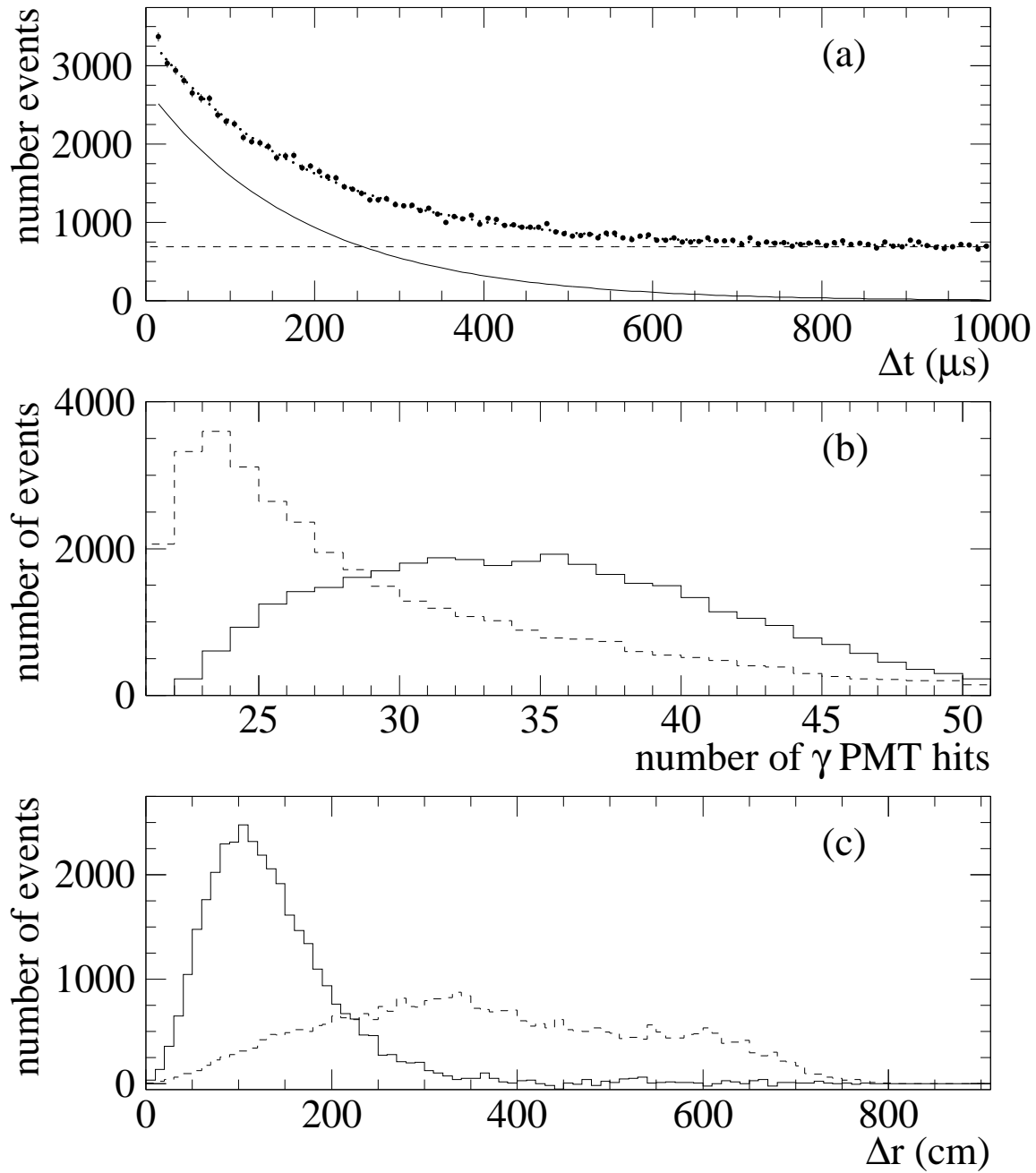}
\caption{Distributions obtained from cosmic-ray neutron data for
$\gamma$s that are correlated (solid) or uncorrelated (dashed)
with the primary event: (a) the time
between the photon and primary event; (b) the number of photon PMT hits;
(c) the distance between the photon and primary event. The raw data points
are also shown in (a).}
\label{Fig. 2}
\end{figure}\newpage

\begin{figure}
\epsfbox{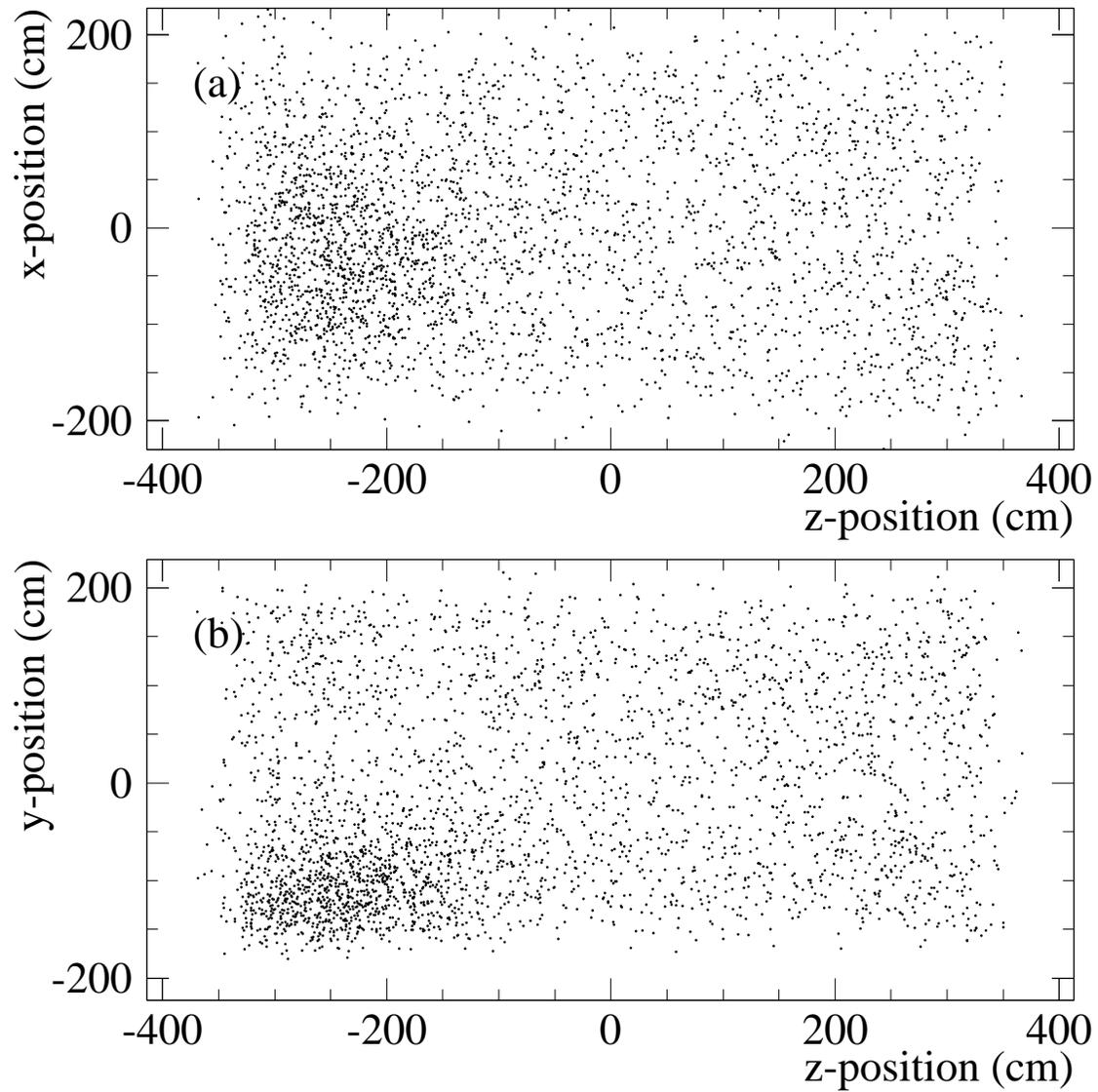}
\caption{Distributions of reconstructed position for accidental $\gamma$s in
the (a) X - Z and (b) Y - Z projections.}
\label{Fig. 3}
\end{figure}\newpage

\begin{figure}
\epsfbox{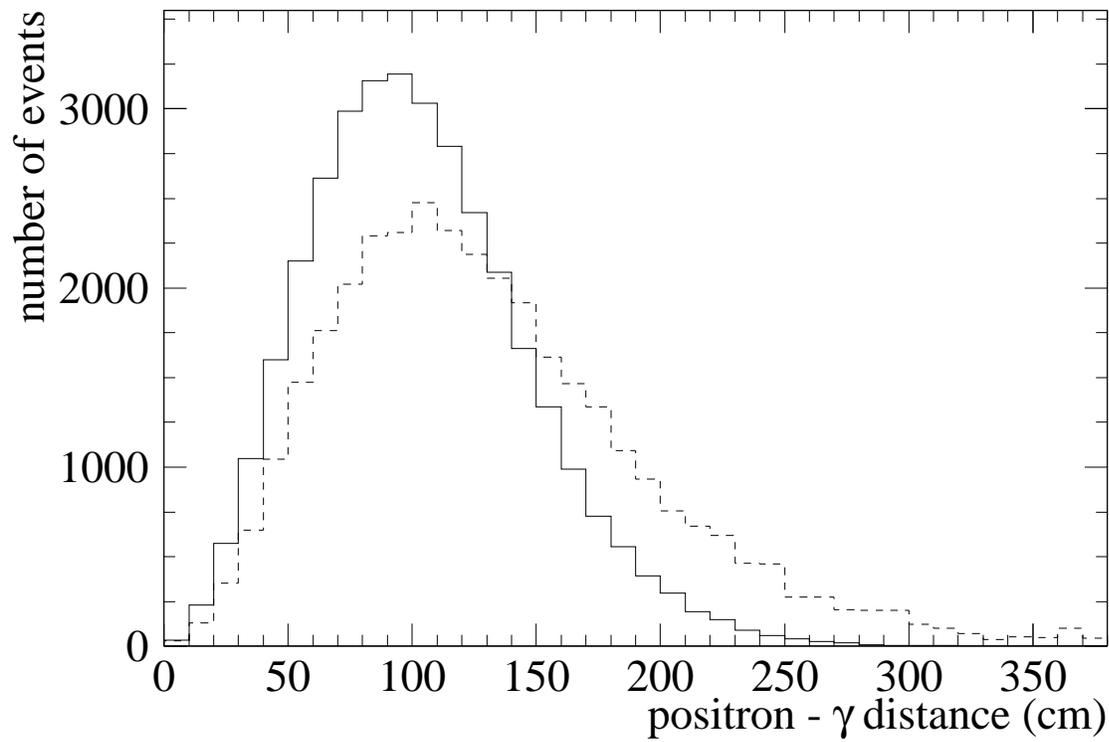}
\caption{Distribution of reconstructed distance between
$e^+$ and a correlated 
$\gamma$ from the Monte Carlo simulation (solid) 
and the cosmic ray neutron sample (dashed).}
\label{Fig. 4}
\end{figure}\newpage

\begin{figure}
\epsfbox{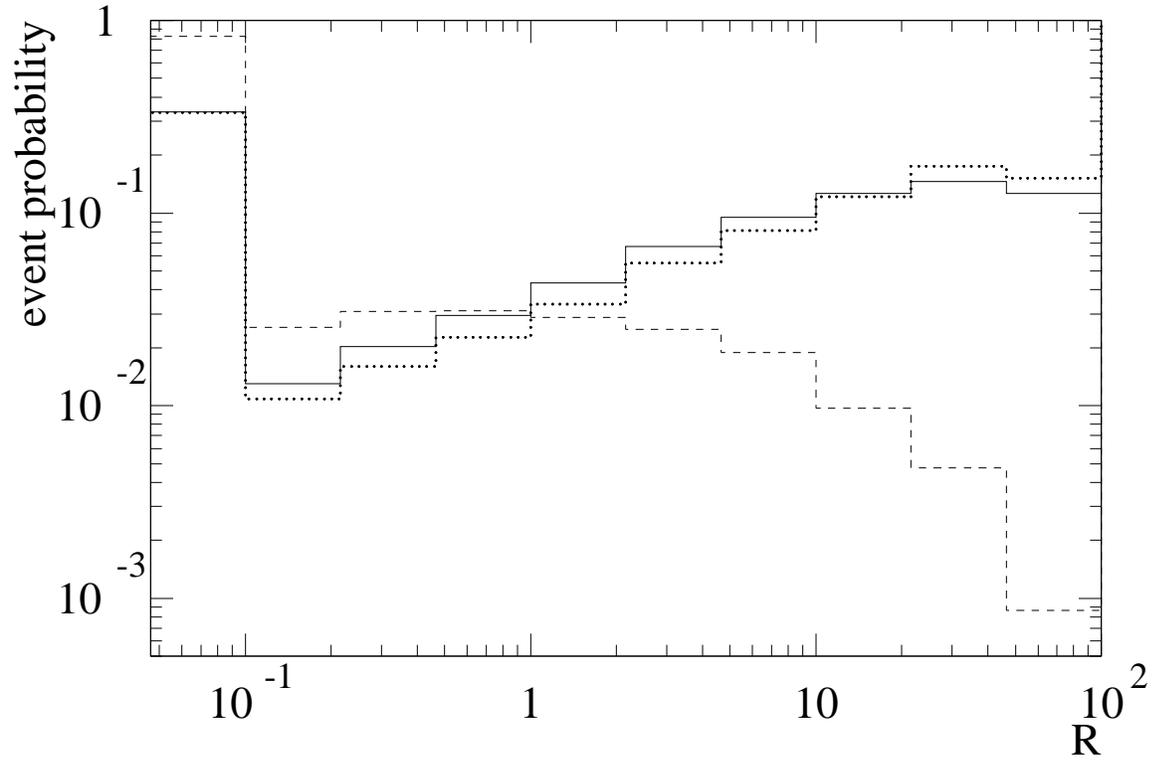}
\caption{Measured R distribution for events
with the $\gamma$ correlated (solid) and uncorrelated (dashed)
with the primary event. 
The dotted curve is also for correlated $\gamma$s,
but with the measured $\Delta r$ values replaced by values distributed
according to the Monte Carlo prediction.}
\label{Fig. 5}
\end{figure}\newpage

\begin{figure}
\epsfbox{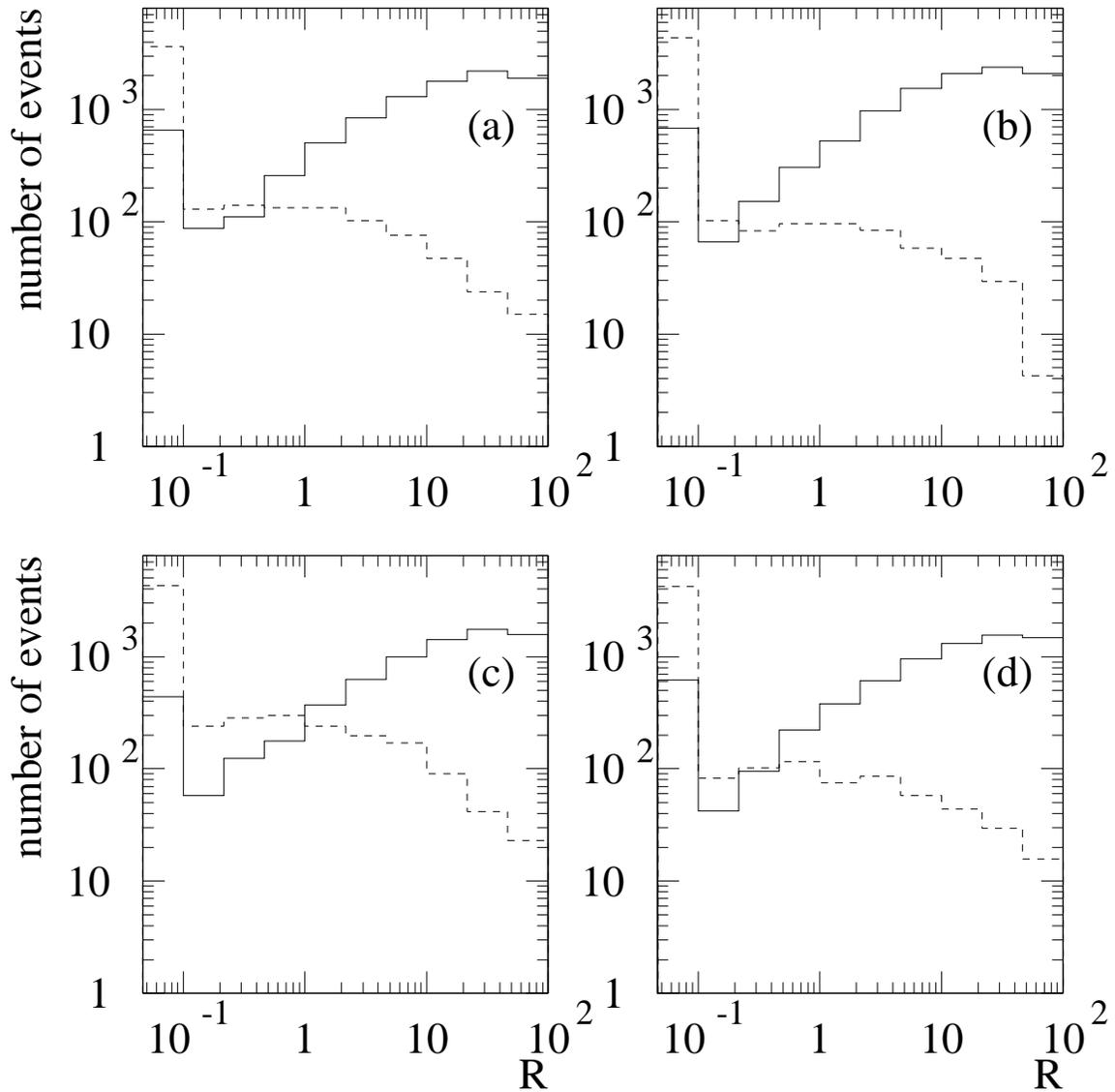}
\caption{The R distributions for correlated $\gamma$s (solid) and 
accidental $\gamma$s (dashed)
for primary events in each of the four quadrants of the Y - Z plane:
(a) $Y>0, Z<0$; (b) $Y>0, Z>0$; (c) $Y<0, Z<0$; (d) $Y<0, Z>0$.}
\label{Fig. 6}
\end{figure}\newpage

\begin{figure}
\epsfbox{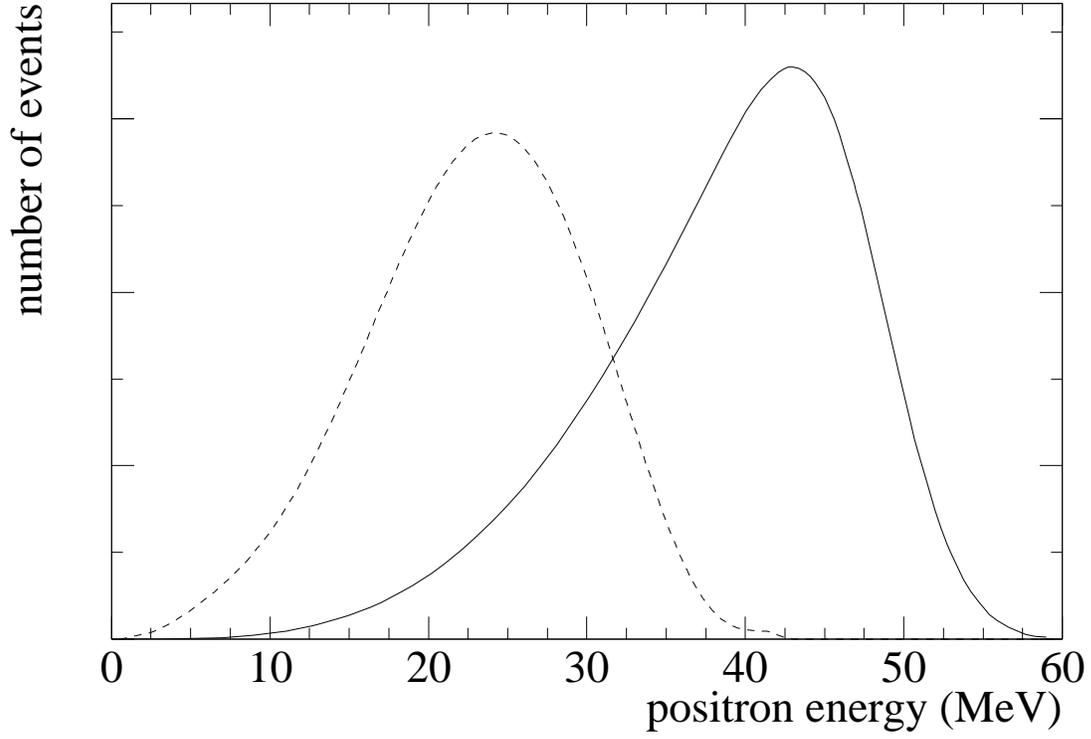}
\caption{Energy distribution expected for oscillation events at large
$\Delta m^2$ ($\Delta m^2 \rightarrow \infty$) (solid) and $\nu_e$C scattering
events (dashed). The distributions include the experimental energy resolution
as determined from the sample of electron events from muon decay.}
\label{Fig. 7}
\end{figure}\newpage

\begin{figure}
\epsfbox{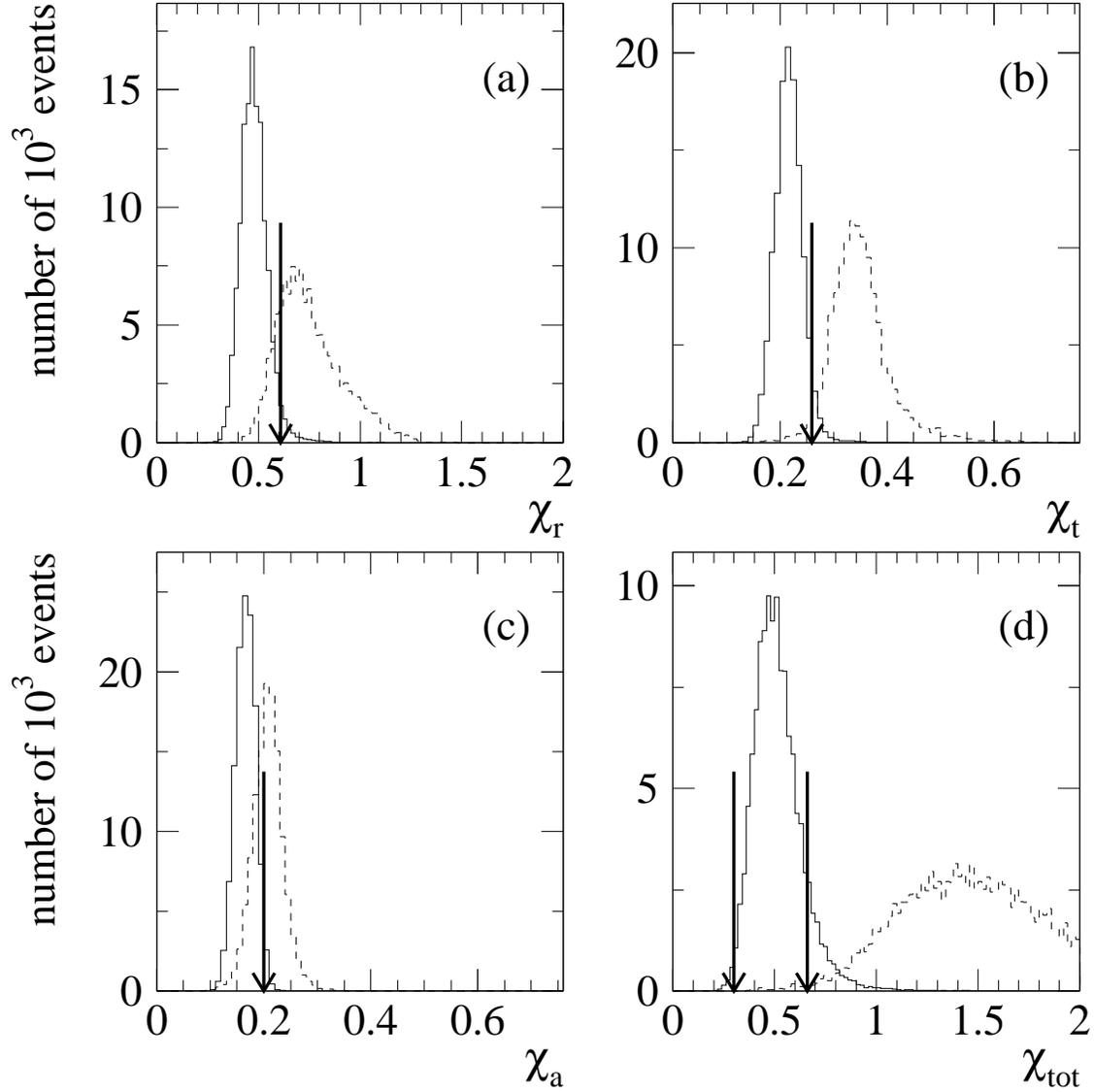}
\caption{Distribution of the PID parameters 
for decay electrons (solid) and neutrons (dashed)
with deposited energy between $36<E_e<60$
$\,{\rm MeV}$. (a) $\chi_r$; (b) $\chi_t$; (c) $\chi_a$; (d) $\chi_{tot}$.
The arrows show the locations of the $\chi$ requirements for
selection VI.}
\label{Fig. 8}
\end{figure}\newpage

\begin{figure}
\epsfbox{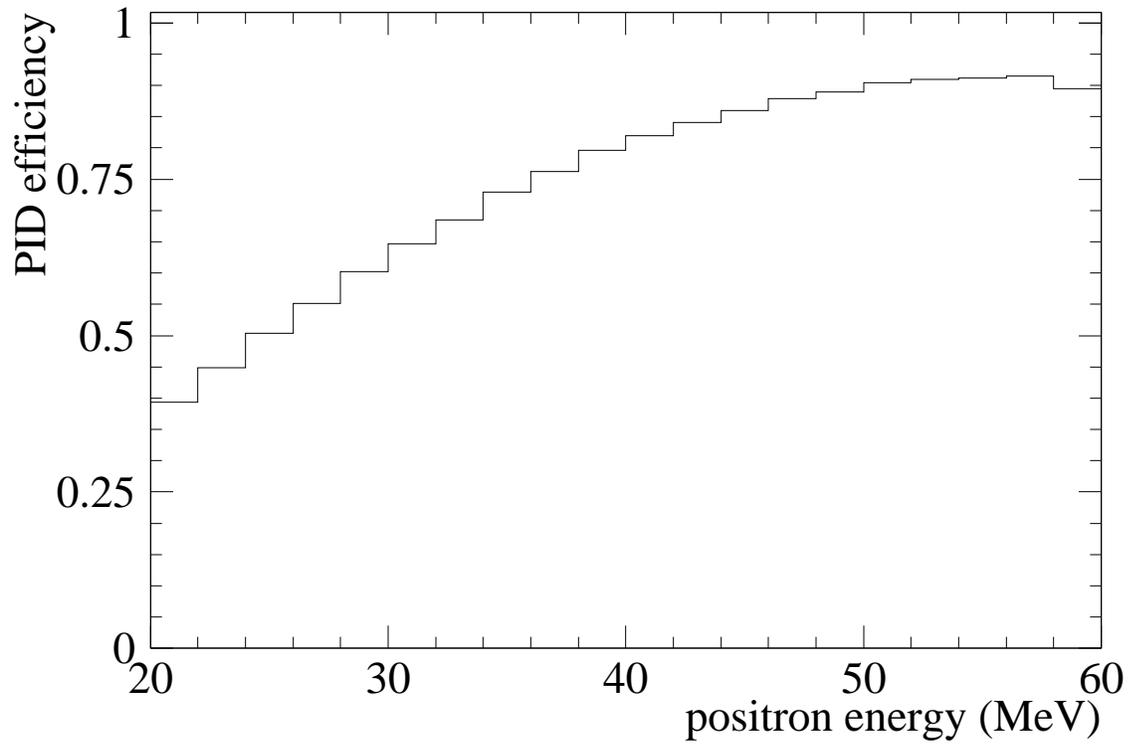}
\caption{The PID efficiency for selection VI as a function
of electron energy.}
\label{Fig. 9}
\end{figure}\newpage

\begin{figure}
\epsfbox{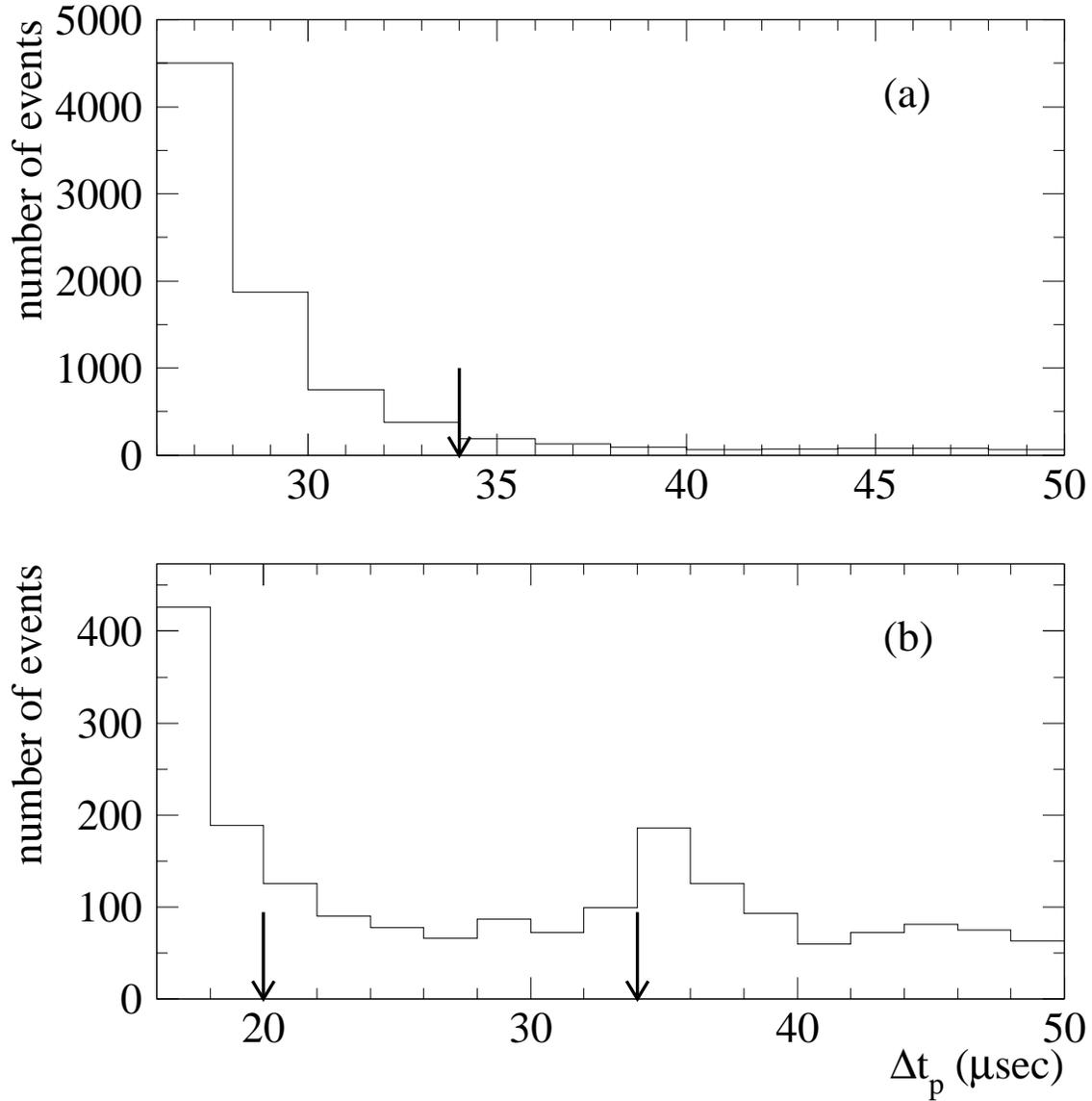}
\caption{Distribution
of $\Delta t_p$ 
for beam-off events that satisfy the other positron selection 
criteria for (a) events
with no $\Delta t_p$ 
requirement and (b) events with no correlated activities within 34 $\mu s$.
The arrows show the locations of the $\Delta t_p$ requirements.}
\label{Fig. 10}
\end{figure}\newpage

\begin{figure}
\epsfbox{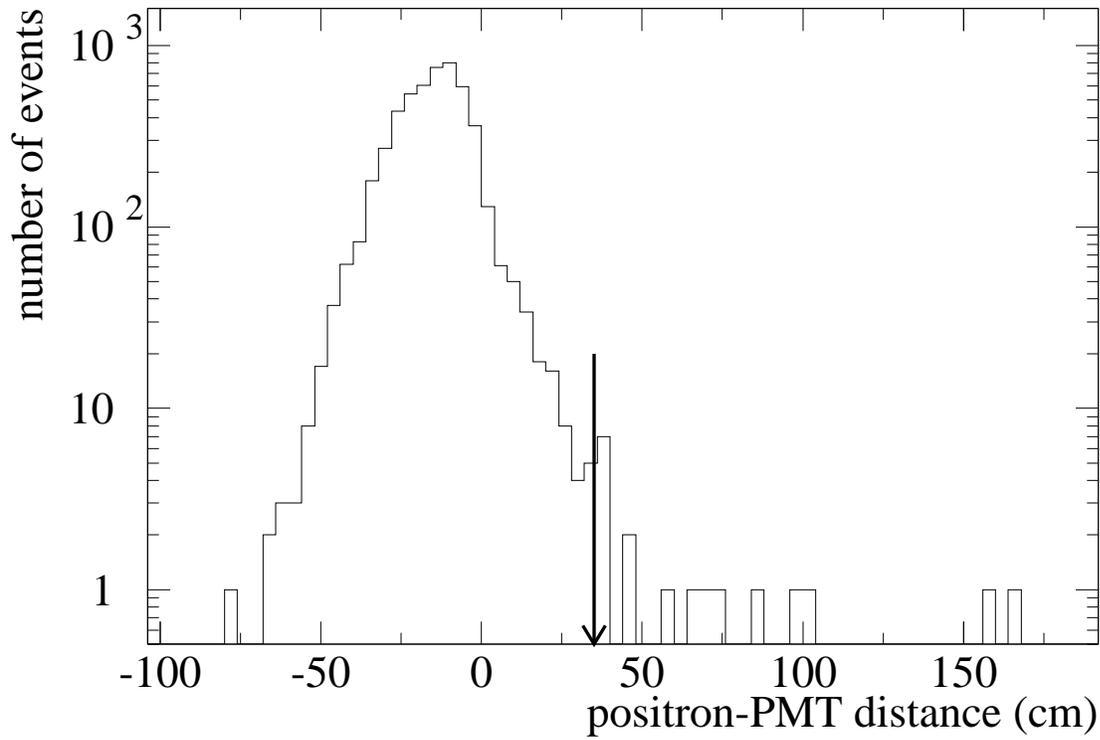}
\caption{The $D$ distribution, the reconstructed distance from the
PMT surfaces, for a sample of Monte Carlo electron events generated behind
the PMT surfaces. The arrow shows the location of the $D>35$ cm cut.}
\label{Fig. 11}
\end{figure}\newpage

\begin{figure}
\epsfbox{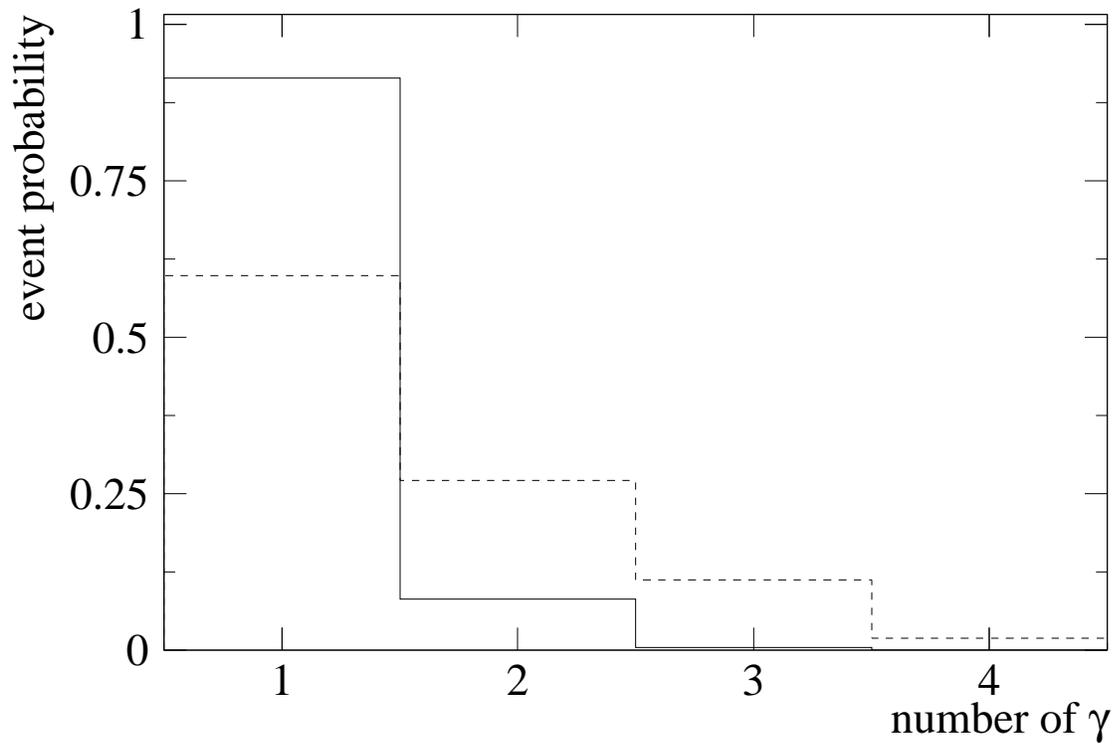}
\caption{Number of associated $\gamma$s ($R>1.5$) distribution expected
for oscillation events (solid)
and for beam-off events (dashed).}
\label{Fig. 12}
\end{figure}\newpage

\begin{figure}
\epsfbox{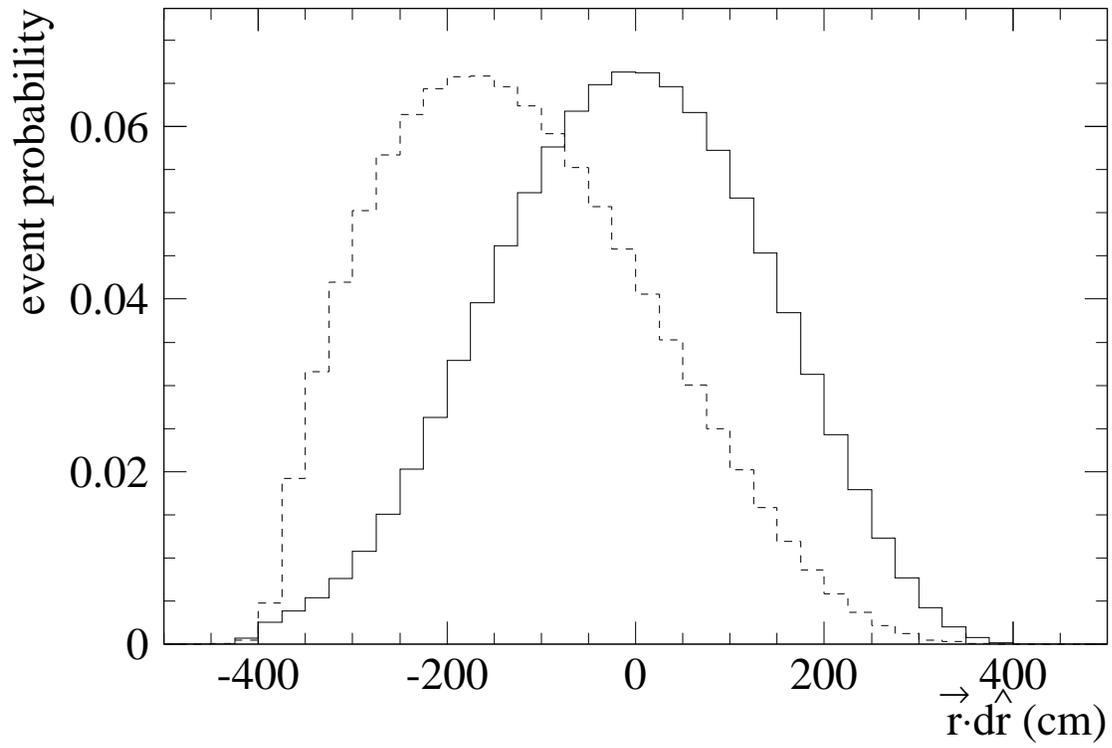}
\caption{Distribution of $\vec{r} \cdot \hat{dr}$ 
for $\nu_e$C events (solid) and beam-off 
events (dashed).}
\label{Fig. 13}
\end{figure}\newpage

\begin{figure}
\epsfbox{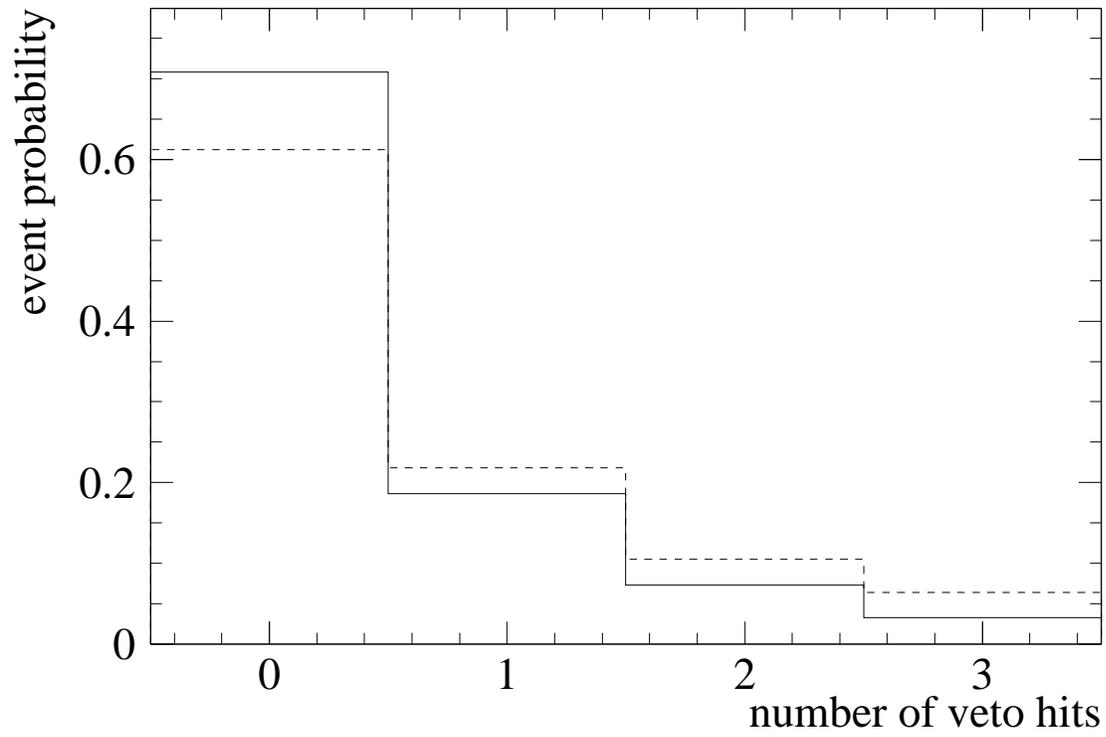}
\caption{Distribution of veto hits for laser events (solid) and 
beam-off events (dashed).}
\label{Fig. 14}
\end{figure}\newpage

\begin{figure}
\epsfbox{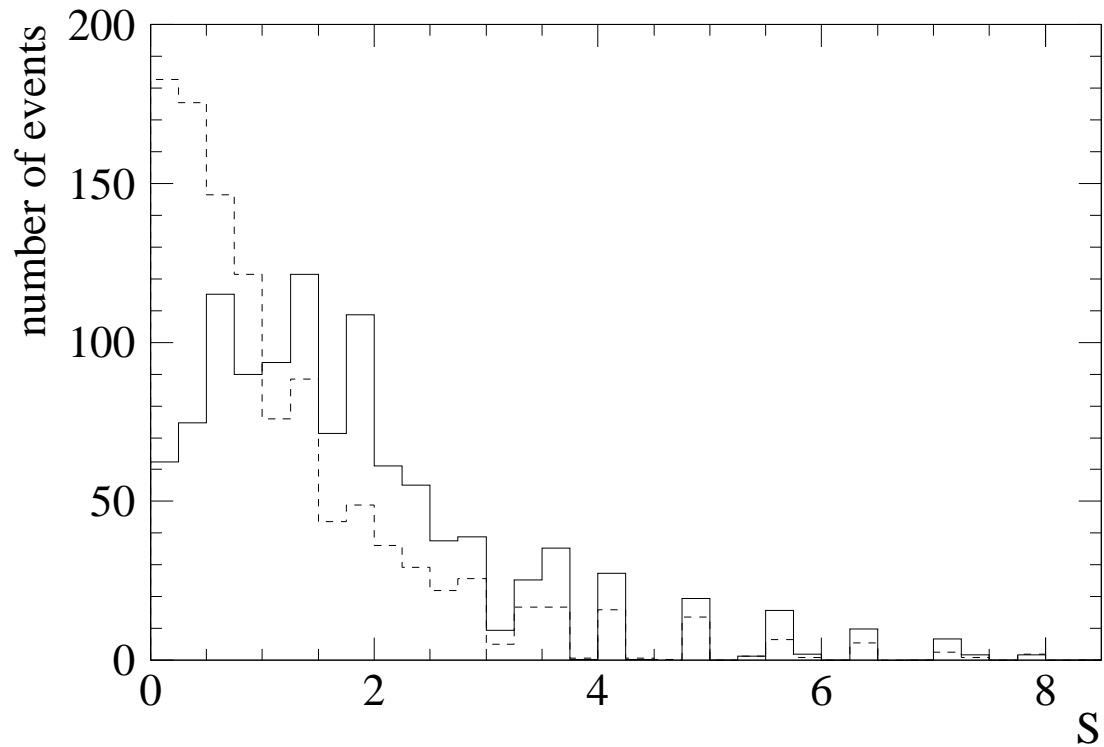}
\caption{Distribution of S for $\nu_e C$ events (solid)
and beam-off events (dashed).}
\label{Fig. 15}
\end{figure}\newpage

\begin{figure}
\epsfbox{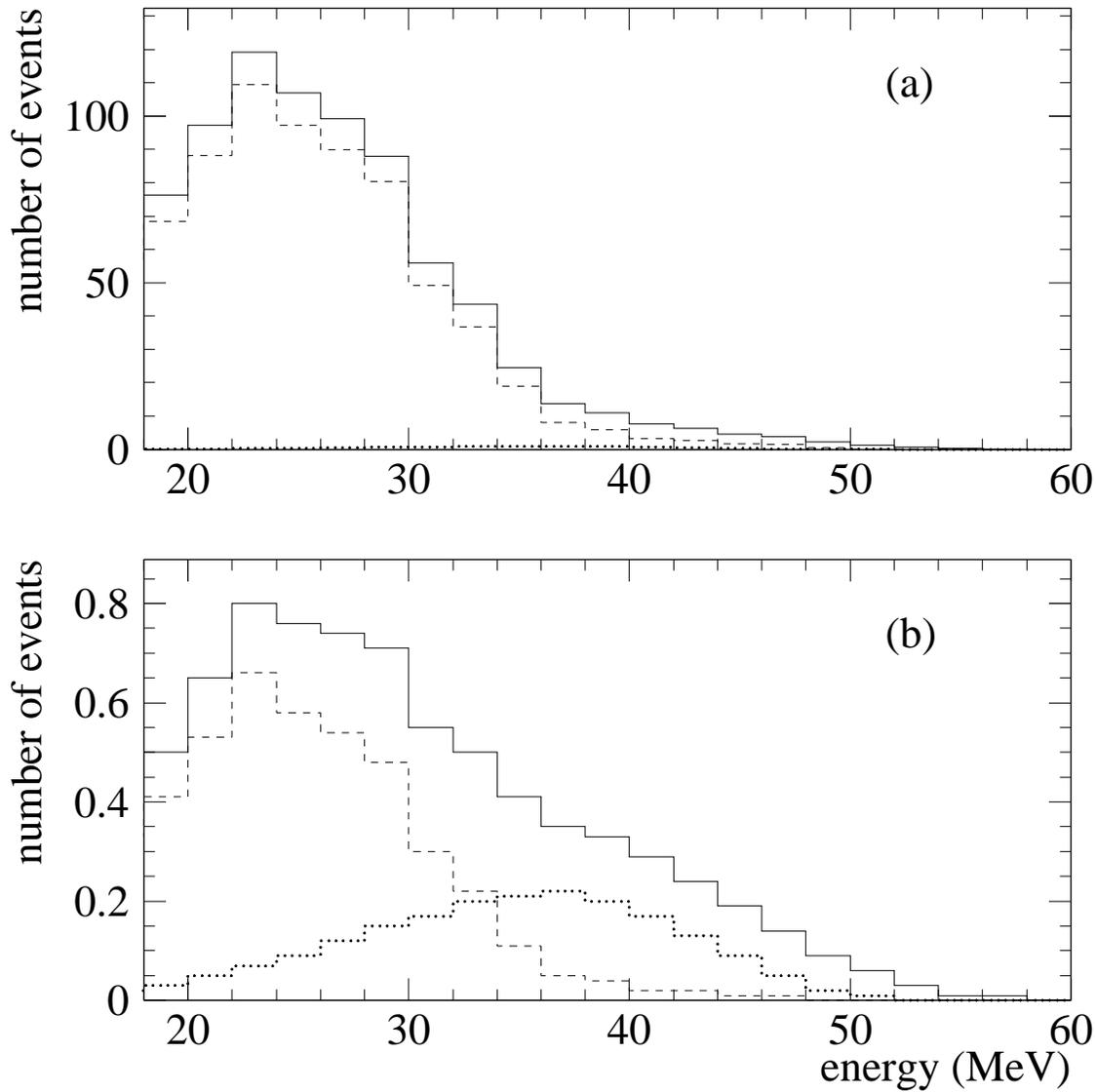}
\caption{Total beam-related background (solid curve) calculated 
as a function of energy for 
(a) $R \ge 0$ and (b) $R>30$. Also shown are the contributions from
the backgrounds
$\nu_e C \rightarrow e^- X$ scattering (dashed curve) and $\mu^-$ DAR
(dotted curve).}
\label{Fig. 16}
\end{figure}\newpage

\begin{figure}
\epsfbox{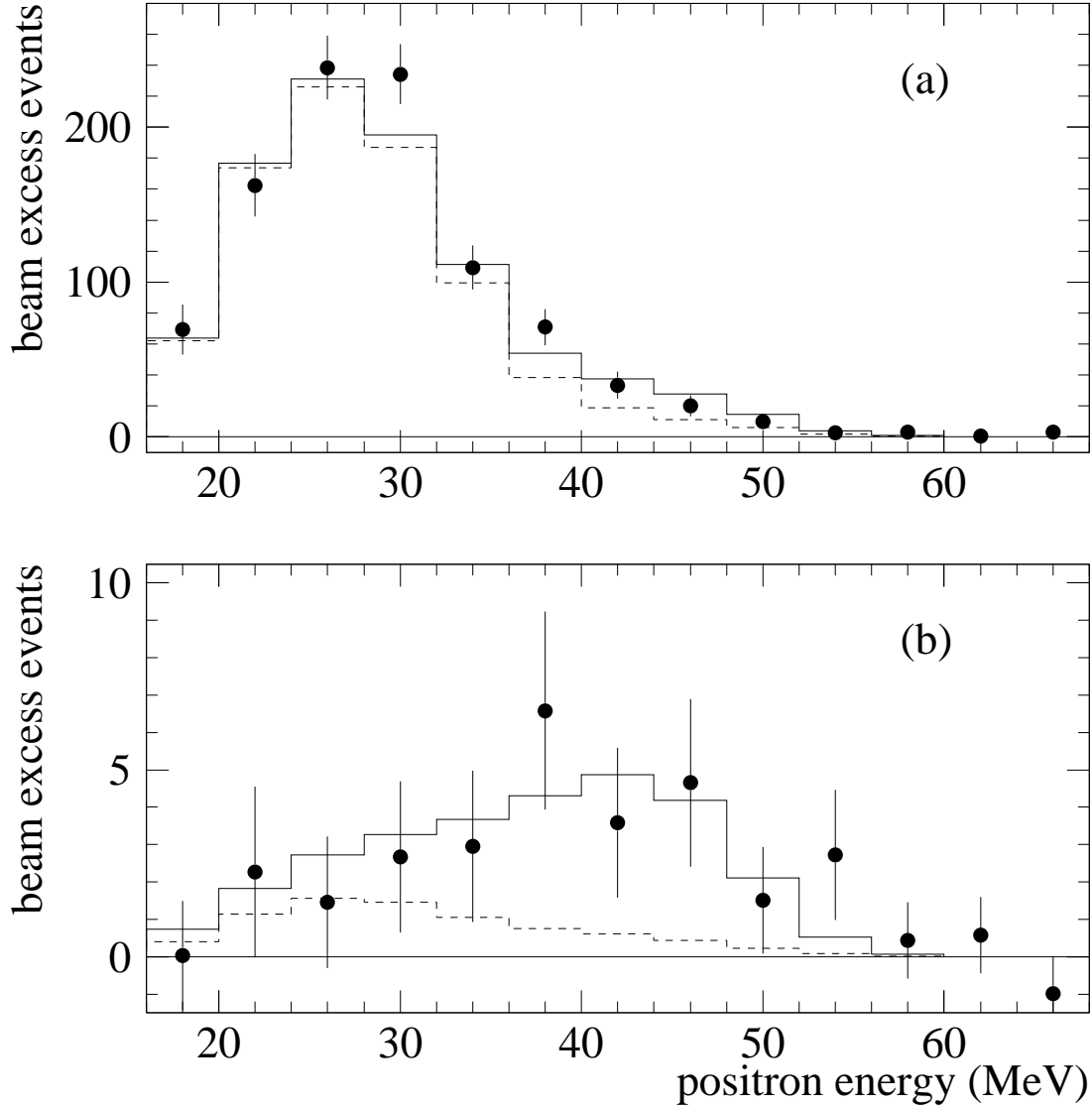}
\caption{The energy distribution for events with (a) $R \ge 0$ and
(b) $R>30$. Shown in the figure are the beam-excess data,
estimated neutrino background (dashed), and expected
distribution for neutrino oscillations at large $\Delta m^2$ plus
estimated neutrino background (solid).}
\label{Fig. 17}
\end{figure}\newpage

\begin{figure}
\epsfbox{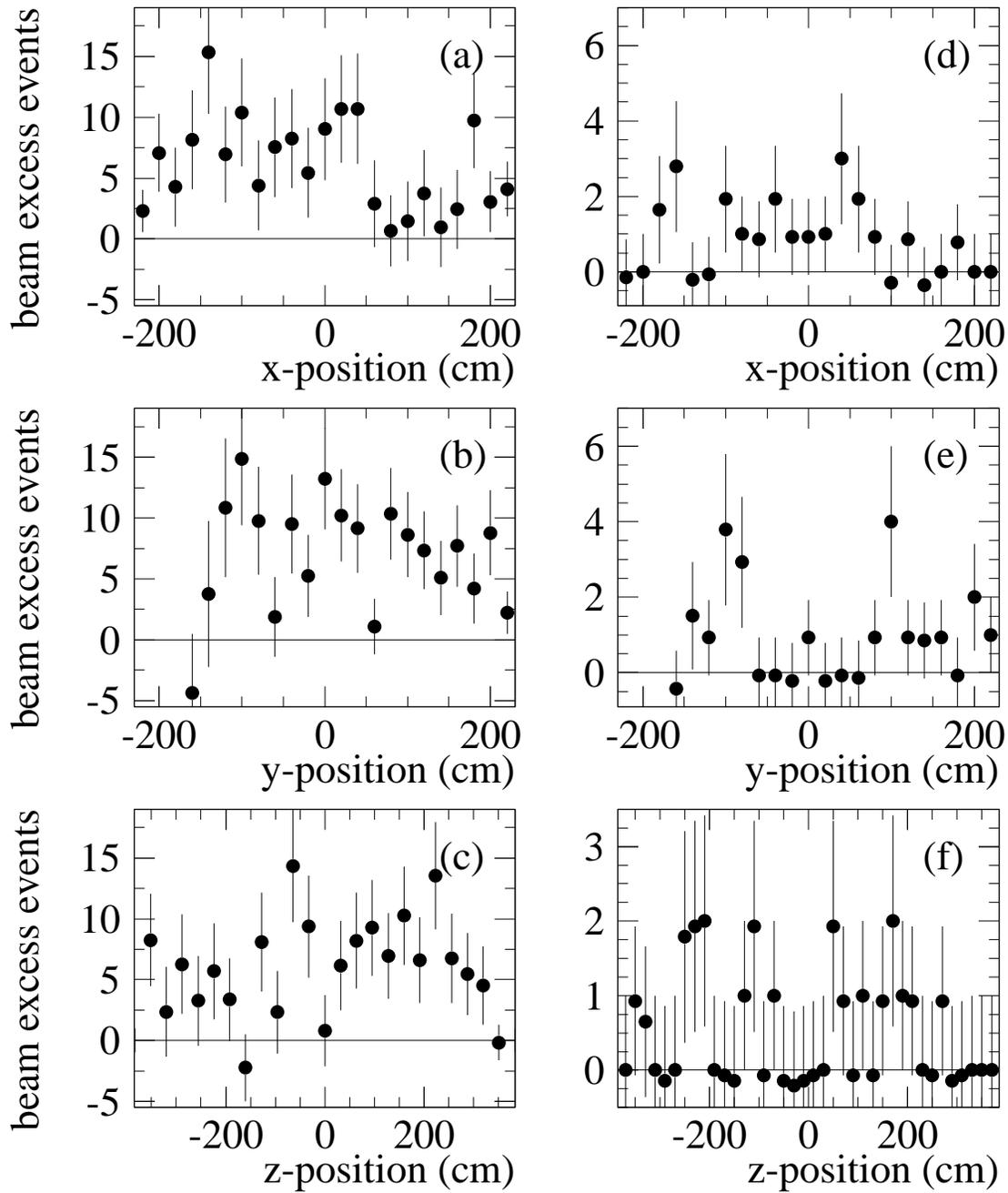}
\caption{The spatial distributions for
beam-excess data events with $36<E_e<60$ MeV. (a) - (c) are for
$R\ge 0$ and (d) - (f) are for $R>30$.}
\label{Fig. 18}
\end{figure}\newpage

\begin{figure}
\epsfbox{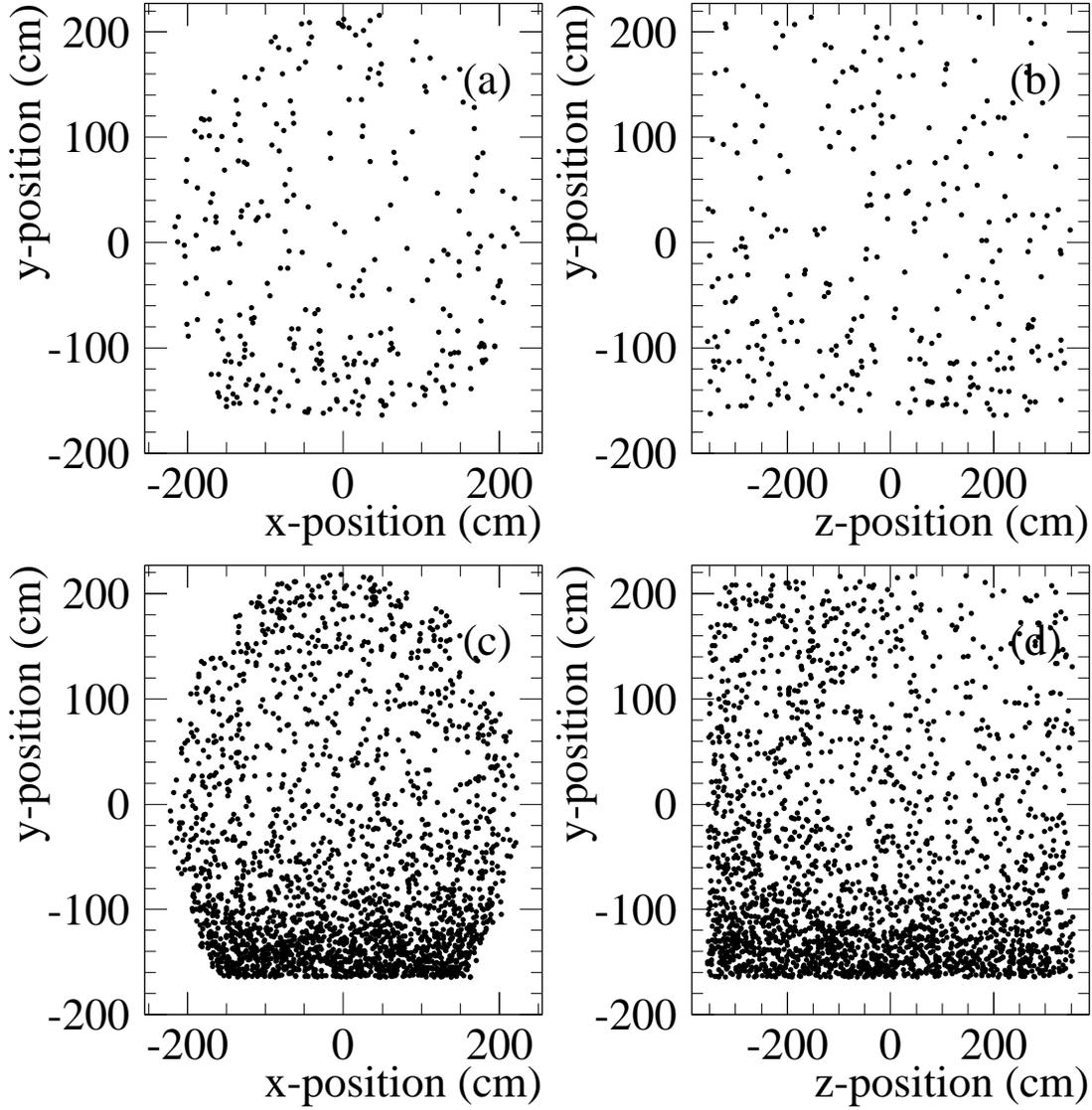}
\caption{Spatial distributions of
positron events with $36<E_e<60$ MeV and
$R\ge 0$ in the y - x and
Y - Z planes for (a,b) the 300 beam-on events and (c,d) the
2293 beam-off events. Note that the beam on-off excess is
139.5 events, so that less than half of the 300
beam-on events are due to neutrino interactions.}
\label{Fig. 19}
\end{figure}\newpage

\begin{figure}
\epsfbox{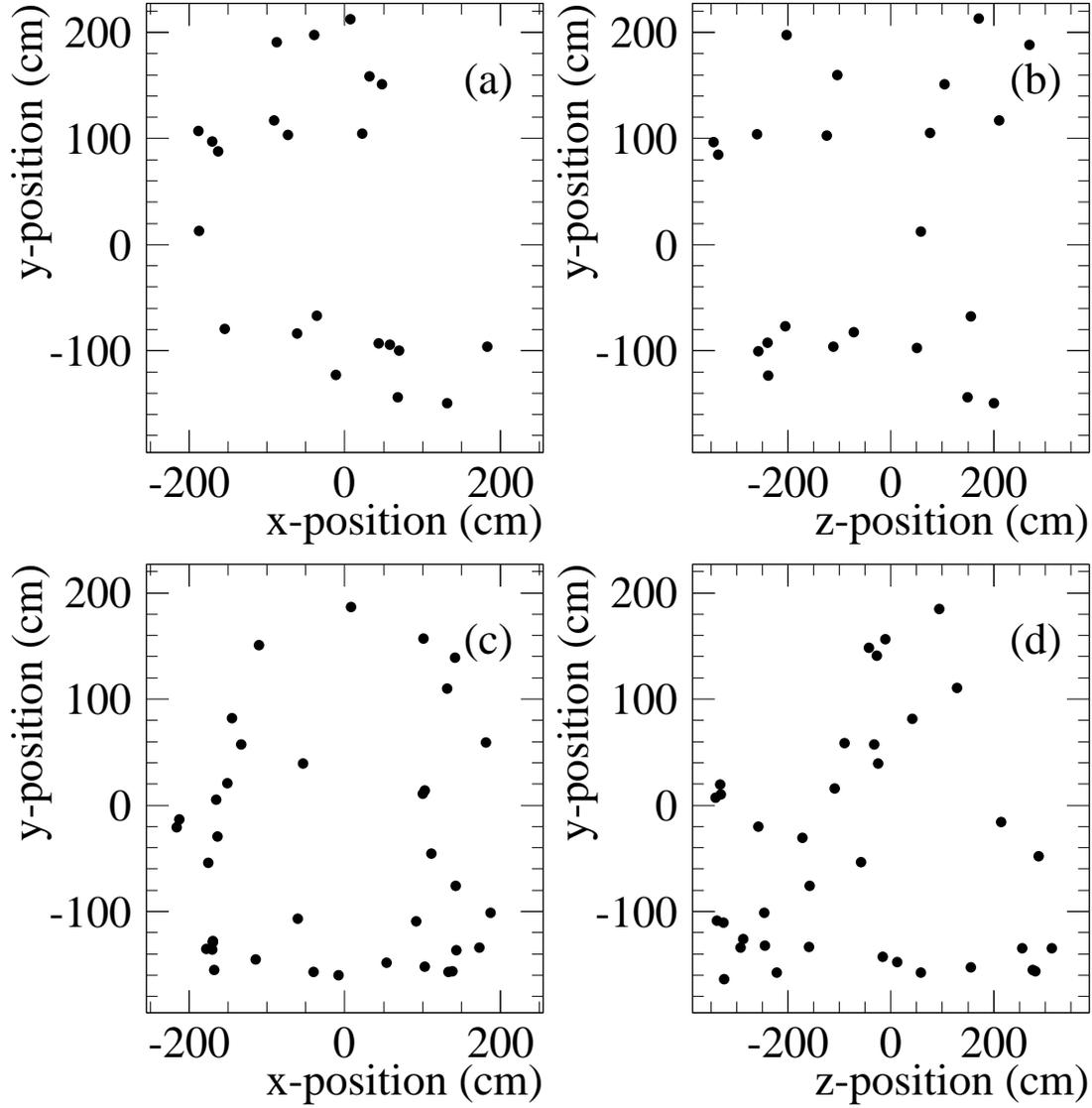}
\caption{Spatial distributions of
data events with $36<E_e<60$ MeV and
$R>30$ in the y - x and
Y - Z planes for (a,b) the 22 beam-on events and (c,d) the
36 beam-off events.}
\label{Fig. 20}
\end{figure}\newpage

\begin{figure}
\epsfbox{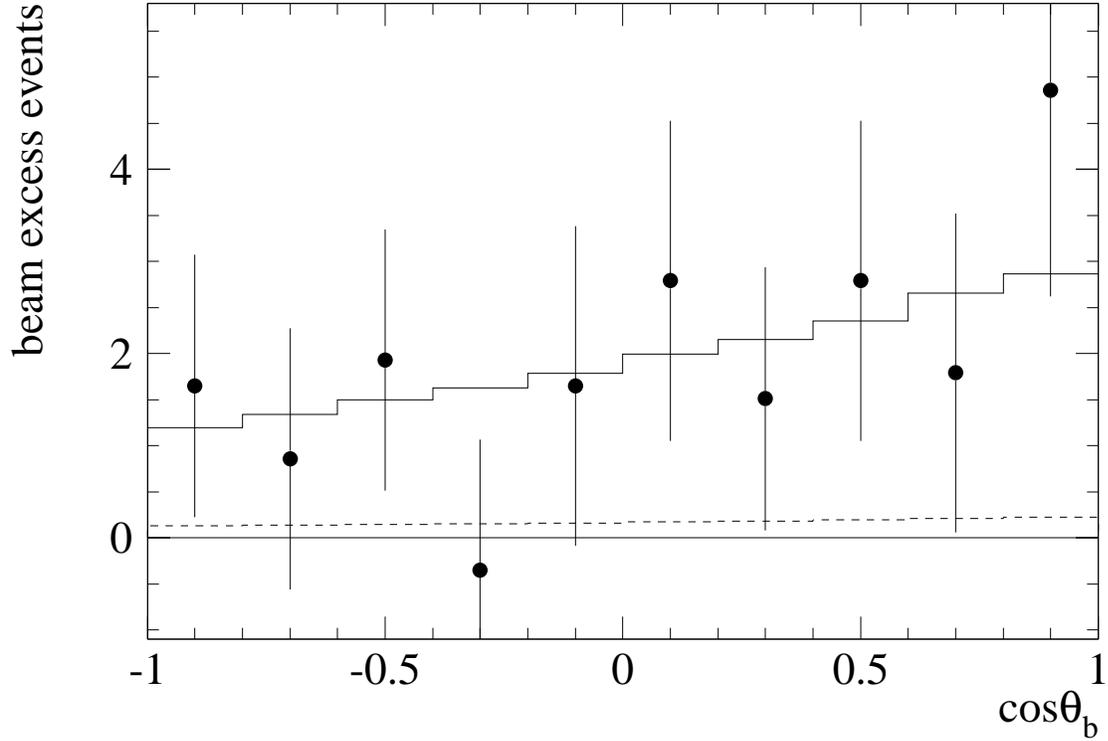}
\caption{The $\cos \theta_b$ distribution for beam-excess 
data events with
$36<E_e<60$ MeV and $R>30$ 
and that expected for
neutrino oscillations at large $\Delta m^2$ (solid).
The dashed curve is the estimated neutrino background.
$\theta_b$ is the $e^+$ angle with respect to the
neutrino direction.}
\label{Fig. 21}
\end{figure}\newpage

\begin{figure}
\epsfbox{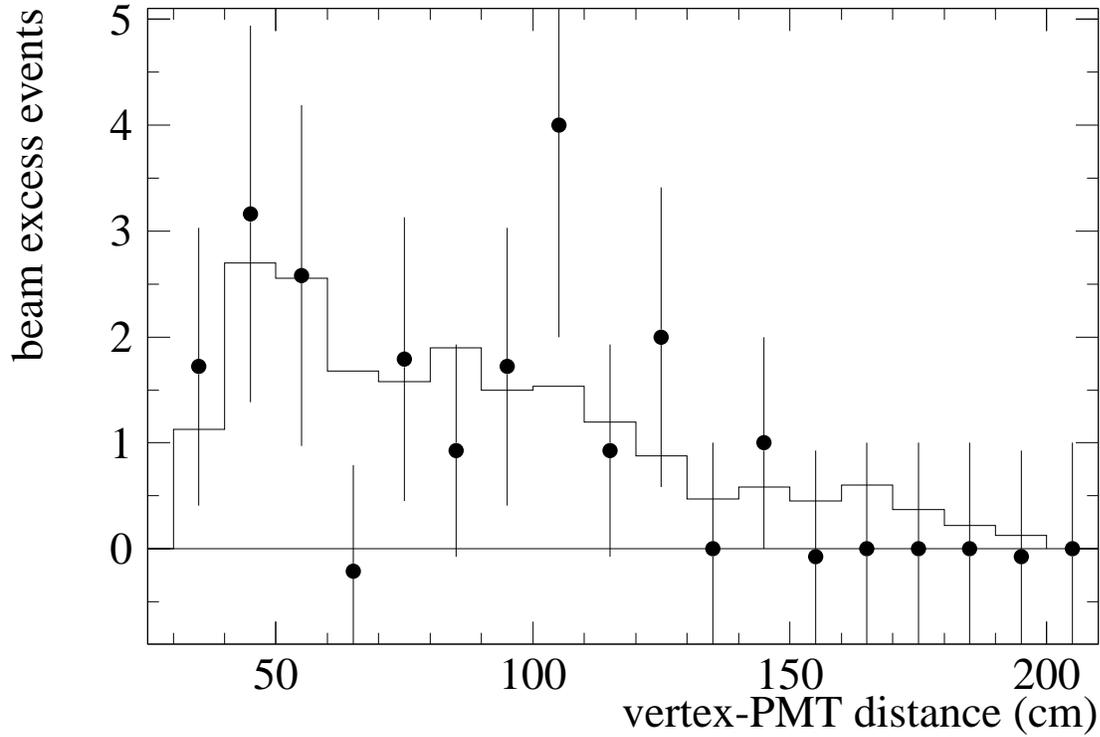}
\caption{Distribution of $D$, the distance of the reconstructed
vertex from the PMT surfaces, for beam-excess
data events with $36<E_e<60$ MeV and
$R>30$. The 
solid histogram is the expected distribution obtained from a
sample of $\nu_e C \rightarrow e^- X$ scattering events.}
\label{Fig. 22}
\end{figure}\newpage

\begin{figure}
\epsfbox{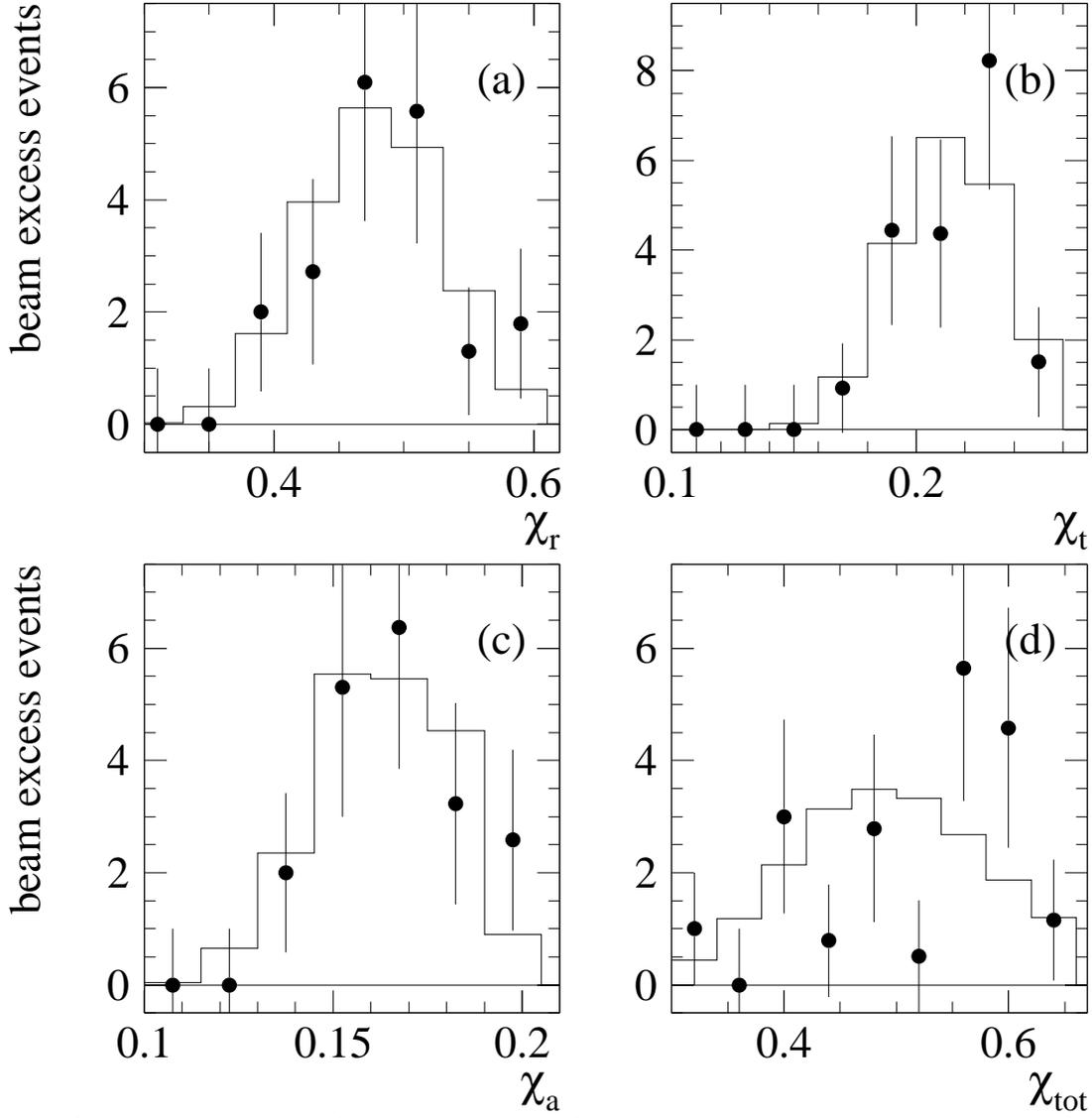}
\caption{Distribution of the $\chi$ parameters for beam-excess 
data events with 
$36<E_e<60$ MeV and $R>30$:
(a) $\chi_r$, (b) $\chi_t$, (c) $\chi_a$, (d) $\chi_{tot}$. The 
solid histograms are the expected distributions obtained from a
sample of electrons from muon decay.}
\label{Fig. 23}
\end{figure}\newpage

\begin{figure}
\epsfbox{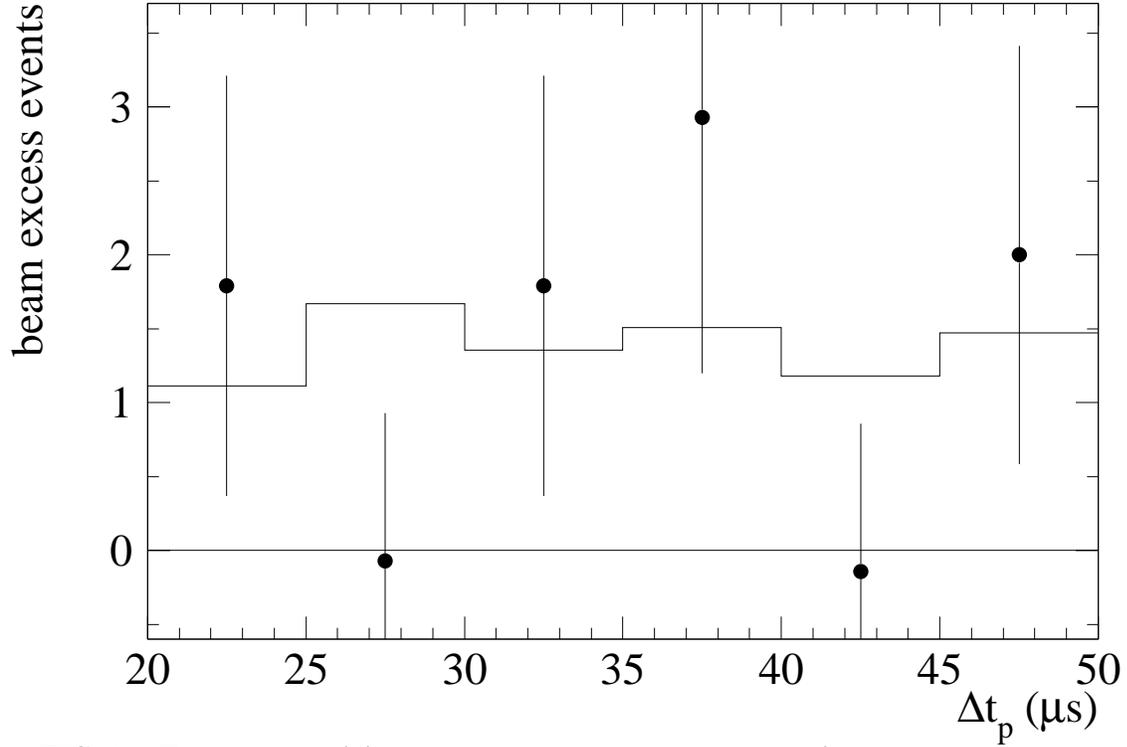}
\caption{Distribution of $\Delta t_p$, the time to the previous event,
for beam-excess data events with $36<E_e<60$ MeV and $R>30$ and
with activities within 50 $\mu$s. The 
solid histogram is the expected distribution obtained from a
sample of $\nu_e C \rightarrow e^-X$ scattering events.}
\label{Fig. 24}
\end{figure}\newpage

\begin{figure}
\epsfbox{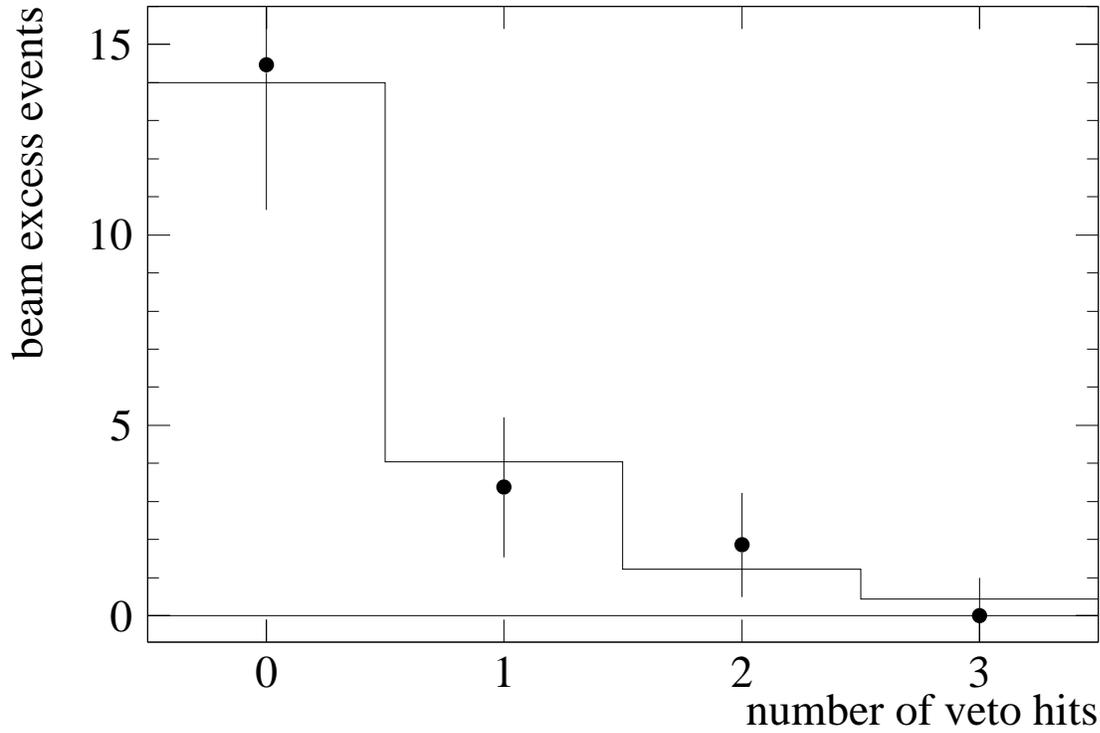}
\caption{Distribution of veto hits for beam-excess data events with 
$36<E_e<60$ MeV and $R>30$. The 
solid histogram is the expected distribution obtained from a
sample of $\nu_e C \rightarrow e^- X$ scattering events.}
\label{Fig. 25}
\end{figure}\newpage

\begin{figure}
\epsfbox{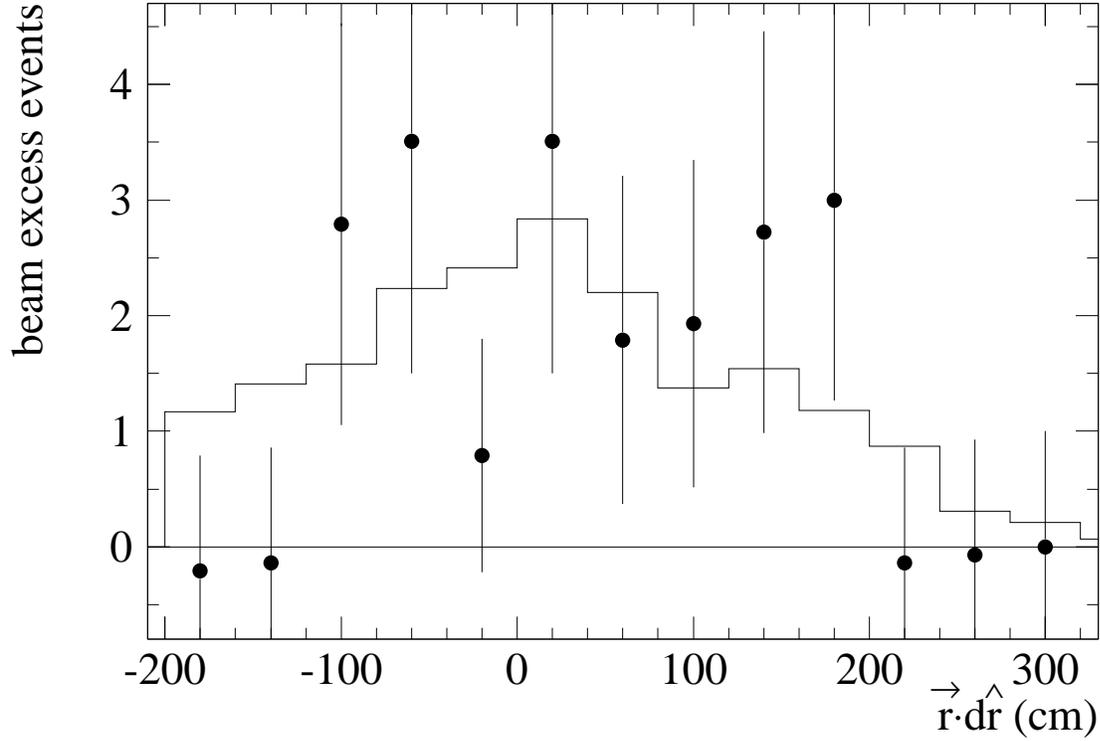}
\caption{Distribution of $\vec r \cdot \hat {dr}$ for beam-excess data 
events with $36<E_e<60$ MeV and $R>30$. The 
solid histogram is the expected distribution obtained from a
sample of $\nu_e C \rightarrow e^- X$ scattering events. The
$S>0.5$ cut eliminates all events with $\vec r \cdot \hat {dr}<-200$ cm.}
\label{Fig. 26}
\end{figure}\newpage

\begin{figure}
\epsfbox{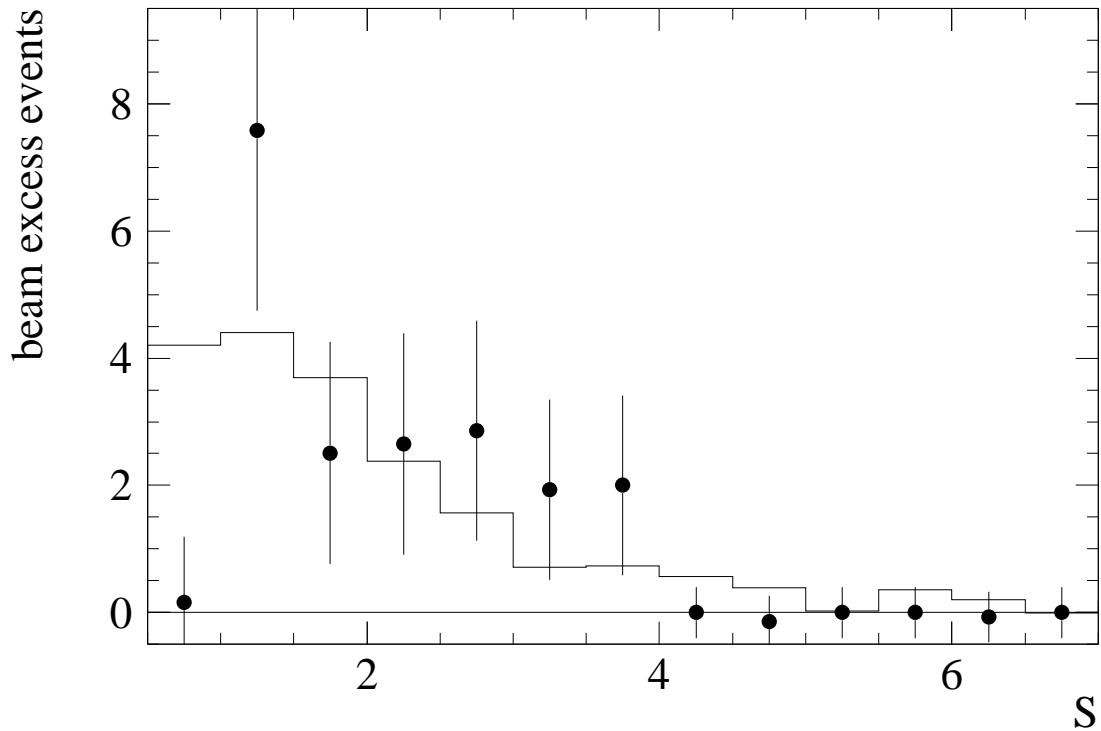}
\caption{Distribution of the likelihood ratio, S, 
for beam-excess data events with $36<E_e<60$ MeV and $R>30$. The 
solid histogram is the expected distribution obtained from a
sample of $\nu_e C \rightarrow e^- X$ scattering events.}
\label{Fig. 27}
\end{figure}\newpage

\begin{figure}
\epsfbox{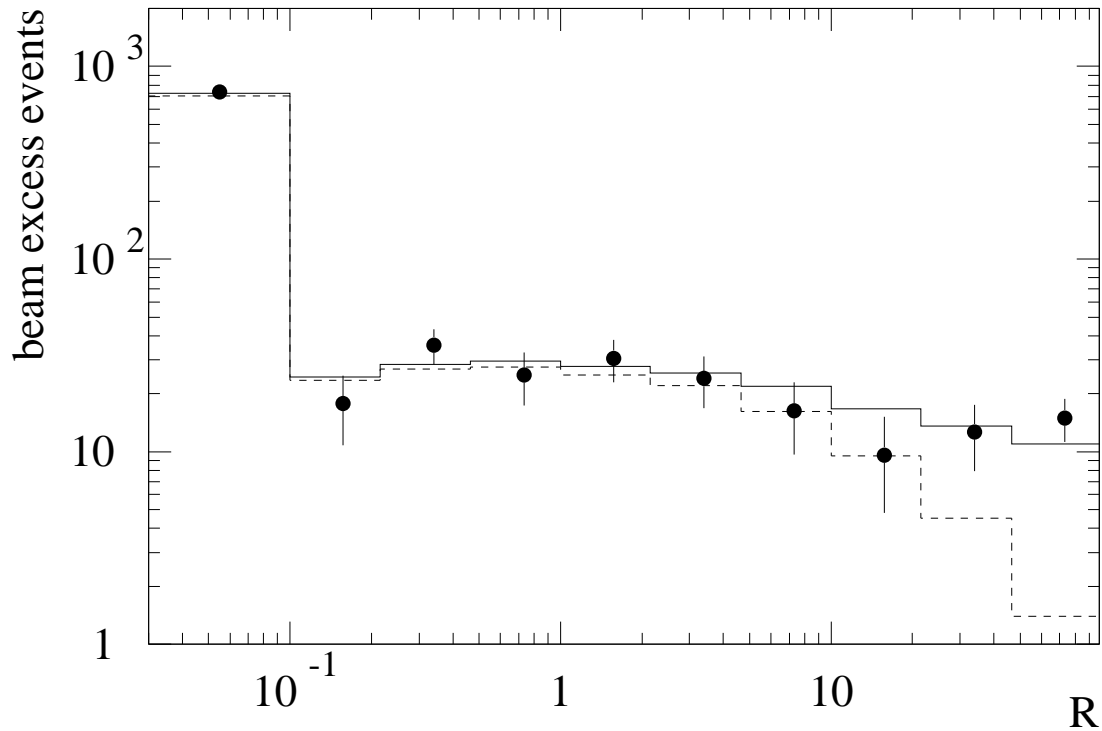}
\caption{The R distribution, beam on minus beam off excess, for
events that satisfy selection VI and that have energies in the range
$20<E_e<60$ MeV. The solid curve is the best fit to the data, while the
dashed curve is the component of the fit with an uncorrelated $\gamma$.}
\label{Fig. 28}
\end{figure}\newpage
   
\begin{figure}
\epsfbox{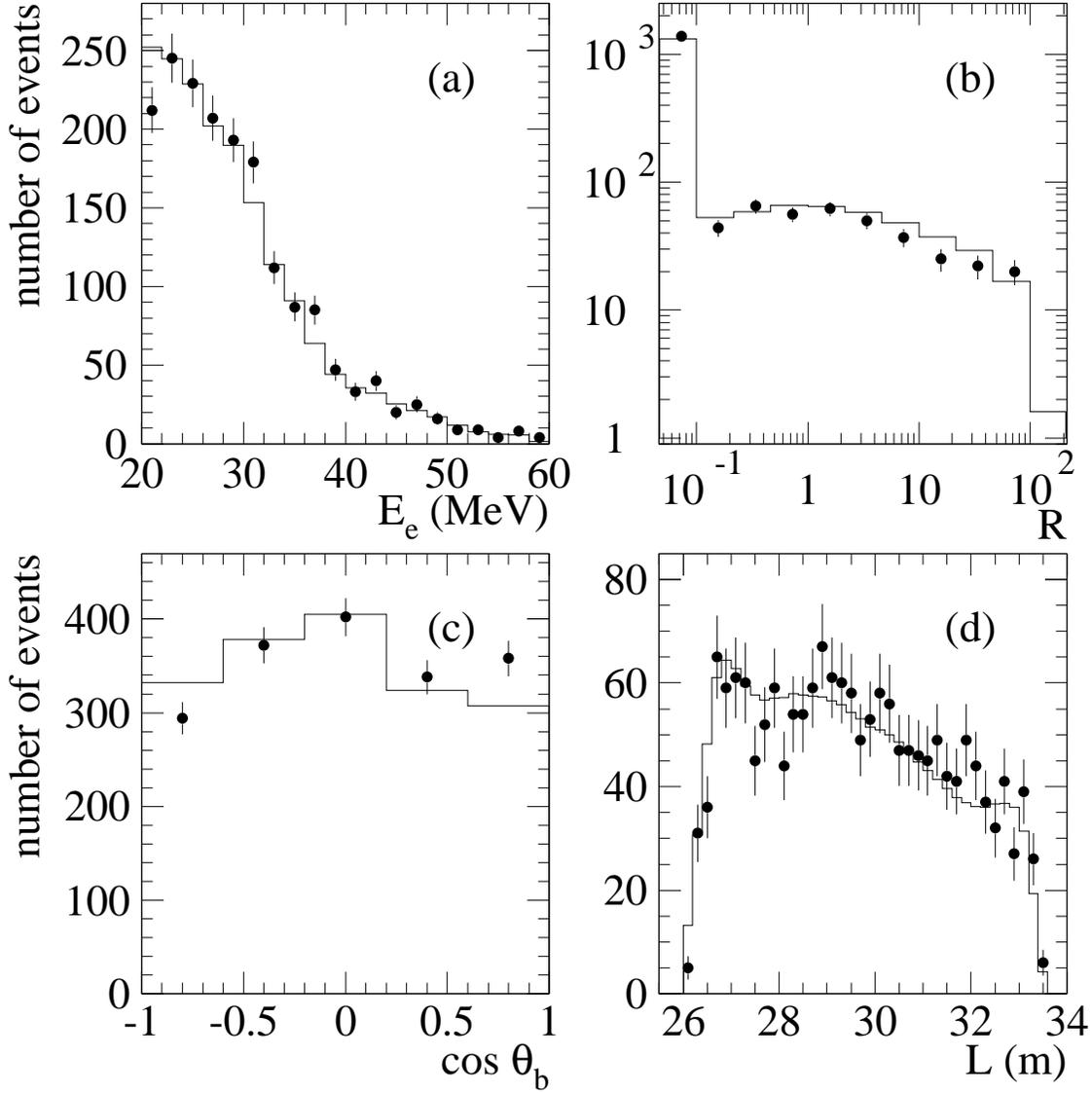}
\caption{Distributions of $E_{e}$, $R$, $\cos \theta_b$, and $L$ for 
the beam-on sample
compared with the expected distributions (including oscillations at $
19 eV^{2}, \sin^2 2\theta = 0.006$).}
\label{Fig. 29}
\end{figure}\newpage

\begin{figure}
\epsfbox{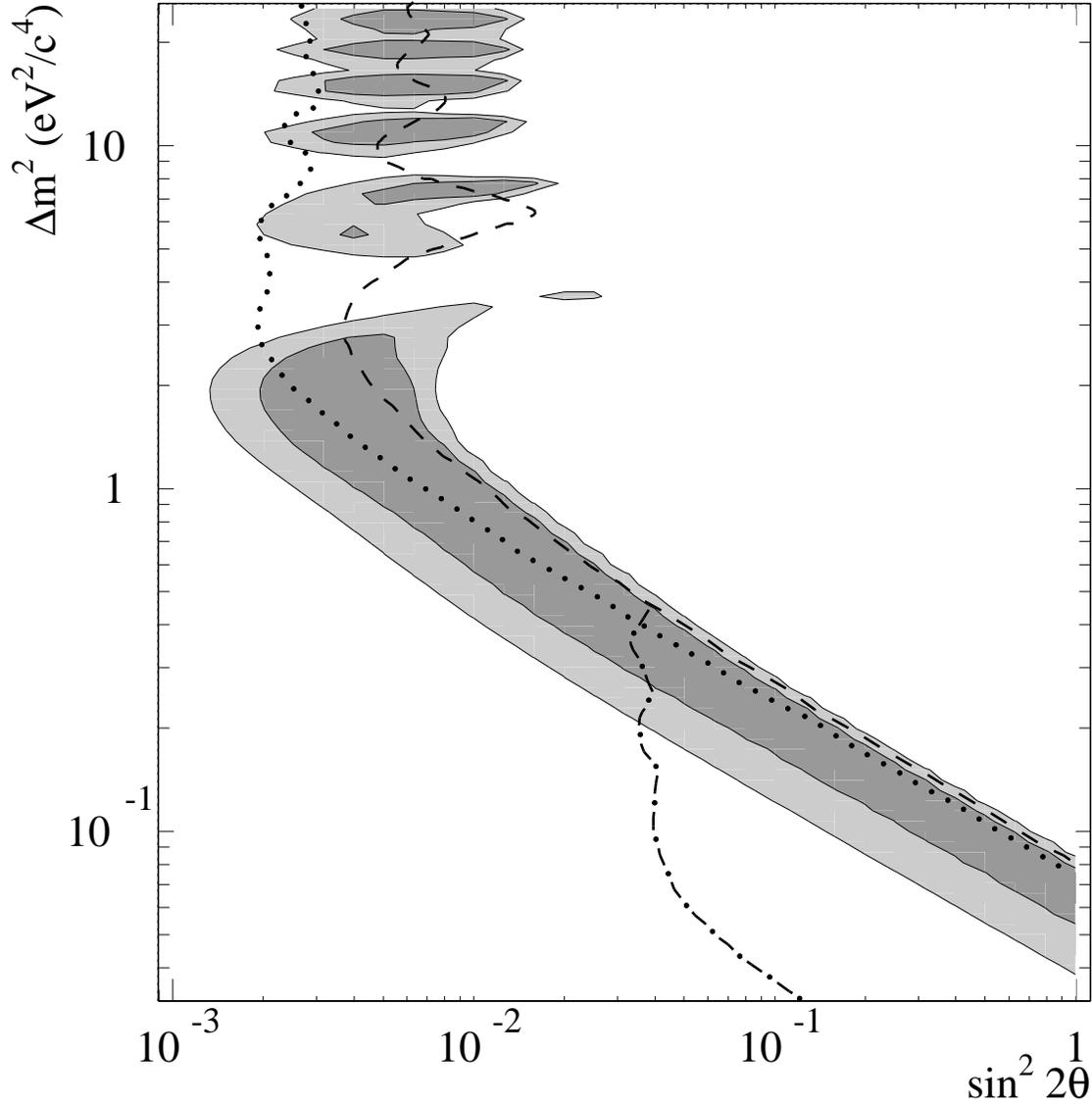}
\caption{Plot of the LSND $\Delta m^{2}~$ $\sl vs$ $\sin^2 2 \theta~$ 
favored regions.
The method used to obtain these contours is described in the text.
The darkly-shaded and lightly-shaded regions
correspond to 90\% and 99\% 
likelihood regions after the inclusion of the effects 
of systematic errors. Also shown are $90\%$ 
C.L. limits from KARMEN at ISIS (dashed curve), E776 at BNL (dotted curve),
and the Bugey reactor experiment (dot-dashed curve).}
\label{Fig. 30}
\end{figure}\newpage

\begin{figure}
\epsfbox{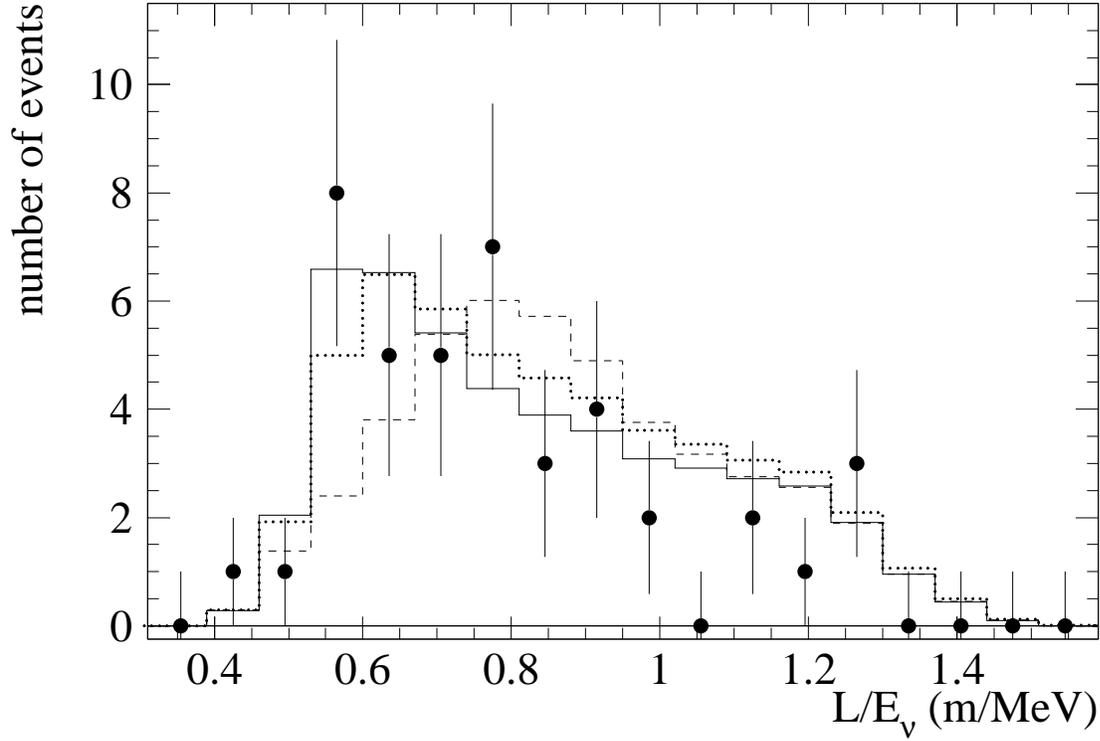}
\caption{Distribution of $L/E_{\nu}$ for the beam-on data with high R
compared with the expected distributions at $(19 eV^2, \sin^2 2\theta=0.006$: 
solid line),
$(4.3 eV^2, \sin^2 2\theta=0.01$: dashed line), and $(0.06 eV^2, 
\sin^2 2\theta=1.$: dotted line).}
\label{Fig. 31}
\end{figure}\newpage
 
\begin{figure}
\epsfbox{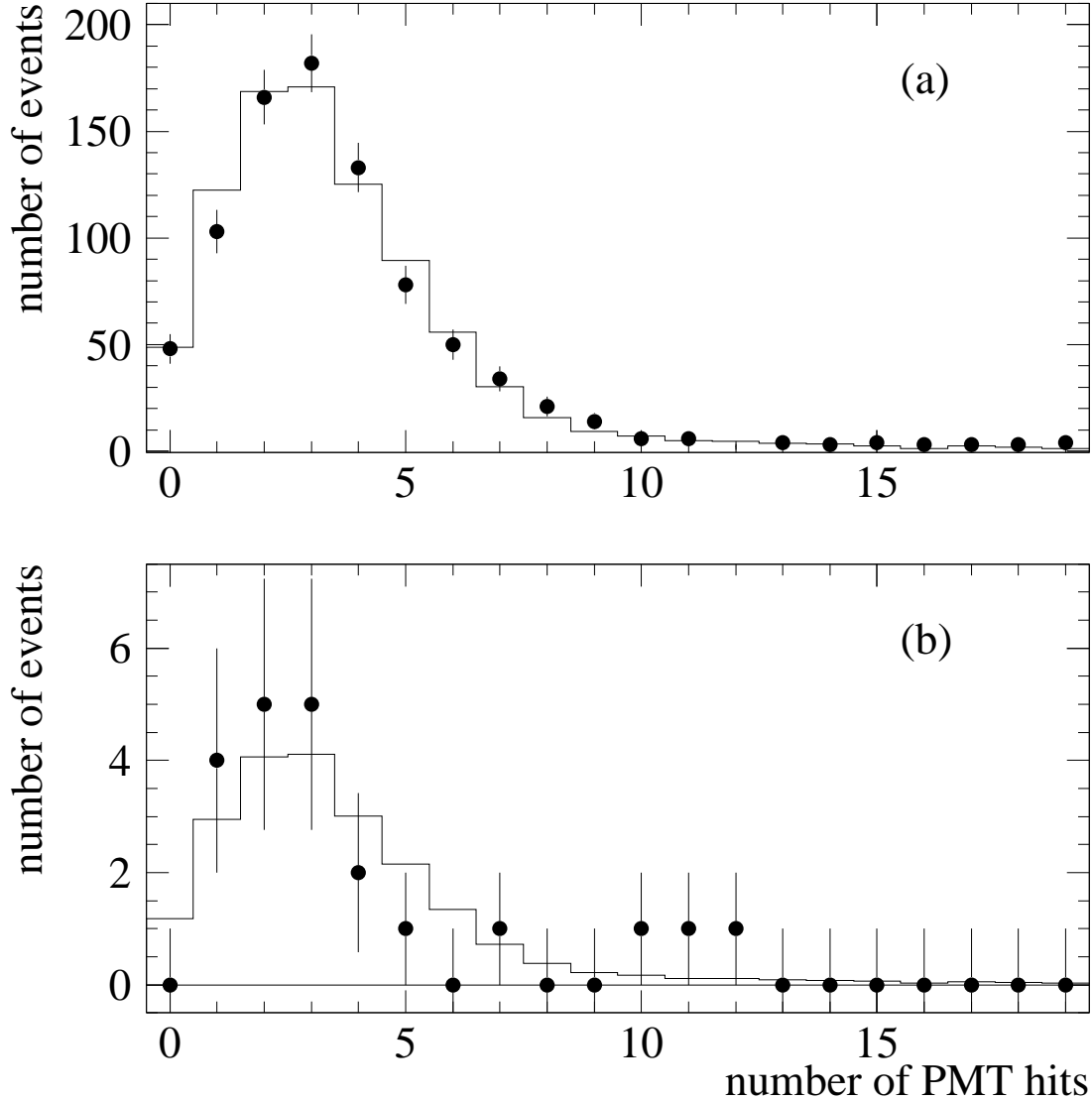}
\caption{The total number of hit
PMTs in the detector tank for the extra events that occur 
$0 - 3 \mu$s and $3 - 6 \mu$s
prior to oscillation candidate 
events. The candidates are in the $25<E_e<60$ MeV energy range
with (a) $R\ge0$ and (b) $R>30$. The data points are the beam-on 
events, while the solid curve is what is expected from random PMT hits
as determined from the sample of laser calibration events.}
\label{Fig. 32}
\end{figure}\newpage

\begin{figure}
\epsfbox{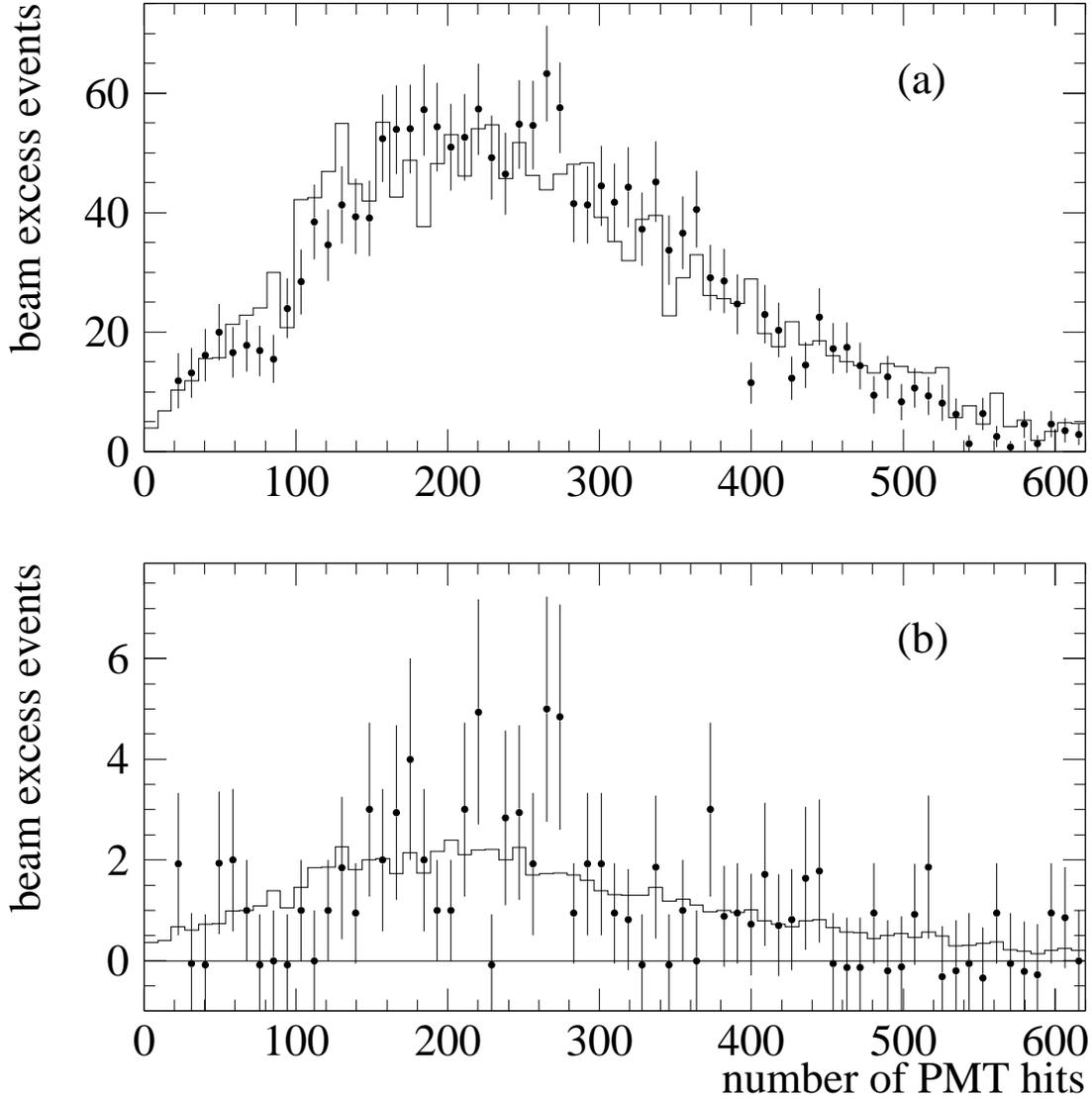}
\caption{The observed hit PMT distribution for all $\nu_{\mu} C$ 
scattering events
(including $\nu_{\mu} C \rightarrow \mu^- X$, $\bar \nu_{\mu} C 
\rightarrow \mu^+ X$,
and $\bar \nu_{\mu} p \rightarrow \mu^+ n$) for events with (a) $R\ge 0$ and
(b) $R>30$. The solid histogram in each case is the prediction from the
Monte Carlo simulation, normalized to the data.}
\label{Fig. 33}
\end{figure}\newpage

\begin{figure}
\epsfxsize6.5in
\epsfbox{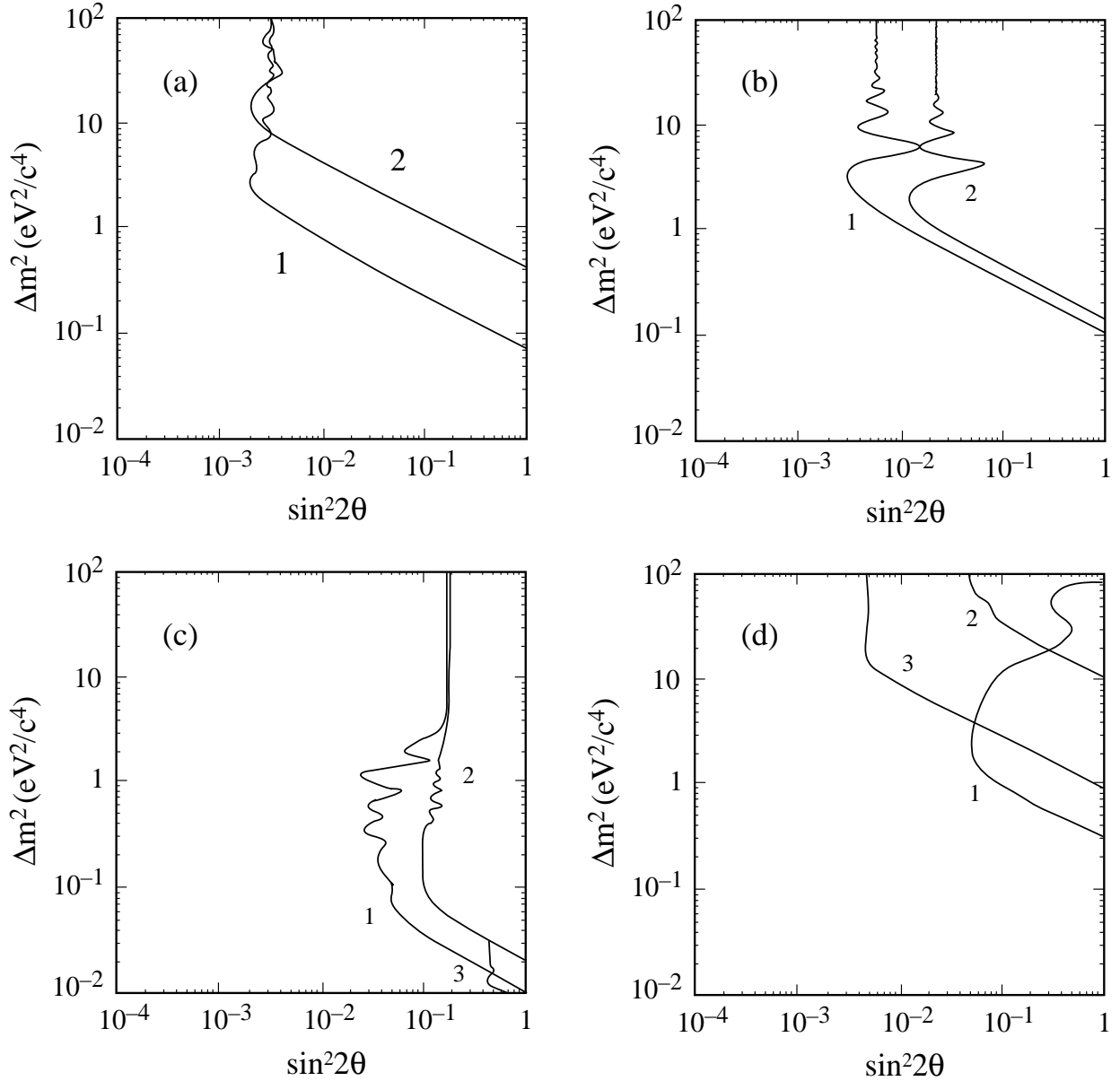}
\caption{Most sensitive limits on neutrino oscillations at 90\% C.L.
(a) $\nu_{\mu} \to \nu_e$ appearance
from the (1) E776 and (2) E734 experiments at BNL.
(b) $\bar\nu_{\mu} \to \bar\nu_e$ appearance
from
the (1) KARMEN and (2) E645 experiments.
(c) $\bar \nu_e$ disappearance
from the (1) Bugey,
(2) Gosgen and (3) Krasnoyarsk reactor experiments.
(d) $\nu_{\mu} $ disappearance
from the
(1) CDHS and (2) CCFR experiments. Also shown is the limit from the (3) E531
$\nu_{\mu} \rightarrow \nu_{\tau}$ appearance experiment.}
\label{Fig. 34}
\end{figure}


\newpage


\clearpage

\begin{references}

\bibitem{bigpaper1}
C.\ Athanassopoulos  {\it et.\ al.\ }, LA-UR-96-1327,
submitted to Phys.\ Rev.\ C.

\bibitem{paper1}
C.\ Athanassopoulos  {\it et.\ al.\ },
Phys.\ Rev.\ Lett. {\bf 75}, 2650 (1995).

\bibitem{bahcall}
J. N. Bahcall and M. M. Pinsonneault, Rev. Mod. Phys. {\bf 64}, 885 (1992).

\bibitem{davis}
R. Davis, Prog. Part. Nucl. Phys. {\bf 32}, 13 (1994).

\bibitem{Kamioka}
Y. Fukuda {\it et.\ al.\ }, Phys.\ Lett.\ B {\bf 335}, 237 (1994).

\bibitem{gallium}
J. N. Abdurashitov {\it et.\ al.\ }, Phys.\ Lett.\ B {\bf 328}, 234 (1994);
P. Anselmanni {\it et.\ al.\ }, Phys.\ Lett.\ B {\bf 328}, 377 (1994).

\bibitem{atmosphere}
R. Becker-Szendy {\it et.\ al.\ }, Phys. Rev. D{\bf 46}, 3720 (1992). 

\bibitem{atmoskam}
Y.\ Hirata {\it et.\ al.\ }, Phys. Lett. B {\bf 335}, 237 (1994).

\bibitem{soudan}
M.\ C.\ Goodman, Nucl. Phys. B (Proc. Suppl.) {\bf 38}, 337 (1995).

\bibitem{gaisser}
T.\ K.\ Gaisser, F.\ Halzen, and T.\ Stanev, Phys. Rep. {\bf 258}, 173 (1995);
G. Barr, T.K. Gaisser, and T. Stanev,
Phys. Rev. D {\bf 39}, 3532(1989);
S. Barr, T.K. Gaisser, S. Tilav, and P. Lipari,
Phys. Lett. B {\bf214}, 147 (1988).

\bibitem{fuller}
C.\ Cardall and G.\ M.\ Fuller, to appear in Phys. Rev. D.

\bibitem{burman} 
R.\ L.\ Burman, M.\ E.\ Potter, and E.\ S.\ Smith, 
Nucl. Instrum. Methods A{\bf 291}, 621 (1990);
R.\ L.\ Burman, A.\ C.\ Dodd, and P.\ Plischke,
Nucl. Instrum. Methods in Phys. Research A{\bf 368}, 416 (1996).

\bibitem{vogel}
C.\ H.\ Llewellyn Smith,
Physics Reports {\bf 3}, 262 (1972); 
P. Vogel, Phys. Rev. D {\bf 29}, 1918 (1984);
E.\ J.\ Beise and R.\ D.\ McKeown,
Comm. Nucl. Part. Phys. {\bf 20}, 105 (1991).

\bibitem{reeder}
R.\ A.\ Reeder {\it et.\ al.\ }, Nucl.\ Instrum.\ Methods A
 {\bf 334}, 353 (1993).

\bibitem{veto}
J. J. Napolitano {\it et.\ al.\ },
Nucl.\ Instrum.\ Methods A {\bf 274}, 152 (1989).

\bibitem{lsndmc}
K.\ McIlhany, {\it et.\ al.\ },
D.\ Whitehouse, A.\ M.\ Eisner, Y-X.\ Wang and D.\ Smith,
{\sl Computing in High Energy Physics '94\/}
(Proceedings of the Conference on
Computing in High Energy Physics
April 1994), LBL Report 35822, 357 (1995).

\bibitem{fuku2}
M.\ Fukugita, Y.\ Kohyama, and K.\ Kubodera,
Phys. Lett.\ B
{\bf 212}, 139 (1988).

\bibitem{albert}
M.\ Albert {\it et.\ al.\ },
Phys. Rev. C {\bf 51}, 1065 (1995).

\bibitem{pi0}
D.\ Rein and L.\ M.\ Sehgal,
Nucl. Phys. {\bf B223}, 29 (1983).

\bibitem{kolbe}
E.\ Kolbe, K.\ Langanke, and S.\ Krewald, Phys.\ Rev.\ C {\bf 49},
1122 (1994); (K.\ Langanke, private communication).
 
\bibitem{fuku} 
M. \ Fukugita, Phys.\ Rev.\ C {\bf 41}, 1359 (1990).

\bibitem{karmen}
B.\ Bodmann {\it et.\ al.\ }, Phys.\ Lett.\ B {\bf 267}, 321 (1991);
B.\ Bodmann {\it et.\ al.\ }, Phys.\ Lett.\ B {\bf 280}, 198 (1992);
B.\ Zeitnitz {\it et.\ al.\ },
Prog. Part. Nucl. Phys., {\bf 32} 351 (1994).

\bibitem{wonyong}
L.\ Borodovsky  {\it et.\ al.\ },
Phys.\ Rev.\ Lett. {\bf 68}, 274 (1992).

\bibitem{bugey}
B.\  Achkar {\it et.\ al.\ }, Nucl. Phys. {\bf B434}, 503 (1995).

\bibitem{e645}
S.\ J.\ Freedman {\it et.\ al.\ },
Phys. Rev. D {\bf 47}, 811 (1993).

\bibitem{karmen2}
K. Eitel, thesis, Universitat und Forschungszeutrum Karlsruhe
(1995); Forschungszeutrum Karlsruhe report FZKA 5684 (1995).

\bibitem{e734}
L. \ A.\ Ahrens {\it et.\ al.\ },
Phys. Rev. D {\bf 31}, 2732 (1985).

\bibitem{e816}
P. Astier {\it et.\ al.\ },
Nucl.\ Phys. {\bf B335}, 517 (1990).

\bibitem{e225}
R.\ C.\ Allen {\it et.\ al.\ },
Phys.\ Rev.\ D {\bf 47}, 11 (1993).

\bibitem{ccfr}
K.\ S.\ McFarland {\it et.\ al.\ }, Phys.\ Rev.\ Lett. {\bf 75}, 3993 (1995);
C.\ Haber {\it et.\ al.\ }, Phys.\ Rev.\ Lett. {\bf 52}, 1384 (1984).

\bibitem{gosgen}
G.\  Zacek {\it et.\ al.\ },
Phys. Rev. D {\bf 34}, 2621 (1986).

\bibitem{krasnoyarsk}
G.\ S.\ Vidyakin {\it et.\ al.\ },
JETP Letters {\bf 59}, 364 (1984).

\bibitem{cdhs}
F.\ Dydak {\it et.\ al.\ }, Phys.\ Lett.\ B {\bf 134}, 281 (1984).

\bibitem{reay}
N.\ Ushida {\it et.\ al.\ }, Phys.\ Rev.\ Lett. {\bf 57}, 2897 (1986).


\end{references}
\end{document}